\documentclass[aps, showpacs, showkeys, nofootinbib, floatfix, superscriptaddress]{revtex4}

\usepackage{amsfonts}
\usepackage{amssymb}
\usepackage{amsmath}
\usepackage{graphicx}
\usepackage{float}
\usepackage{epstopdf}
\usepackage{soul}
\usepackage{makecell}
\usepackage{hyperref}
\usepackage{multirow}  
\usepackage{subcaption}
\usepackage{soulpos}       
\usepackage{xcolor}        
\usepackage{amsmath}     
\usepackage{hyperref}     

\begin{document}

\title{Particle dynamics and optical appearance of charged spherically symmetric black holes in bumblebee gravity}
\author{Mou Xu}
\affiliation{Department of Physics, Liaoning Normal University, Dalian 116029, P. R. China}
\author{Jianbo Lu}
\email{lvjianbo819@163.com}
\affiliation{Department of Physics, Liaoning Normal University, Dalian 116029, P. R. China}
\author{Ruonan Li}
\affiliation{Department of Physics, Liaoning Normal University, Dalian 116029, P. R. China}
\author{Yu Liu}
\affiliation{Department of Physics, Liaoning Normal University, Dalian 116029, P. R. China}
\author{Shu-Min Wu}
\affiliation{Department of Physics, Liaoning Normal University, Dalian 116029, P. R. China}

\begin{abstract}

In this paper, we study the particle dynamics, shadow, and optical appearance of charged black holes (BHs) in bumblebee gravity. Firstly, we find that the Lorentz-violation parameter $l$ and charge parameter $Q$ have opposite effects on the peak of the effective potential by analyzing timelike geodesics, and the radius of the innermost stable circular orbit (ISCO) decreases as the BH parameters $l$ and $Q$ increase. We also explore the behaviors of particle energy, angular momentum, and Keplerian frequency. Secondly, for null geodesics, both the photon sphere radius and the shadow radius decrease with increasing $l$ and $Q$, and are consistently smaller than those of the Reissner–Nordström black hole (RNBH) and Schwarzschild-like BH. And based on observational data reported by the Event Horizon Telescope (EHT) Collaboration, we constrain the parameters $l$ and $Q$ by using the shadow radius data of Sgr A*. Thirdly, we explore the observation characteristics of charged BHs under three thin disk accretion models. The results show that, compared to RNBH, the increase of $l$ leads to a greater thickness of the photon rings and lensed rings. However, due to their extremely narrow ranges, the contributions are small, and the observed intensities are mainly contributed by the direct emissions. Moreover, when the parameter $l$ is fixed, the peaks of the observed intensities of rings decrease with increasing $Q$ for the same emission model, and it is always lower than the corresponding value for Schwarzschild-like BH. These findings contribute to distinguishing bumblebee charged black holes (BCBHs) from other types of BHs based on their optical appearance.
\end{abstract}

\keywords{bumblebee gravity; black hole; particle dynamics; optical appearance}
\maketitle

\section{$\text{Introduction}$}

Einstein's General Relativity (GR), established in 1915 as the standard model of gravity, has received extensive experimental support. In recent years, the detection of gravitational waves by the Laser Interferometer Gravitational-Wave Observatory (LIGO) \cite{1GW} and the release of BH images by the EHT collaboration \cite{1BHshadow1,1BHshadow2,1BHshadow3,1BHshadow4,1BHshadow5,1BHshadow6} have provided substantial support for this theory. However, it also faces some significant challenges, such as its inability to address gravitational phenomena at the quantum level and the lack of renormalizability in high energy regimes. In contrast, the Standard Model of particle physics successfully describes fundamental particles and their interactions at the quantum level. In order to solve the problem of quantizing gravity, researchers have constructed several modified gravity theories, including loop quantum gravity \cite{1loop1,1loop2,1loop3}, non-commutative field theory \cite{1non1,1non2,1non3}, and string theory \cite{1st1,1st2,1st3}. Although the properties of these gravity theories can only be directly probed at the Planck scale, some quantum gravity theories suggest that Lorentz symmetry breaking (LSB) may manifest at lower energy scales \cite{1LIV1,1LIV2}. Experimental and observational studies related to LSB have also received much attention, which provide possible ways to probe quantum gravity under current technological conditions \cite{1OE1,1OE2,1OE5,1OE6}.

Studying LSB allows us to explore the nature of spacetime from multiple perspectives. The concept of spontaneous LSB originated from string theory \cite{1st1,1LIV1}. Based on this idea, Colladay and Kostelecky proposed a general framework to study spontaneous LSB \cite{2.3,2.4}, i.e., the Standard Model Extension (SME) \cite{1LIV1,2.3,2.4}, which is an effective field theory framework. The SME has been extensively studied \cite{2.5,2.6,2.7,2.8,2.9,2.12}, providing an important research foundation for probing gravity and LSB. In this context, Kostelecky and Samuel proposed the bumblebee model \cite{1st1,2.15}, which is one of the simplest field theory models and has been widely studied both in curved spacetime and Minkowski spacetime \cite{2.16,2.17,2.18,2.20,2.21,2.23}. In the bumblebee gravity framework, a vector field $B_{\mu}$ called the bumblebee field is introduced. This vector field acquires a non-zero vacuum expectation value through a self-interaction potential $V(B^{\mu}B_{\mu})$, which generates preferred directionality in spacetime and triggers spontaneously breaking local Lorentz invariance for particles \cite{1LIV1,2.24}. Within this theoretical framework, Casana et al. considered the non-minimal coupling of the bumblebee vector field with the Ricci scalar and constructed an exact Schwarzschild-like BH solution \cite{1LIV2}. Since then, researchers have obtained several exact BH solutions within bumblebee gravity framework, including (anti-) de Sitter Schwarzschild BH solutions \cite{2.26}, static spherically symmetric charged BH solutions \cite{2.27}, slowly rotating Kerr-like BH solutions \cite{2.28}, the solutions with global monopole \cite{2.29}, and with Gauss-Bonnet terms \cite{2.30}. Meanwhile, the related wormhole solutions were also discussed in \cite{2.31}. Various BH properties have been extensively investigated within the bumblebee gravity framework, including thermodynamic properties \cite{add1,2.32,2.33}, Hawking radiation \cite{2.34}, field propagation and quasi-normal modes \cite{2.35,2.36,add2}, gravitational waves \cite{2.37}, shadows \cite{2.26,2.38,2.39}, gravitational lensing \cite{2.40}, particle motion \cite{add3} and greybody factors \cite{2.42,2.43} etc. These astrophysical phenomena provide potential experimental avenues for studying LSB. Recent studies have shown that introducing the non-minimal coupling between the bumblebee vector field and matter fields can lead to physically meaningful BH solutions \cite{2505.01374}. In particular, considering that compact astrophysical objects with charge are common, \cite{2.27} investigated such coupling between the bumblebee field and the electromagnetic field within the framework of bumblebee gravity. The coupling provides source terms and leads to significant modifications of Maxwell equation \cite{2503.02323}. In this context, we focus on a static, spherically symmetric charged BH solution recently obtained in \cite{2.27} within bumblebee gravity. Various modified gravity theories have been proposed from different physical motivations, and studying phenomena in strong gravitational fields is considered one of the key approaches for testing their feasibility. In recent years, several methods have been developed to distinguish between different gravity theories. For instance, gravitational lensing has been widely employed for this purpose \cite{GL1,GL2}. Especially, \cite{2408.02195,2408.02195ref} proposed a global Gaussian bending measure based on differential geometry, rather than on specific gravity theories, offering a novel perspective for distinguishing between gravity theories. In addition, the optical appearance of BHs can also serve as a discriminating tool \cite{BHS1,BHS2}. Thus, this paper further investigates the particle dynamics and optical appearance of charged static spherically symmetric BHs in bumblebee gravity theory, in order to gain a deeper understanding of LSB related properties.

In recent years, the studies of test particle dynamics around BHs have played an important role in verifying gravitational theories and explaining astrophysical phenomena, providing crucial information for investigating the gravitational field around BHs. The nature of spacetime geometry in the vicinity of BHs strongly affects the geodesic structures of massive particles and photons, which changes some observable features \cite{3.1,3.5,3.9,3.12,3.18,3.19}, such as the orbital features of massive particles, the radius of photon rings, and the size of the BH shadow. Null geodesics are fundamental to studying phenomena such as BH shadows, gravitational lensing, and optical imaging. Timelike geodesic, on the other hand, provide a theoretical basis for understanding the structure and evolution of BH accretion disks. Extensive research has been conducted on these properties, and significant progress has been made \cite{3.24,3.25,3.35,3.36,3.39,3.45,3.49,3.50}. Furthermore, the presence of external magnetic fields or other matter fields can also lead to changes in the geodesic structure of test particles, and thus the motion of test particles in different gravitational frameworks has become one of the hot topics of current research \cite{3.52,3.54,3.56}. These researches provide possibilities for distinguishing GR and modified gravity theories in observations. Specifically, LSB may lead to significant changes in BH dynamics, which affects the spacetime structure around the BHs. Based on the above background, we explore the timelike and null geodesic structures of charged BHs within bumblebee gravity, in order to better understand the spacetime structure around BHs.

The supermassive BH images observed by the EHT consist of a dark central region and a bright ring, which are commonly referred to as the "shadow" and "photon ring" of BH \cite{4.1,4.2}. The optical appearance of BH provides a valuable window for a deeper understanding of these mysterious compact objects. Shadow images of BHs can provide information about the geometric structure and physical properties near the BH event horizon \cite{4.3,4.4}. The photon ring, formed by photons moving around the BH, is a crucial component of the BH shadow and can be used to explore the strong gravitational field effects of BHs \cite{4.1}. In recent years, methods to analyze BH shadows and photon rings have been proposed \cite{4.1,4.6,4.7}, which can distinguish various BHs under different gravity theories. It is generally believed that luminous accretion flows around BHs are essential for obtaining BH images. Bardeen was the first to systematically analyze the optical properties of supermassive objects and study the BH shadow and photon ring \cite{4.7}. Subsequently, the optical appearance of various static BHs has also been widely discussed, such as hairy BHs \cite{4.10}, Schwarzschild-MOG BHs \cite{4.11}, and so on. Studies on the shadows of rotating BHs can be found in the literature \cite{4.12,4.13,4.14,4.16}. Given the challenges faced by GR, it makes sense to explore the nature of gravity under different theories. BH shadows in some modified gravity have been investigated, such as charged BHs in Rastall theory \cite{BHS1}, charged BHs in Horndeski theory \cite{BHS2}, and magnetically charged BHs in Born-Infeld electrodynamics theory \cite{4.19}.

The structure of this paper is arranged as follows: Section 2 briefly reviews the static spherically symmetric charged BH solutions within bumblebee gravity framework and discusses the effects of the LSB parameter and charge parameter on the BH event horizon radius. In section 3, we derive the timelike and null geodesics under bumblebee gravity and use the geodesics to analyze the spacetime structure in the vicinity of the BH, focusing on properties such as the circular orbits of massive particles and the photon ring radius. Furthermore, we constrain the BCBH model parameters using observational data released by the EHT collaboration. Assuming the existence of an optically thin accretion disk around the BH, in section 4, we investigate the effects of three types of thin disk accretion models on the optical appearance of the BCBHs. Finally, section 5 contains the main conclusions of this study. We adopt the metric signature as $(-+++)$ and use geometric units with $G=c=M=1$ in the calculations throughout the paper.

\section{$\text{Static spherically symmetric charged black hole solutions in bumblebee gravity}$}

In this section, we briefly describe the static spherically symmetric charged BH solutions under the bumblebee gravity obtained in \cite{2.27}. In the bumblebee model, GR is extended by introducing a vector field $B_{\mu}$ with a non-zero vacuum expectation value. The dynamics term of this vector field leads to the spontaneous breaking of Lorentz symmetry. Specifically, the bumblebee field nonminimally couples with gravity and its action is given by \cite{2.27,1LIV1}:
\begin{equation}
S=\int d^{4}x\sqrt{-g}\left[\frac{1}{2\kappa}(R-2\Lambda)+\frac{\xi}{2\kappa}B^{\mu}B^{\nu}R_{\mu\nu}-\frac{1}{4}B_{\mu\nu}B^{\mu\nu}-V\left(B^{\mu}B_{\mu}\pm \bar{b}^{2}\right)\right] + \int d^{4}x\sqrt{-g}\mathcal{L}_{M}, 
\label{eq21}
\end{equation}
where $\Lambda$ is the cosmological constant, $\kappa = \frac{8 \pi G}{c^4}$ is the gravitational coupling constant, and $\xi$ represents the non-minimal coupling constant between gravity and the bumblebee field. The bumblebee field strength $B_{\mu \nu}$ is defined as $B_{\mu \nu} = \partial_{\mu} B_{\nu}-\partial_{\nu} B_{\mu}$. The potential $V$ provides a non-zero vacuum expectation value to the vector field $B_{\mu}$, thereby resulting in the spontaneous LSB in the gravitational sector. Consider the potential function of the form $V(B^{\mu} B_{\mu} \pm \bar{b}^2)$, where $\bar{b}^2$ is a real number, and the potential $V$ must attain a minimum when $B^{\mu} B_{\mu} \pm \bar{b}^2 = 0$. Consequently, the bumblebee field acquires a non-zero vacuum expectation value $\langle B_{\mu} \rangle = \bar{b}_{\mu}$, where $\bar{b}_{\mu}$ is a function of spacetime coordinates, satisfying $\bar{b}^{\mu} \bar{b}_{\mu} = \mp \bar{b}^2 = \text{constant}$. For convenience, we define $X = B^{\mu} B_{\mu} \pm \bar{b}^2$ and $V' = \frac{\partial V}{\partial X}$.

In order to obtain the charged BH solution within the framework of bumblebee gravity, the electromagnetic field, which is considered to be non-minimally coupled to the bumblebee vector field, is the matter field. The Lagrangian density expression is given by \cite{actionLm}:
\begin{equation}
\mathcal{L}_{M}=\frac{1}{2 \kappa}\left(F^{\mu \nu} F_{\mu \nu}+\gamma B^{\mu} B_{\mu} F^{\alpha \beta} F_{\alpha \beta}\right),
\label{eq22}
\end{equation}
where $F_{\mu \nu} = \partial_{\mu} A_{\nu} - \partial_{\nu} A_{\mu}$ is the electromagnetic field tensor, with $A = (\phi(r), 0, 0, 0)$, and $\gamma$ is the coupling constant between the electromagnetic field and the vector field. By varying the action (\ref{eq21}) with respect to the metric tensor $g_{\mu \nu}$, the gravitational field equations within bumblebee gravity framework are obtained:
\begin{equation}
G_{\mu \nu}+\Lambda g_{\mu \nu}=\kappa T_{\mu \nu}^{B}+\kappa T_{\mu \nu}^{M},
\label{eq23}
\end{equation}
where
\begin{equation}
\begin{aligned}
T_{\mu \nu}^{B}=&\frac{\xi}{\kappa}\left[\frac{1}{2} B^{\alpha} B^{\beta} R_{\alpha \beta} g_{\mu \nu}-B_{\mu} B^{\alpha} R_{\alpha \nu}-B_{\nu} B^{\alpha} R_{\alpha \mu}+\frac{1}{2} \nabla_{\alpha} \nabla_{\mu}\left(B^{\alpha} B_{\nu}\right)+\frac{1}{2} \nabla_{\alpha} \nabla_{\nu}\left(B^{\alpha} B_{\mu}\right)-\frac{1}{2} \nabla^{2}\left(B_{\mu} B_{\nu}\right)\right.\\
&\left.-\frac{1}{2} g_{\mu \nu} \nabla_{\alpha} \nabla_{\beta}\left(B^{\alpha} B^{\beta}\right)\right]+ 2 V^{\prime} B_{\mu} B_{\nu}+B_{\mu}{ }^{\alpha} B_{\nu \alpha}-\left(V+\frac{1}{4} B_{\alpha \beta} B^{\alpha \beta}\right) g_{\mu \nu},
\end{aligned}
\label{eq24}
\end{equation}
\begin{equation}
T_{\mu \nu}^{M}=\frac{1}{ \kappa}\left[\left(1+\gamma \bar{b}^{2}\right)\left(2 F_{\mu \alpha} F_{\nu}^{\alpha}-\frac{1}{2} g_{\mu \nu} F^{\alpha \beta} F_{\alpha \beta}\right)+\gamma B_{\mu} B_{\nu} F^{\alpha \beta} F_{\alpha \beta}\right].
\label{eq25}
\end{equation}
Similarly, by varying the action with respect to the bumblebee vector field and the electromagnetic field, the corresponding field equations are derived as, respectively:
\begin{equation}
\nabla_{\mu} B^{\mu \nu}-2\left(V^{\prime} B^{\nu}-\frac{\xi}{2 \kappa} B_{\mu} R^{\mu \nu}-\frac{1}{2 \kappa} \gamma B^{\nu} F^{\alpha \beta} F_{\alpha \beta}\right)=0,
\label{eq26}
\end{equation}
\begin{equation}
\nabla_{\mu}\left(F^{\mu \nu}+\gamma B^{\alpha} B_{\alpha} F^{\mu \nu}\right)=0.
\label{eq27}
\end{equation}

\begin{figure}[H]
\centering
\includegraphics[width=8cm]{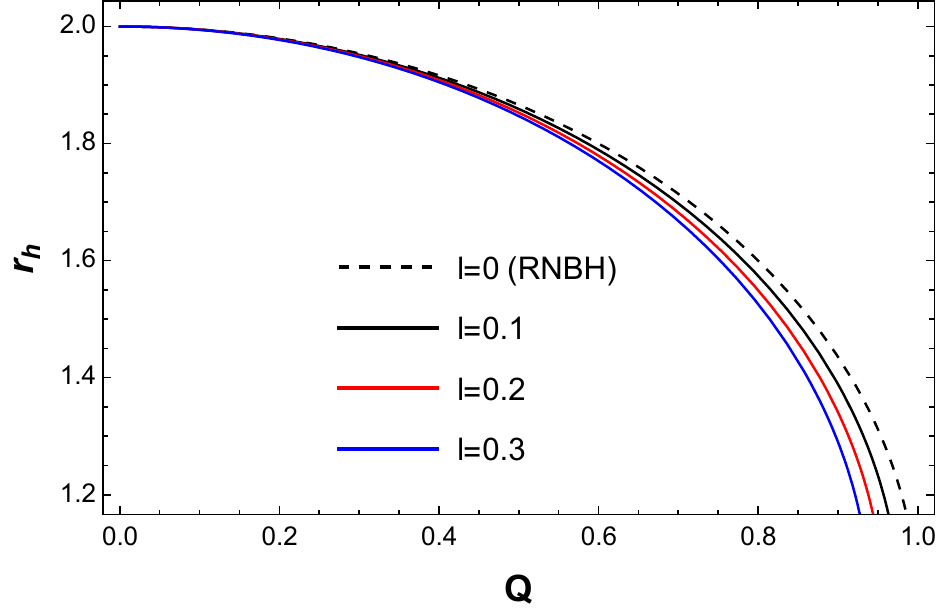}
\includegraphics[width=8cm]{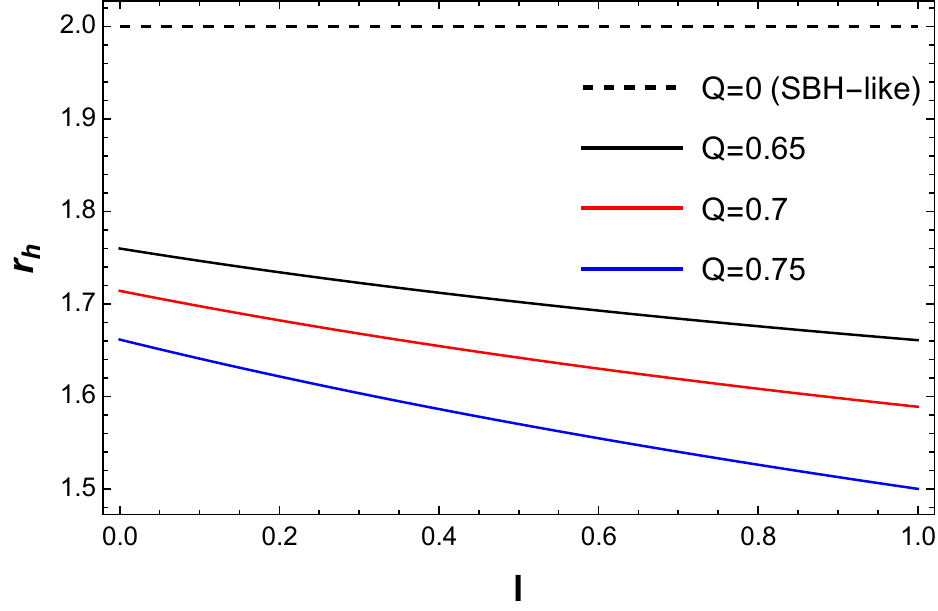}
\caption{The variation of event horizon radius $r_h$ of BCBHs with respect to the parameters $l$ and $Q$.}
\label{fig22}
\end{figure}

In the static spherically symmetric spacetime, the line element can be expressed as:
\begin{equation}
ds^{2}=-A(r)dt^{2}+B(r)dr^{2}+r^{2}d\theta^{2}+r^{2}\sin^{2}\theta d\phi^{2}.
\label{eq28}
\end{equation}
Literature \cite{1LIV2} derived the vacuum BH solution in the framework of the bumblebee theory. The \cite {2.27} considers a specific potential function form and calculates the charged BH solution under this theoretical framework as follows:
\begin{equation}
A(r)=1-\frac{2M}{r}+\frac{2(1+l)Q^{2}}{(2+l)r^{2}},
\label{eq29}
\end{equation}
\begin{equation}
B(r)=\frac{1+l}{1-\frac{2M}{r}+\frac{2(1+l)Q^{2}}{(2+l)r^{2}}},
\label{eq210}
\end{equation}
\begin{equation}
\phi (r)=\frac{Q}{r},
\label{eq211}
\end{equation}
where $Q$ is the charge of the BH and $l=\xi \bar{b}^2$ is the LV parameter. Throughout this paper, we adopt geometric units ($G=c=M=1$), so both of these parameters $Q=Q/M$ and $l=l/M$ are dimensionless. When $l \rightarrow 0 $, the solution degenerates to the standard RNBH solution. Clearly, the spacetime described by this solution is not asymptotically flat. Moreover, by setting $A(r)=0$, we obtain the expression for the event horizon radius $r_{h}$ of the BCBH as:
\begin{equation}
r_{h}=\frac{4 M+2 l M+\sqrt{(-4 M-2 l M)^{2}-4(2+l)\left(2 Q^{2}+2 l Q^{2}\right)}}{2(2+l)}.
\label{eq212}
\end{equation}
It is obvious that the charge parameter $Q$ and the LV parameter $l$ must satisfy the condition $Q^2 \leq \frac{2+l}{2(1+l)}$. Using Eq.(\ref{eq212}) and taking some values of BH parameters $l$ and $Q$, Fig.\ref{fig22} shows the dependence of the event horizon radius of the BCBH on the LV parameter $l$ and the charge $Q$. One can notice that the values for the parameters $l$ and $Q$ used in the numerical calculations throughout this paper are all consistent with the constraints derived in Section III.B, based on the shadow radius data of Sgr A* reported by the EHT Collaboration. According to Fig.\ref{fig22}, we can see that the charge parameter $Q$ has a more significant effect on the BH event horizon radius compared to the LV parameter $l$. Moreover, in contrast to the Schwarzschild-like BH and RNBH, the event horizon radius of the BCBH decreases with increasing parameters $l$ and $Q$, and is always smaller than the corresponding values for the Schwarzschild-like BH and RNBH.

\section{$\text{The geodesic structure of test particles in bumblebee spacetime}$}

This section focuses on properties related to the test particles dynamics in the vicinity of a charged BH in the framework of bumblebee gravity. In the four-dimensional spacetime described by Eqs. (\ref{eq28}-\ref{eq211}), the behavior of the test particle can be expressed by the following Lagrangian function \cite{Lag}:
\begin{equation}
\mathcal{L}=\frac{1}{2} g_{\mu\nu } \dot{x}^{ \mu } \dot{x}^{ \nu }=\frac{1}{2}\left[-\left(1-\frac{2M}{r}+\frac{2(1+l)Q^{2}}{(2+l)r^{2}}\right)\dot{t}^{2}+\left(\frac{1+l}{1-\frac{2M}{r}+\frac{2(1+l)Q^{2}}{(2+l)r^{2}}}\right)\dot{r}^{2}+r^2\dot{\theta}^2+r^{2}sin^2\theta \dot{\phi}^{2}\right],
\label{eq31}
\end{equation}
where $\dot{x}^{\mu}$ represents the four-velocity of test particle, and ''dot'' denotes the derivative of the general coordinate with
respect to the affine parameter $\lambda$. Due to the spherical symmetry of the spacetime geometry, we consider particles moving on the equatorial plane without loss of generality, where $\theta = \pi/2$ and $\dot{\theta} = 0$. Additionally, in static spherically symmetric spacetime, the Lagrangian of the particle is independent of time $t$ and azimuthal angle $\phi$. Therefore, we can derive two corresponding conserved quantities along with the motion of particles in bumblebee spacetime:
\begin{equation}
E=-\frac{\partial \mathcal{L}}{\partial \dot{t}}=\left(1-\frac{2M}{r}+\frac{2(1+l)Q^{2}}{(2+l)r^{2}}\right) \dot{t},
\label{eq32}
\end{equation}
\begin{equation}
L=-\frac{\partial \mathcal{L}}{\partial \dot{\phi}}=r^{2} \dot{\phi}.
\label{eq33}
\end{equation}
Here, $E$ and $L$ represent the energy and angular momentum of test particle, respectively. Based on the normalization condition $g_{\mu\nu} \dot{x}^{\mu} \dot{x}^{\nu} = \epsilon$, we can further derive:
\begin{equation}
-\left(1-\frac{2M}{r}+\frac{2(1+l)Q^{2}}{(2+l)r^{2}}\right)\dot{t}^2+\left(\frac{1+l}{1-\frac{2M}{r}+\frac{2(1+l)Q^{2}}{(2+l)r^{2}}}\right)\dot{r}^{2}+r^{2}\dot{\phi}^{2}=\epsilon, 
\label{eq34}
\end{equation}
where $\epsilon =-1$ corresponds to timelike geodesics and $\epsilon = 0$ corresponds to null geodesics. Using Eq.(\ref{eq32}) and Eq.(\ref{eq33}), Eq.(\ref{eq34}) can be rewritten as:
\begin{equation}
\dot{r}^{2}=\mathcal{E}-V_{eff},
\label{eq35}
\end{equation}
where $\mathcal{E}$ represents the effective energy of the particle, and $V_{eff}$ is the effective potential for the radial motion of the particle, defined as:
\begin{equation}
\mathcal{E} \equiv \frac{E^{2}}{(1+l)},
\label{eq36}
\end{equation}
\begin{equation}
V_{eff} \equiv \frac{1-\frac{2M}{r}+\frac{2(1+l)Q^{2}}{(2+l)r^{2}}}{1+l}\left(\frac{L^{2}}{r^{2}}-\epsilon\right).
\label{eq37}
\end{equation}
Next, we will discuss the orbital behavior of massive particles ($\epsilon = -1$) and the circular orbit properties of photons ($\epsilon = 0$), respectively.

\subsection{$\text{The geodesic structure of massive particles}$}

\begin{figure}[H]
\centering
\includegraphics[width=8cm]{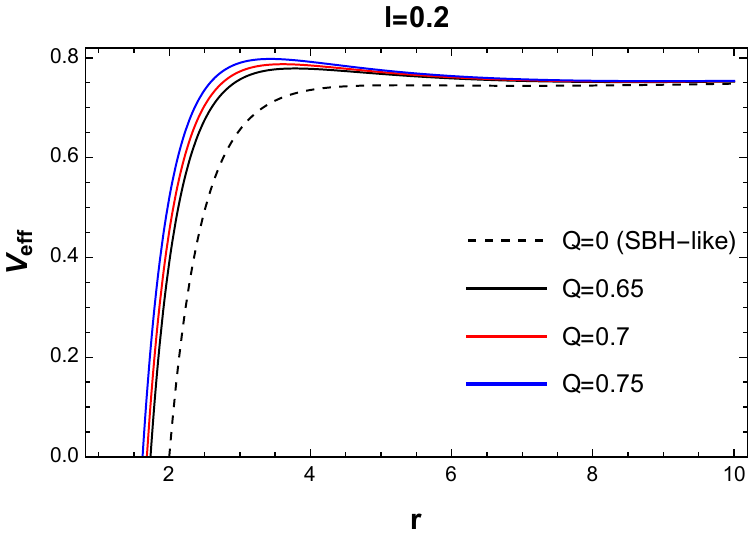}
\includegraphics[width=8cm]{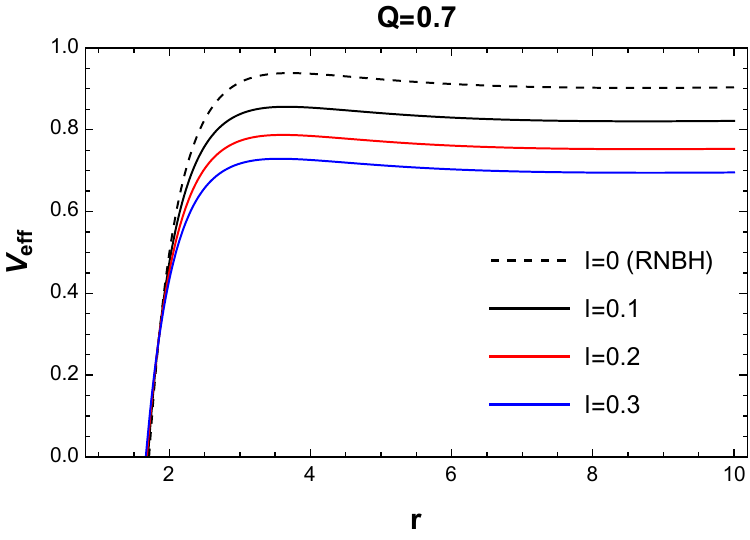}
\caption{The variation of the effective potential $V_{eff}$ of massive particles around BCBHs with respect to the radial coordinate $r$. The left panel corresponds to $l=0.2$, and the right panel corresponds to $Q=0.7$, with $L=3.5$.}
\label{fig311}
\end{figure}

\begin{figure}[H]
\centering
\includegraphics[width=8cm]{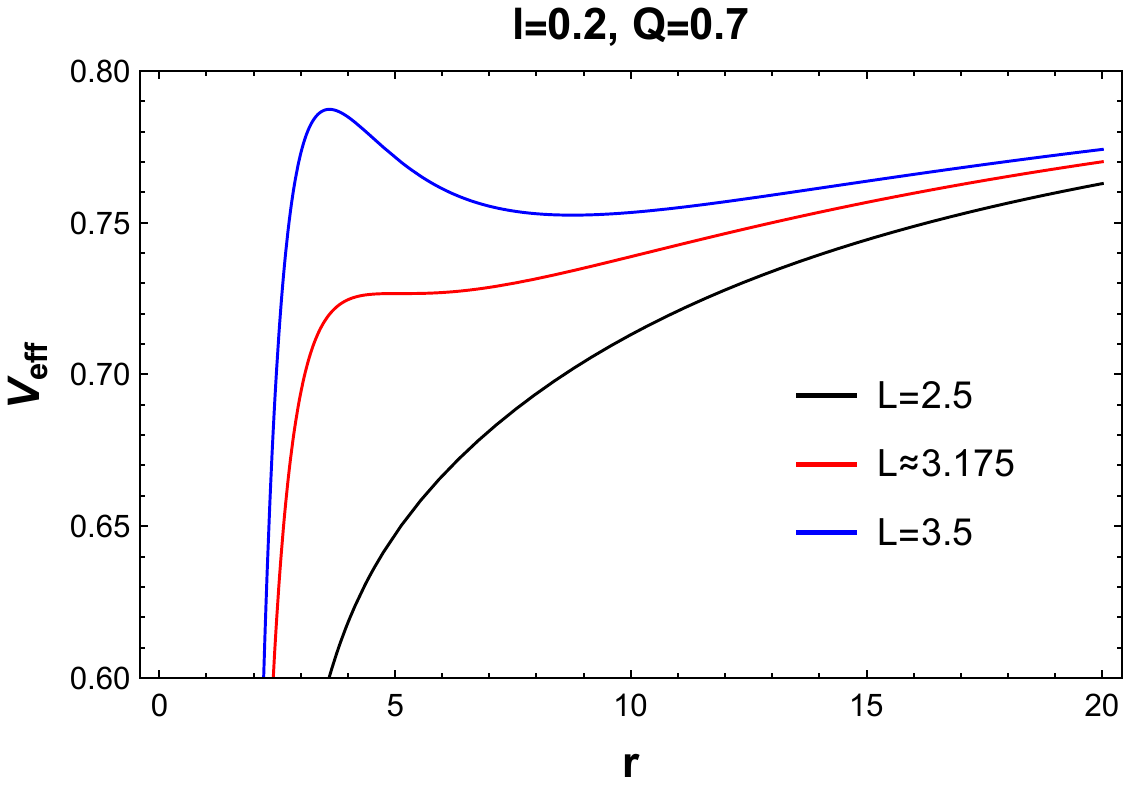}
\caption{The effective potential $V_{eff}$ of massive particles as a function of the radial coordinate $r$ under different angular momentum. We set $l=0.2$, $Q=0.7$.}
\label{fig312}
\end{figure}

Studying the motion of massive particles is crucial for understanding the profile of BH accretion material. In this section, we analyze the motion properties of massive particles around charged BHs within the framework of bumblebee gravity. According to Eq.(\ref{eq37}), the effective potential for timelike test particles can be expressed as:
\begin{equation}
V_{eff} \equiv \frac{1-\frac{2M}{r}+\frac{2(1+l)Q^{2}}{(2+l)r^{2}}}{1+l}(\frac{L^2}{r^2}+1).
\label{eq311}
\end{equation}
From Eq.(\ref{eq311}), it can be seen that effective potential depends on the BH parameters $l$ and $Q$, as well as the angular momentum $L$. In Fig.\ref{fig311}, we plot the variation of the effective potential $V_{eff}$ with radial coordinate $r$ with angular momentum $L = 3.5$, under different BH parameters. In general, when the test particle moves along a circular orbit around BH, the effective potential must satisfy the condition $V_{eff}^{\prime} = 0$, where “$\prime $” denotes differentiation with respect to the radial coordinate. If the effective potential has a local minimum, the corresponding circular orbit is stable, and if it has a local maximum, the orbit is unstable. From Fig.\ref{fig311}, we observe that compared to the charge parameter $Q$, the LV parameter $l$ has a more significant effect on the peak value of the effective potential. As $l$ increases, the peak of effective potential decreases, indicating that less energy is required for a particle to move on an unstable circular orbit. And the increase of the charge parameter $Q$ has a positive effect on the peak of effective potential. In Fig.\ref{fig312}, we choose the parameters $l = 0.2$ and $Q = 0.7$ and display the variation of effective potential for particles in bumblebee spacetime under different angular momentum. Fig.\ref{fig312} shows that angular momentum of the particle plays a decisive role in whether circular orbits exist. When the angular momentum $L = 2.5$, effective potential has no extremum, meaning that no circular orbit exists. As the angular momentum increases to $L \approx 3.175$, effective potential curve exhibits an extremum, corresponding to a stable circular orbit. When $L = 3.5$, the effective potential has two extrema, indicating the presence of two circular orbits. Specifically, (1) when the radius $r_{c1} \approx 3.610$, the extremum of effective potential corresponds to an unstable circular orbit. If the particle is slightly perturbed, when the radius is smaller than $r_{c1}$, the particle will fall into BH; when the radius is larger than $r_{c1}$, the particle will move away from BH. (2) When the radius is $r_{c2} \approx 8.759$, the extremum of effective potential corresponds to a stable circular orbit.
 
Furthermore, we explore circular orbits properties of BCBHs by analyzing the effective potential. By applying the conditions:
\begin{equation}
\dot{r}=0, \quad \frac{dV_{eff}}{dr}=0, 
\label{eq312}
\end{equation}
we derive expressions for energy and angular momentum of particles undergoing circular motion on the equatorial plane, respectively:
\begin{equation}
\mathcal{E}=\sqrt{\frac{\left(-2(1+l) Q^{2}+(2+l)(2 M-r) r\right)^{2}}{(1+l)(2+l) r^{2}\left(4(1+l) Q^{2}-(2+l)(3 M-r) r\right)}},
\label{eq313}
\end{equation}
\begin{equation}
L=\sqrt{\frac{r^{2}\left(2 Q^{2}+2 l Q^{2}-2 M r-l M r\right)}{-4 Q^{2}-4 l Q^{2}+6 M r+3 l M r-2 r^{2}-l r^{2}}}.
\label{eq314}
\end{equation}
\begin{figure}[H]
\centering
\includegraphics[width=8cm]{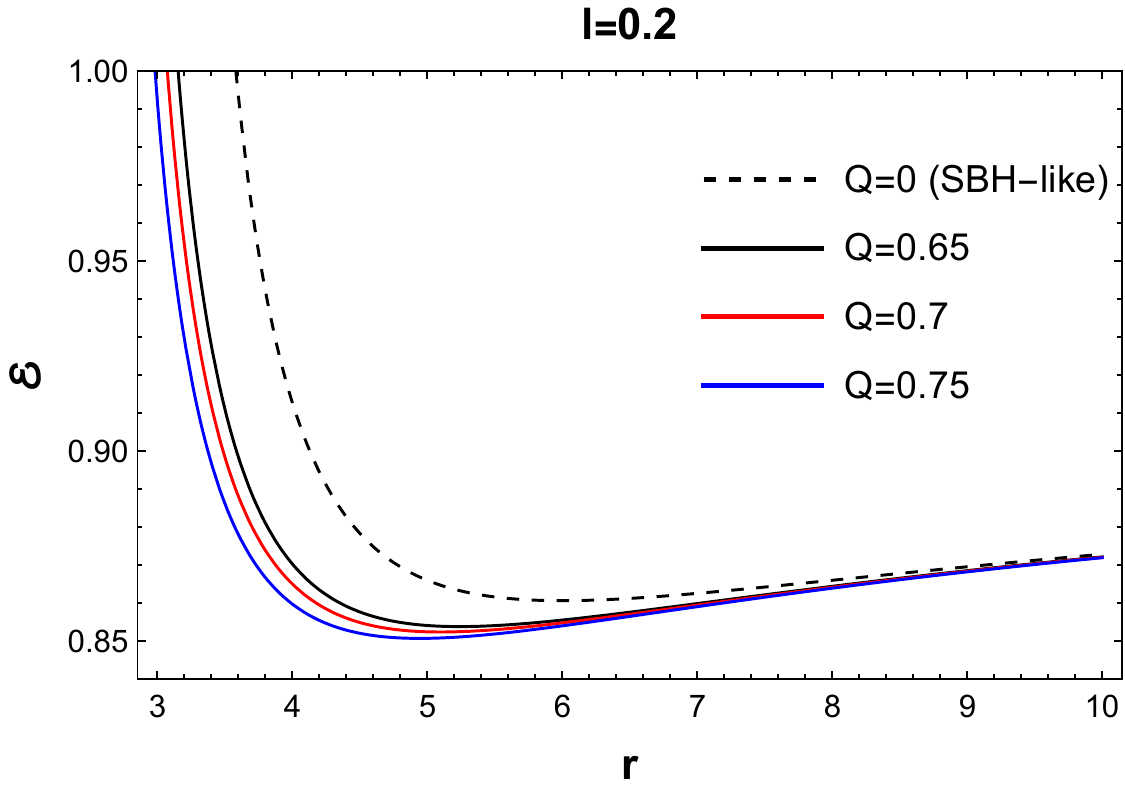}
\includegraphics[width=8cm]{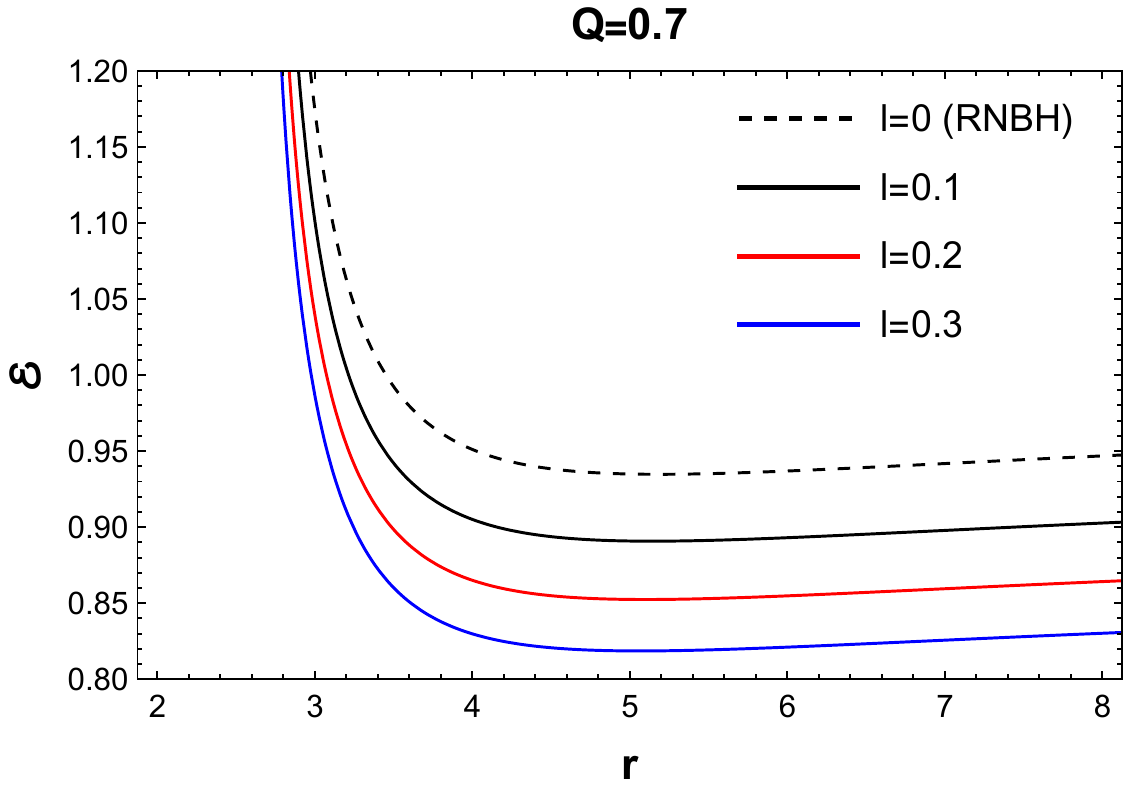}
\includegraphics[width=8cm]{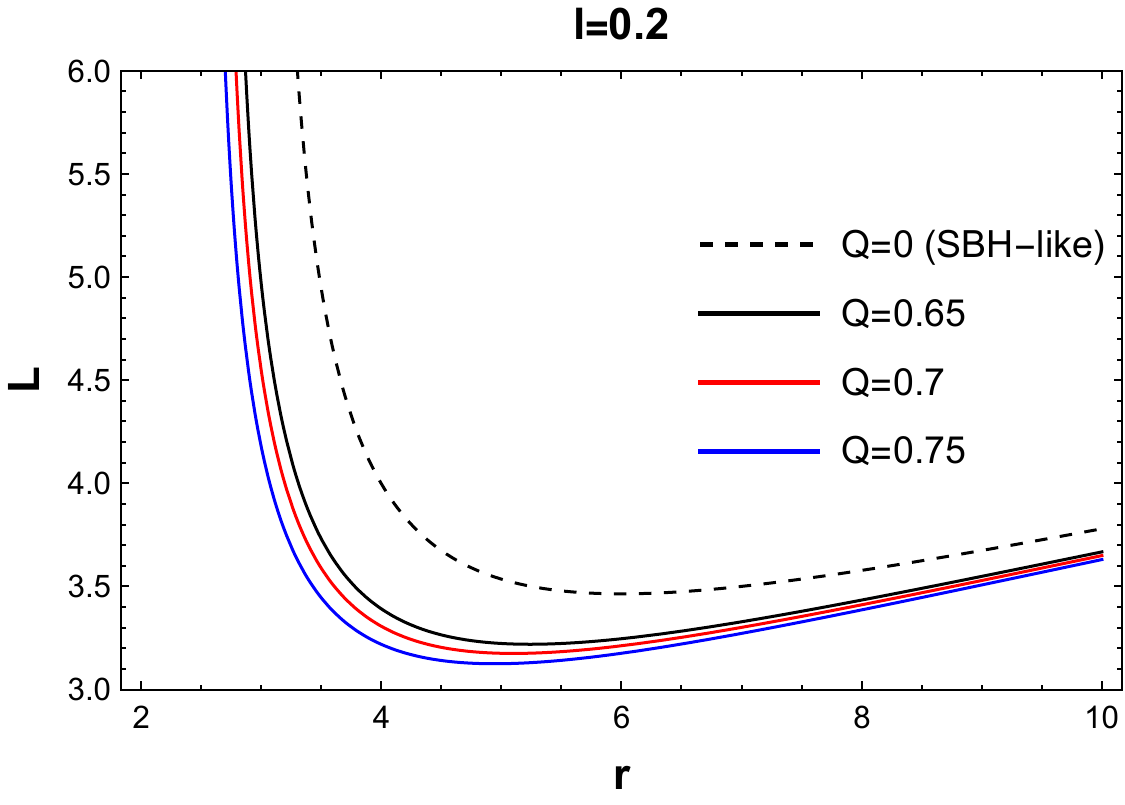}
\includegraphics[width=8cm]{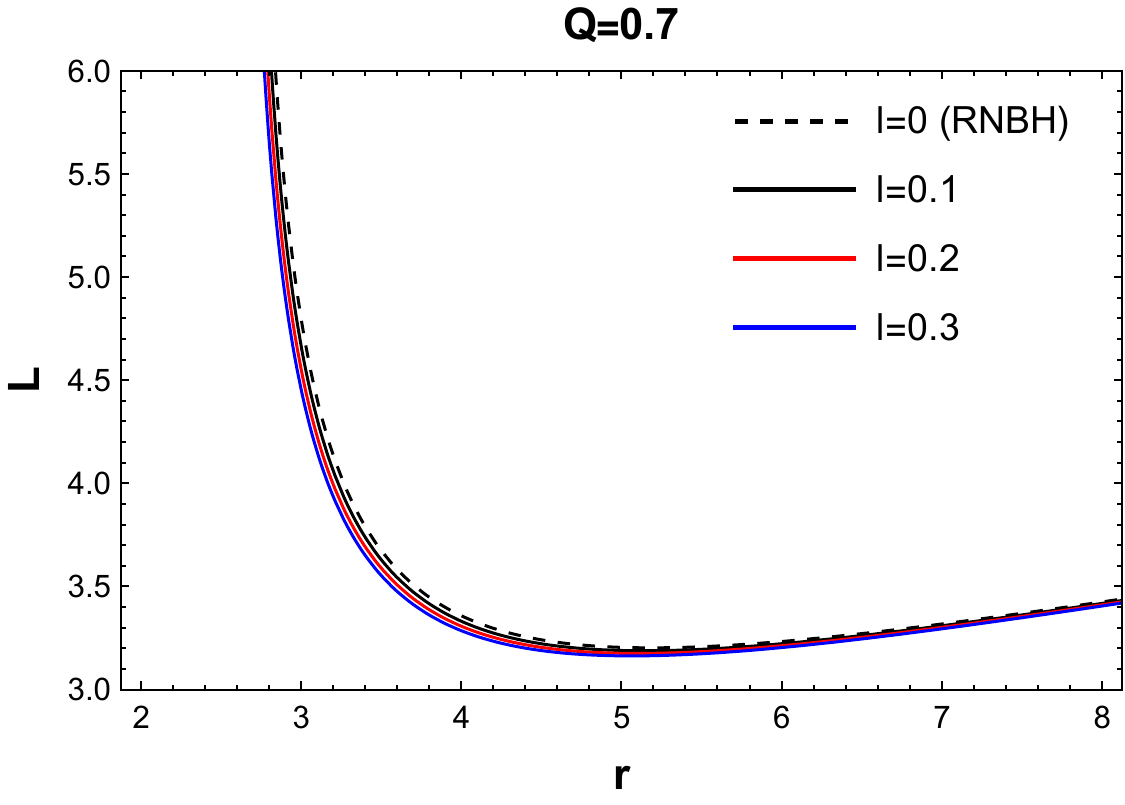}
\caption{The variation of the energy $\mathcal{E}$ and angular momentum $L$ of massive particles with respect to the radial coordinate $r$. The left panel corresponds to $l=0.2$, and the right panel corresponds to $Q=0.7$.}
\label{fig313}
\end{figure}
\noindent Based on Eqs.(\ref{eq313}-\ref{eq314}), we plot the energy and angular momentum of the particles as functions of the radial coordinate. Fig.\ref{fig313} shows that, as the charge parameter $Q$ and the LV parameter $l$ increase, both the energy and angular momentum of particles on circular orbits decrease. As previously mentioned, stable circular orbits correspond to the following conditions:
\begin{equation}
\dot{r}=0,\quad \frac{dV_{eff}}{dr}=0,\quad \frac{d^{2}V_{eff}}{dr^{2}}  \geq 0,
\label{eq315}
\end{equation}
that is, the effective potential has a local minimum. The ISCO is the smallest stable orbit around BH, corresponding to the boundary condition of the particle orbit. The ISCO has important physical significance, as it encodes crucial information about the dynamics of matter and energy in extreme gravitational fields. Furthermore, by using the conditions:
\begin{equation}
\dot{r}=0, \quad \frac{dV_{eff}}{dr}=0, \quad \frac{d^{2}V_{eff}}{dr^{2}}  = 0,
\label{eq316}
\end{equation}
we can derive the expression satisfied by the ISCO radius $r_{ISCO}$ of massive particles in bumblebee gravity as follows:
\begin{equation}
\frac{2 r_{ISCO}^{2}\left(-16(1+l)^{2} Q^{4}+18\left(2+3 l+l^{2}\right) M Q^{2} r_{ISCO}-(2+l)^{2} M(6 M-r_{ISCO}) r_{ISCO}^{2}\right)}{4(1+l) Q^{2}-(2+l)(3 M-r_{ISCO}) r_{ISCO}}=0.
\label{eq317}
\end{equation}
Through numerical calculations, we reveal the correlation between $r_{ISCO}$ and the BH parameters, as shown in Fig.\ref{fig314}. From the figure, it is evident that increases in the parameters $Q$ and $l$ both lead to a decrease in $r_{ISCO}$, indicating that in bumblebee gravity, both the charge and LV parameters result in the ISCO closer to the BH.

\begin{figure}[H]
\centering
\includegraphics[width=8cm]{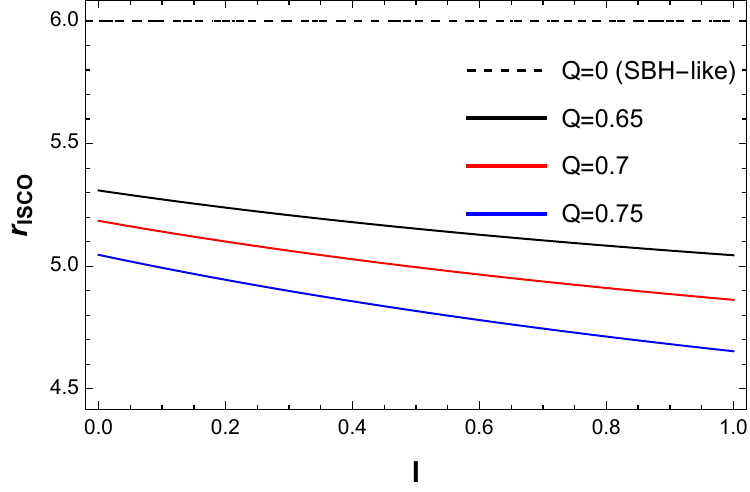}
\includegraphics[width=8cm]{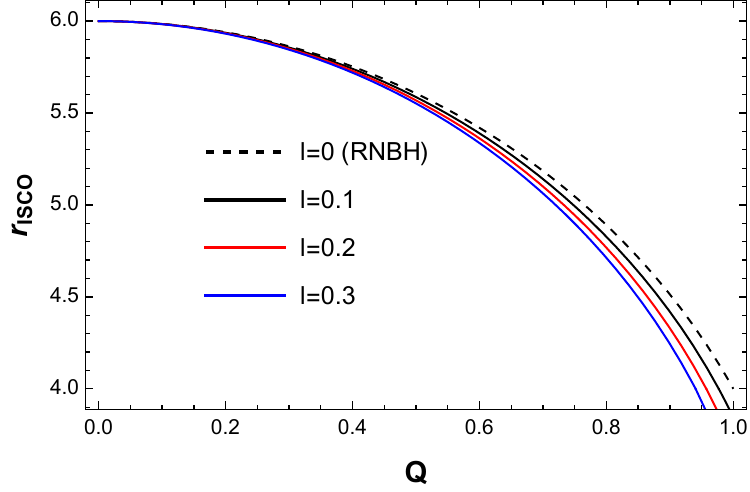}
\caption{The dependence of the ISCO radius $r_{ISCO}$ of BCBHs on the parameters $l$ and $Q$.}
\label{fig314}
\end{figure}

\begin{figure}[H]
\centering
\includegraphics[width=8cm]{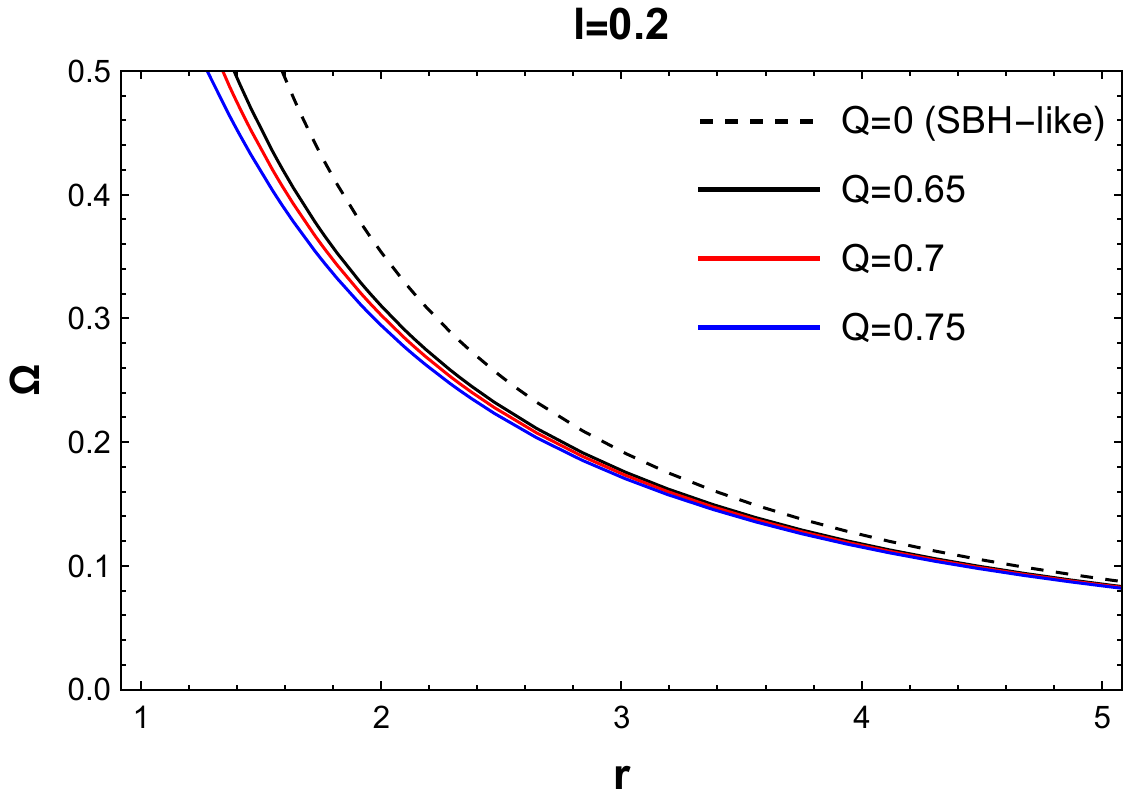}
\includegraphics[width=8cm]{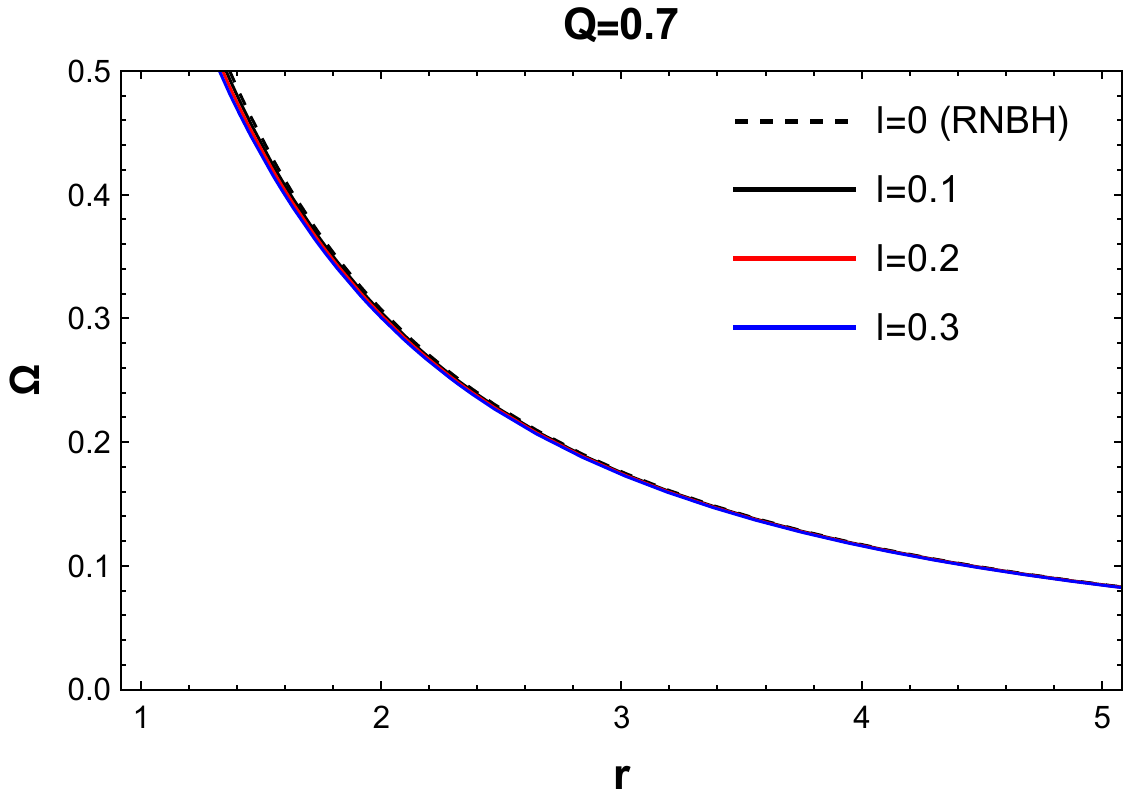}
\caption{The variation of the Keplerian frequency $\Omega$ of massive particles with respect to the radial coordinate $r$. The left panel corresponds to $l=0.2$, and the right panel corresponds to $Q=0.7$.}
\label{fig316}
\end{figure}

The angular velocity $\Omega = \frac{d\phi}{dt}$ of particles moving on circular orbits is the Keplerian frequency. In bumblebee gravity, we derive the expression for the Keplerian frequency of particles near the charged BHs as:
\begin{equation}
\Omega=\frac{d\phi}{dt}=\sqrt{\frac{-2 Q^{2}-2 l Q^{2}+2 M r+l M r}{(2+l) r^{4}}}.
\label{eq3121}
\end{equation}
According to Eq.(\ref{eq3121}), Fig.\ref{fig316} shows the variation of the Keplerian frequency with radial coordinates of the massive test particle moving around BCBHs for different parameters. From Fig.\ref{fig316}, it can be observed that for circular orbits with the same radius, the Keplerian frequency of the particles decreases as the parameters $Q$ or $l$ increase.

\subsection{$\text{The null geodesics}$}

\begin{figure}[H]
\centering
\includegraphics[width=8cm]{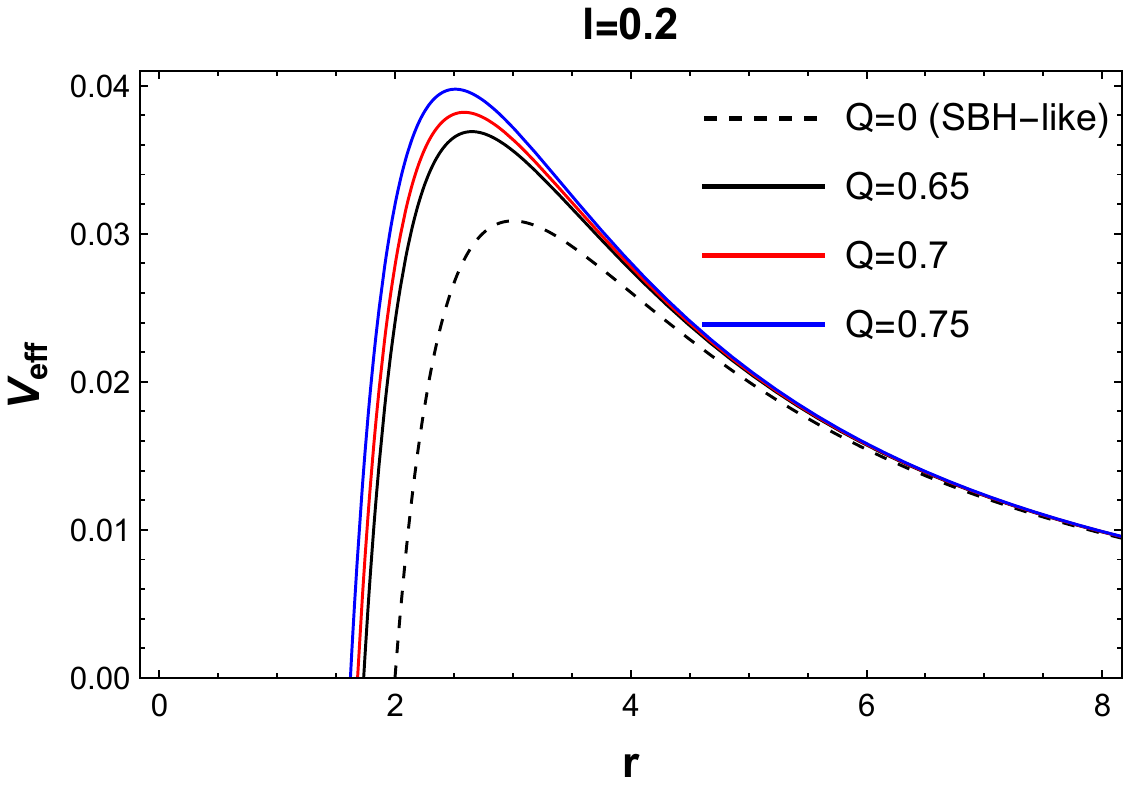}
\includegraphics[width=8cm]{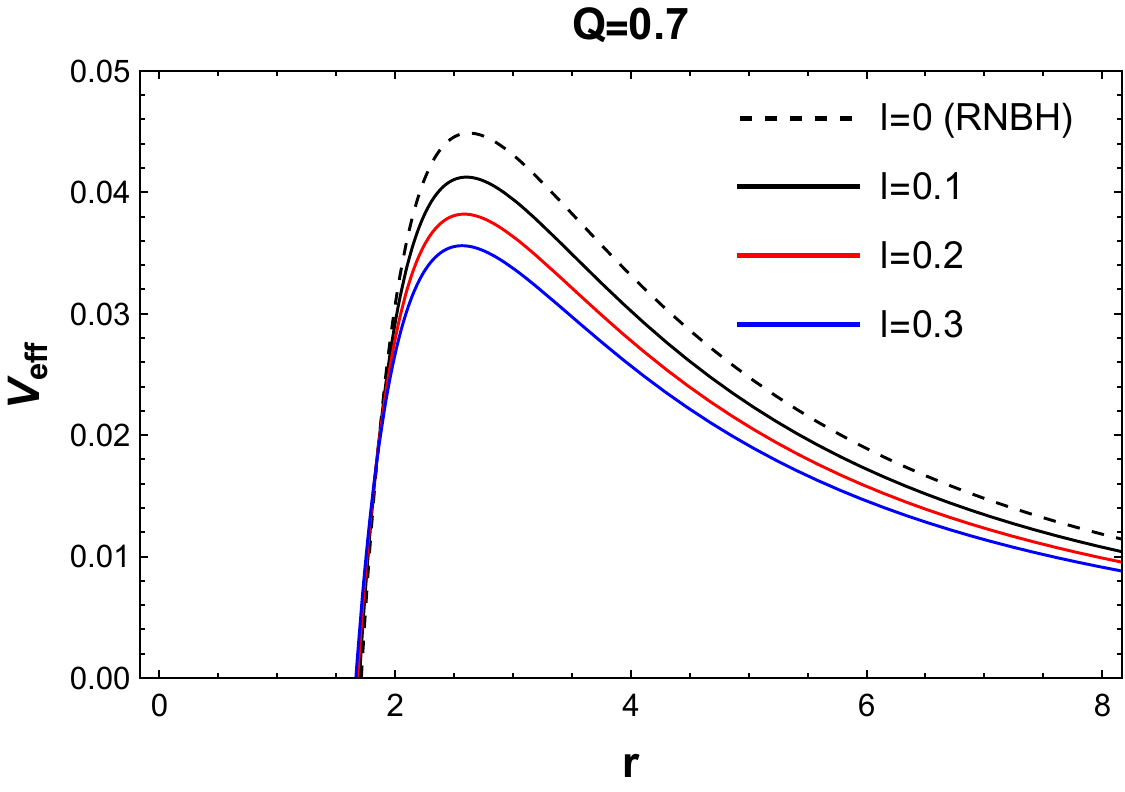}
\caption{The variation of the effective potential $V_{eff}$ of photons around BCBHs with respect to the radial coordinate $r$. The left panel corresponds to $l=0.2$, and the right panel corresponds to $Q=0.7$.}
\label{fig321}
\end{figure}

Due to the strong gravitational field around the BH, light will be strongly deflected when approaching the BH, and may even move along a circular orbit. This strong deflection may prevent photons from escaping from the BH \cite{4.1, phon2, phon4}, creating a dark region called the “black hole shadow”. In this section, the motion of photons around the BCBHs is investigated in terms of the null geodesics. According to Eqs.(\ref{eq32}-\ref{eq34}), we define the impact parameter as $b = L/E$, and use $\lambda L$ instead of $\lambda$. According to this definition, we derive the motion equations of photon in bumblebee gravity as:
\begin{equation}
\dot{t}=\frac{1}{b\left(1-\frac{2M}{r}+\frac{2(1+l)Q^{2}}{(2+l)r^{2}}\right)},
\label{eq321}
\end{equation}
\begin{equation}
\dot{\phi}=\frac{1}{r^2},
\label{eq322}
\end{equation}
\begin{equation}
\dot{r} = \sqrt {\frac{1}{(1+l)b^2}-\frac{2Q^2+2lQ^2-4Mr-2lMr+2r^2+lr^2}{(1+l)(2+l)r^4}}.
\label{eq323}
\end{equation}
From Eq.(\ref{eq323}), the effective potential for photons can be expressed as:
\begin{equation}
V_{eff} \equiv \frac{1-\frac{2M}{r}+\frac{2(1+l)Q^{2}}{(2+l)r^{2}}}{(1+l)r^2}.
\label{eq324}
\end{equation}
In order to visually observe the properties of the effective potential of photons around BCBHs, we plot its variation with respect to the radial coordinate $r$ ($r > r_h$) under different parameter conditions in Fig.\ref{fig321}. At the event horizon, the effective potential is zero. As $r$ increases, the potential gradually reaches a peak (at the photon sphere) and then decays. From Fig.\ref{fig321}, we can clearly observe that the LV parameter has a significant impact on the photon effective potential, with the peak of the effective potential decreasing as $l$ increases.

\begin{figure}[H]
\centering
\includegraphics[width=8cm]{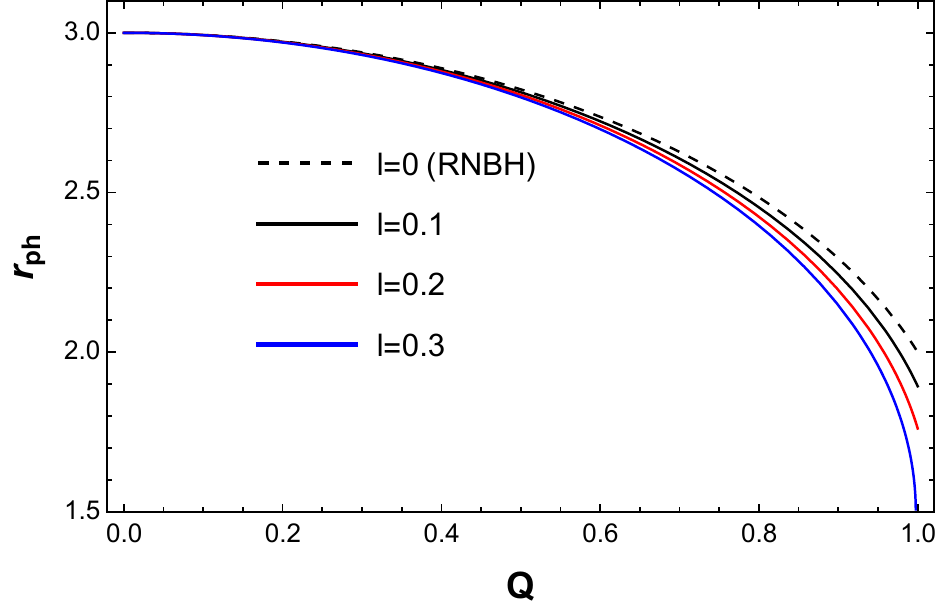}
\includegraphics[width=8cm]{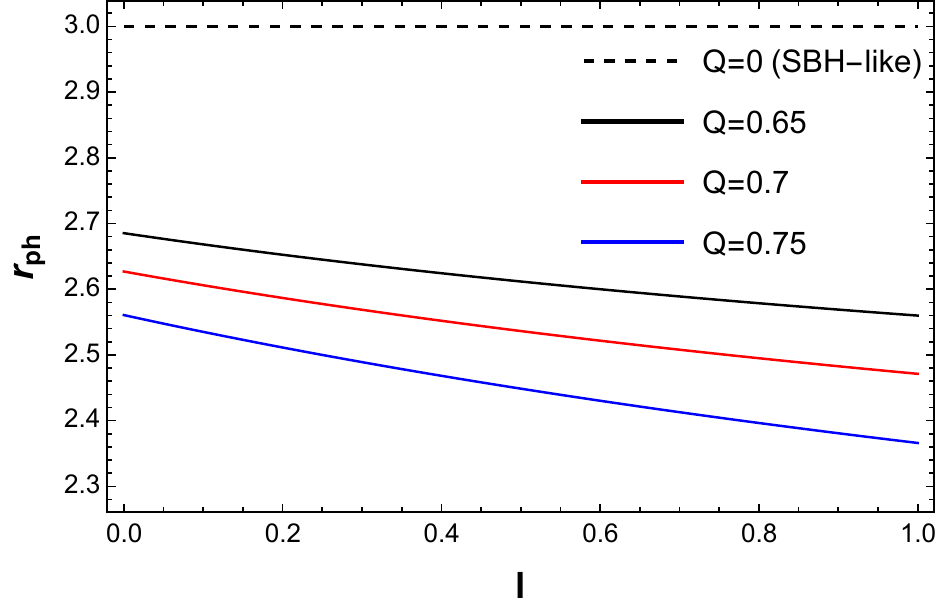}
\caption{The variation of the photon sphere radius $r_{ph}$ in bumblebee spacetime with respect to parameters $l$ and $Q$.}
\label{fig322}
\end{figure}

The photon sphere is a key concept used to describe the critical boundary region between captured and uncaptured photons in the gravitational field of the BH. The radius of the photon sphere can be determined by the conditions $\dot{r} = 0$ and $\ddot{r} = 0$. Under these conditions, the effective potential for photons satisfies:
\begin{equation}
V_{eff}(r_{ph})=\frac{1}{b_{ph}^2(1+l)}, \quad  \frac{dV_{eff}}{dr} \Big|_{r=r_{ph}}=0,
\label{eq325}
\end{equation}
where $r_{ph}$ is the radius of the photon sphere and $b_{ph}$ is the critical impact parameter. For a distant observer, $b_{ph}$ corresponds to the radius of the BH shadow. Based on the above conditions, the specific expressions for $r_{ph}$ and $b_{ph}$ of the charged BH in bumblebee gravity frame are derived as:
\begin{equation}
r_{ph}=\frac{1}{2}\left(3 M+\frac{\sqrt{18 M^{2}+9 l M^{2}-16 Q^{2}-16 l Q^{2}}}{\sqrt{2+l}}\right),
\label{eq326}
\end{equation}
\begin{equation}
b_{ph}= \frac{3 \sqrt{2+l} M+\sqrt{9(2+l) M^{2}-16(1+l) Q^{2}}}{2 \sqrt{2+l} \sqrt{\frac{6(2+l) M^{2}-8(1+l) Q^{2}+2 \sqrt{2+l} M \sqrt{9(2+l) M^{2}-16\left(1+l Q^{2}\right.}}{\left(3 \sqrt{2+l}M^{2} \sqrt{9(2+l) M^{2}-16(1+l) Q^{2}}\right)^{2}}}}.
\label{eq327}
\end{equation}
Fig.\ref{fig322} shows the variation of the photon sphere radius $r_{ph}$ as a function of BCBH parameters. From the figure, it can be observed that the photon sphere tends to shrink inward as the charge parameter $Q$ and the LV parameter $l$ increase. When the LV parameter $l = 0$, the case of RNBH is recovered. Clearly, as the parameter $l$ increases, the differences between RNBHs and BCBHs become increasingly significant. And when the charge parameter $Q = 0$, we obtain the results of the Schwarzschild-like BH, where $r_{ph}$ of the BCBHs are significantly smaller than its Schwarzschild-like BH counterpart. Meanwhile, the BCBHs results show a further decreasing trend as $Q$ increases.

In order to improve the reliability of theoretical studies, it is necessary and meaningful to apply observational data to constrain model parameters \cite{obs1,obs2,obs3}. In some literature, the BH shadow data published by the EHT, such as shadow angular diameter, shadow radius, and fractional differences, etc., have been widely used to constrain model parameters in different BH theories \cite{4.11,Lag,obs5,obs6,obs7,obs8,obs9,obs10}, as well as to test the validity of GR or modified gravity theories. Following the methods used in some works \cite{con1,con2,con3}, where the authors applied the shadow radius data and fractional difference data of Sgr A* from EHT to constrain model parameters of BH solutions in the framework of relevant theories (e.g. lorentz symmetry violation and loop quantum gravity theory), here we utilize the observational data of Sgr A* published by the EHT \cite{1BHshadow4} to constrain the parameters of the BCBHs. When the observer is located at infinity ($r_o \to \infty$), there is the condition $A(r_o) = 1$, and then the BH critical impact parameter is the BH shadow radius:
\begin{equation}
R_{sh}=r_{ph}\sqrt{\frac{A(r_o)}{A(r_{ph})}}=b_{ph}.
\label{eq328}
\end{equation}
\begin{figure}[H]
\centering
\includegraphics[width=8cm]{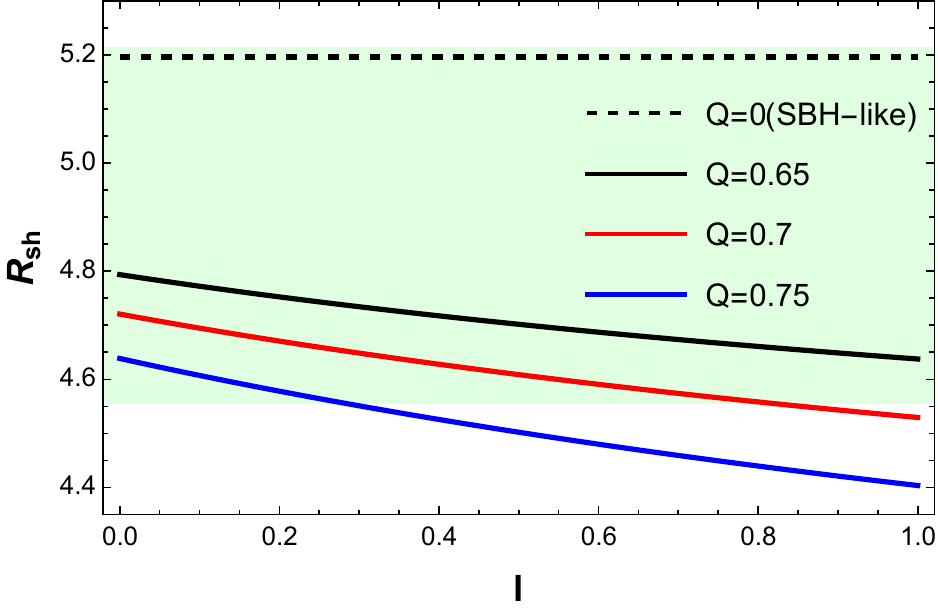}
\includegraphics[width=8cm]{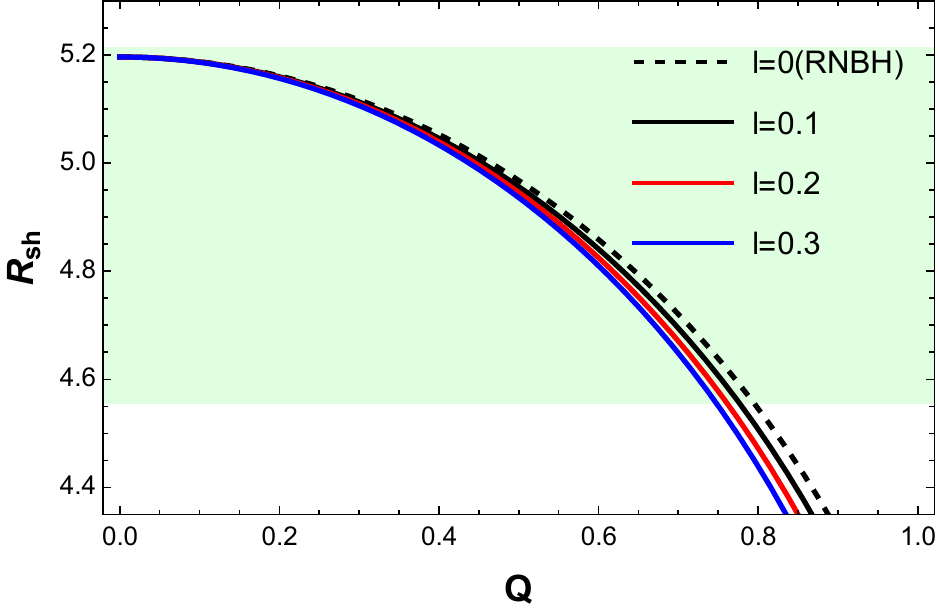}
\caption{The variation of BCBHs shadow radius $R_{sh}$ with respect to the parameters $l$ and $Q$, as well as the constraints on the parameters $l$ and $Q$ from the BH shadow radius data of Sgr A*.}
\label{fig32rsh}
\end{figure}
\noindent Using Eqs.(\ref{eq327}) and (\ref{eq328}), we present the trend of the BH shadow radius $R_{sh}$ with respect to parameter variations, as well as the constraints on the charge parameter and LV parameter by the Sgr A* shadow radius data within the $1\sigma$ confidence interval. The results indicate that as the charge parameter $Q$ and LV parameter $l$ increase, the BCBHs shadow radius decreases and is always smaller than the corresponding values for Schwarzschild-like BHs and RNBHs. The specific constraint results for the two parameters are summarized in Table \ref{table1} and Table \ref{table2}. In the numerical calculations and analyses of this study, the parameters values are chosen to satisfy the constraints of the observational data.
\begin{table}[h]
\centering
\caption{The constraints on the LV parameter $l$ of the BCBHs from the observational data of Sgr A* under different charge parameter $Q$.}
\begin{tabular}{|c|c|c|c|c|c|c|}
\hline
\multirow{2}{*}{} & \multicolumn{2}{c|}{$Q = 0.65$} & \multicolumn{2}{c|}{$Q = 0.7$} & \multicolumn{2}{c|}{$Q = 0.75$} \\ \cline{2-7} 
& lower & upper & lower & upper & lower & upper \\ 
\hline
$l$ & --    & 2.0573 & --    & 0.8549 & --    & 0.3040 \\ 
\hline
\end{tabular}
\label{table1}
\end{table}
\begin{table}[h]
\centering
\caption{The constraints on the charge parameter $Q$ of the BCBHs from the observational data of Sgr A* under different LV parameters $l$.}
\begin{tabular}{|c|c|c|c|c|c|c|}
\hline
\multirow{2}{*}{} & \multicolumn{2}{c|}{$l = 0.1$} & \multicolumn{2}{c|}{$l = 0.2$} & \multicolumn{2}{c|}{$l = 0.3$} \\ \cline{2-7} 
& lower & upper & lower & upper & lower & upper \\ 
\hline
$Q$& --    & 0.7796 & --    & 0.7639 & --    & 0.7505 \\ 
\hline
\end{tabular}
\label{table2}
\end{table}
 
\begin{figure}[H]
\centering
\includegraphics[width=4.4cm]{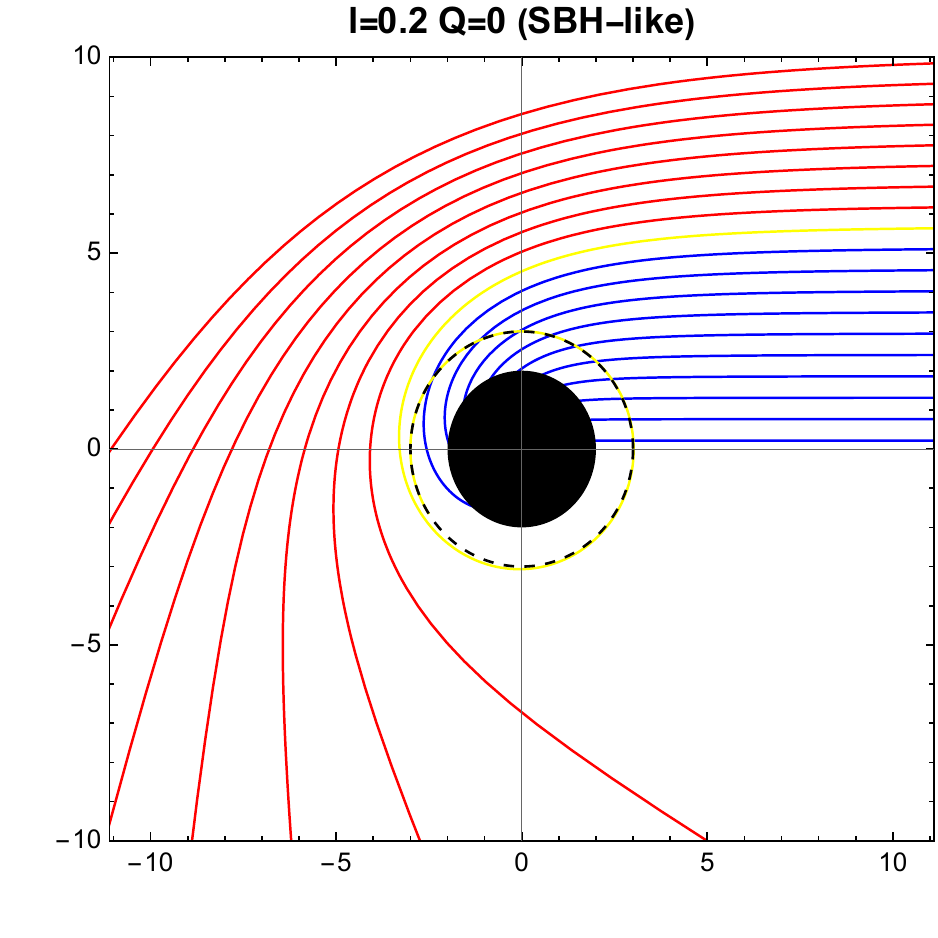}
\includegraphics[width=4.4cm]{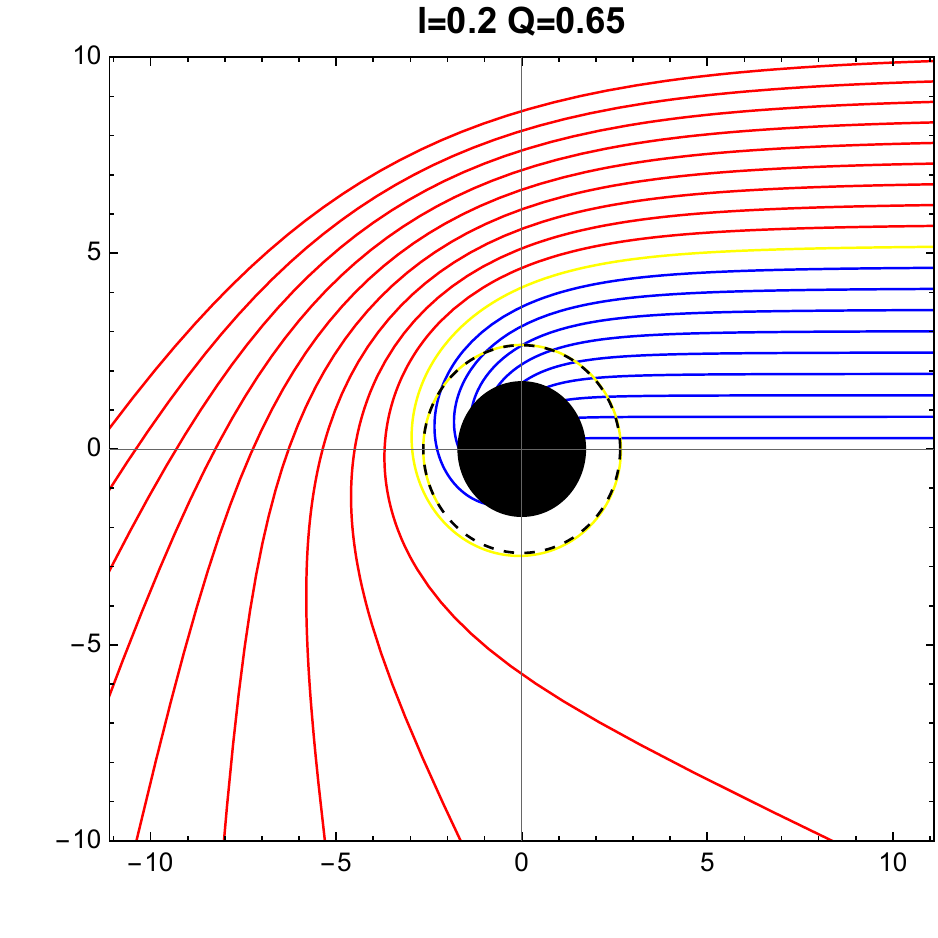}
\includegraphics[width=4.4cm]{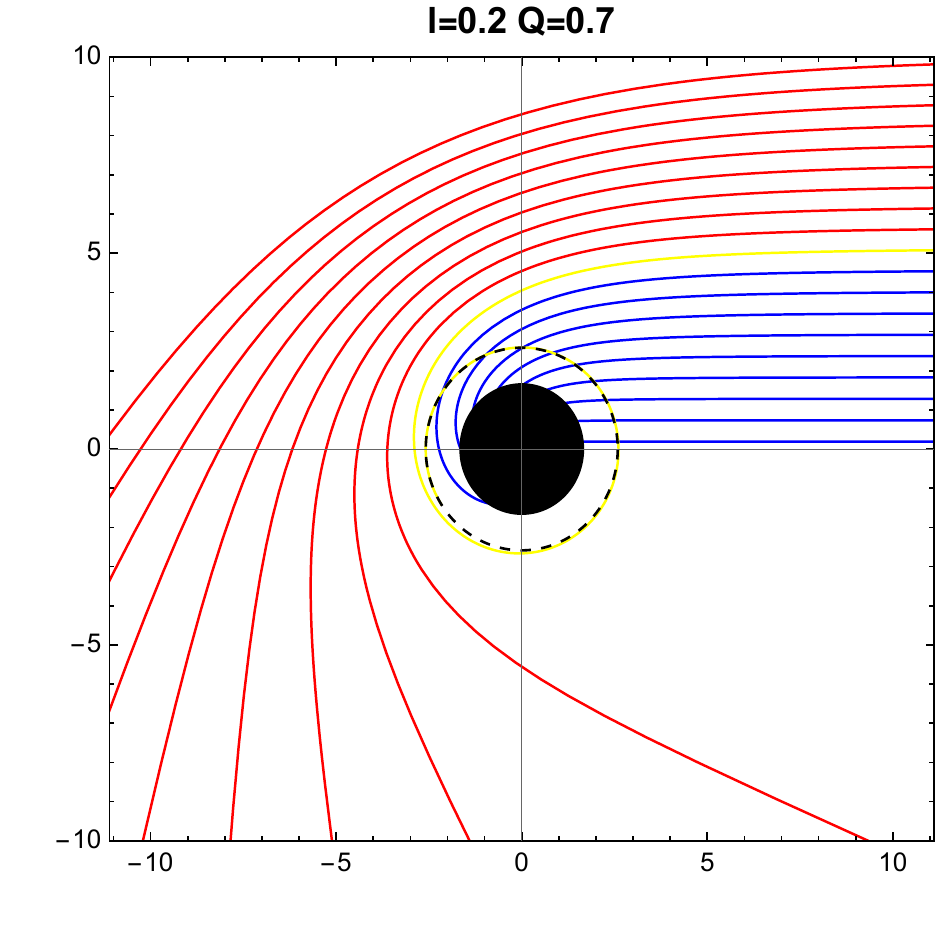}
\includegraphics[width=4.4cm]{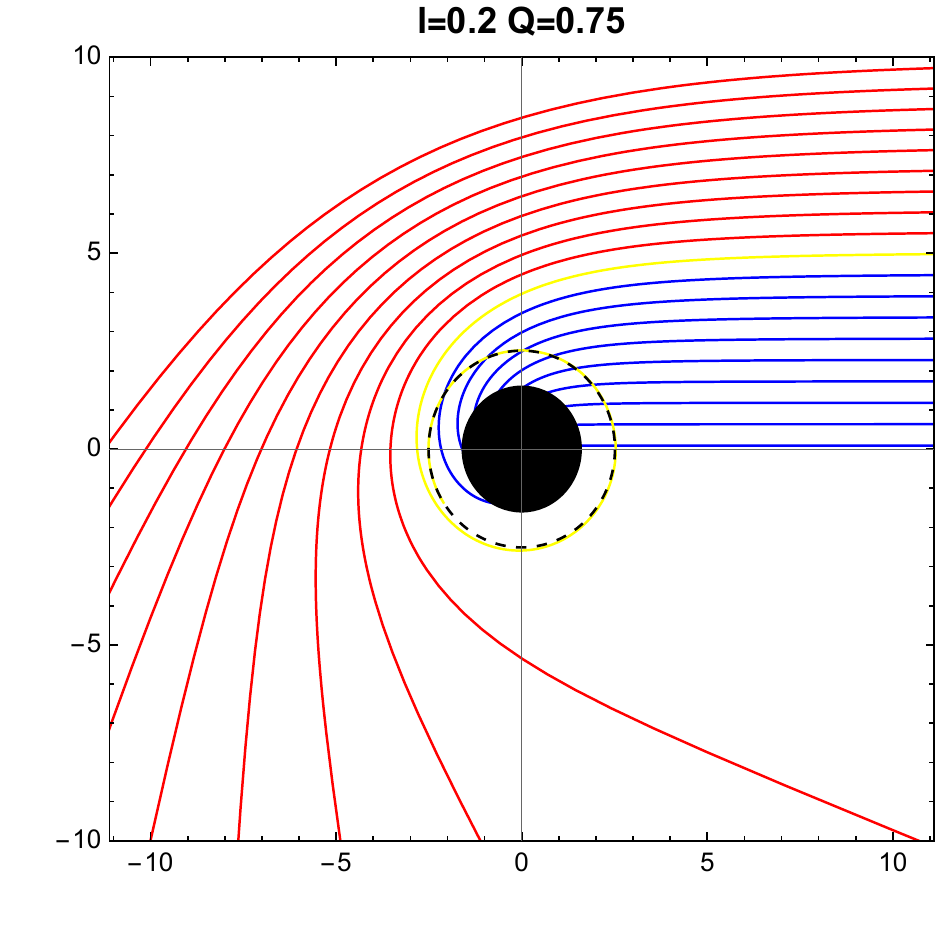}
\caption{The trajectories of photons around BCBHs of different charge parameter $Q$ in the polar coordinates ($r$, $\phi$). We choose $l=0.2$.}
\label{fig324}
\end{figure}

\begin{figure}[H]
\centering
\includegraphics[width=4.4cm]{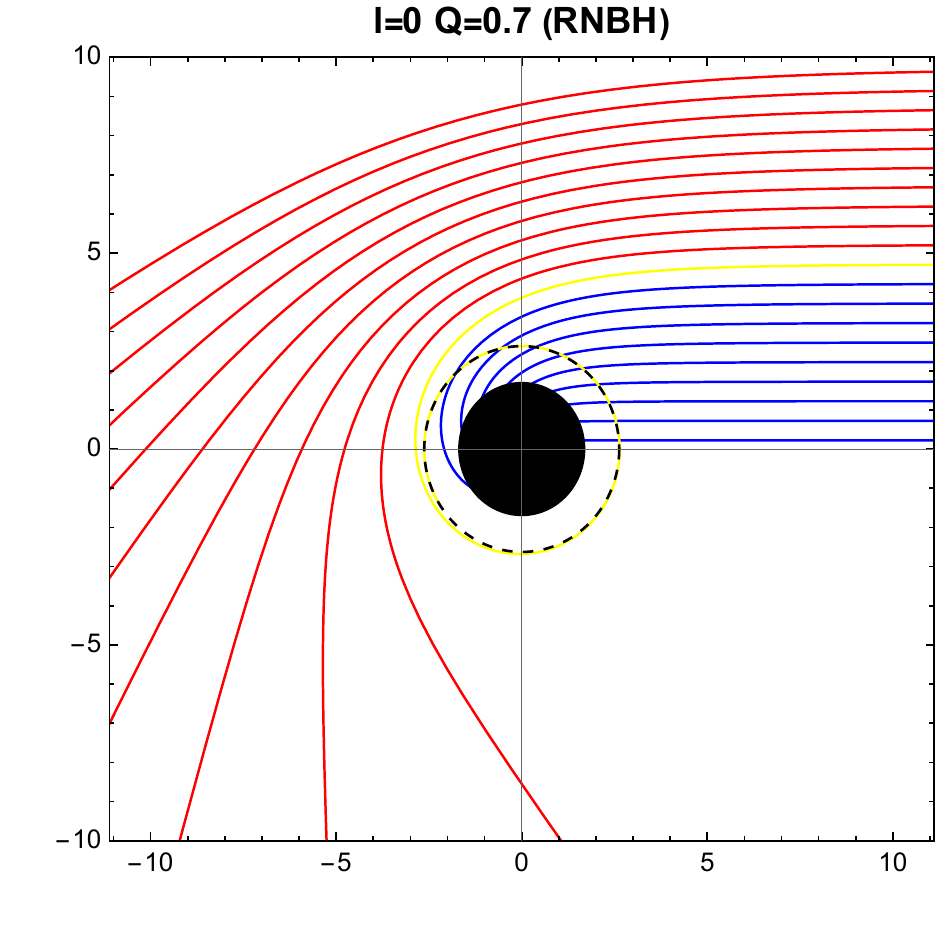}
\includegraphics[width=4.4cm]{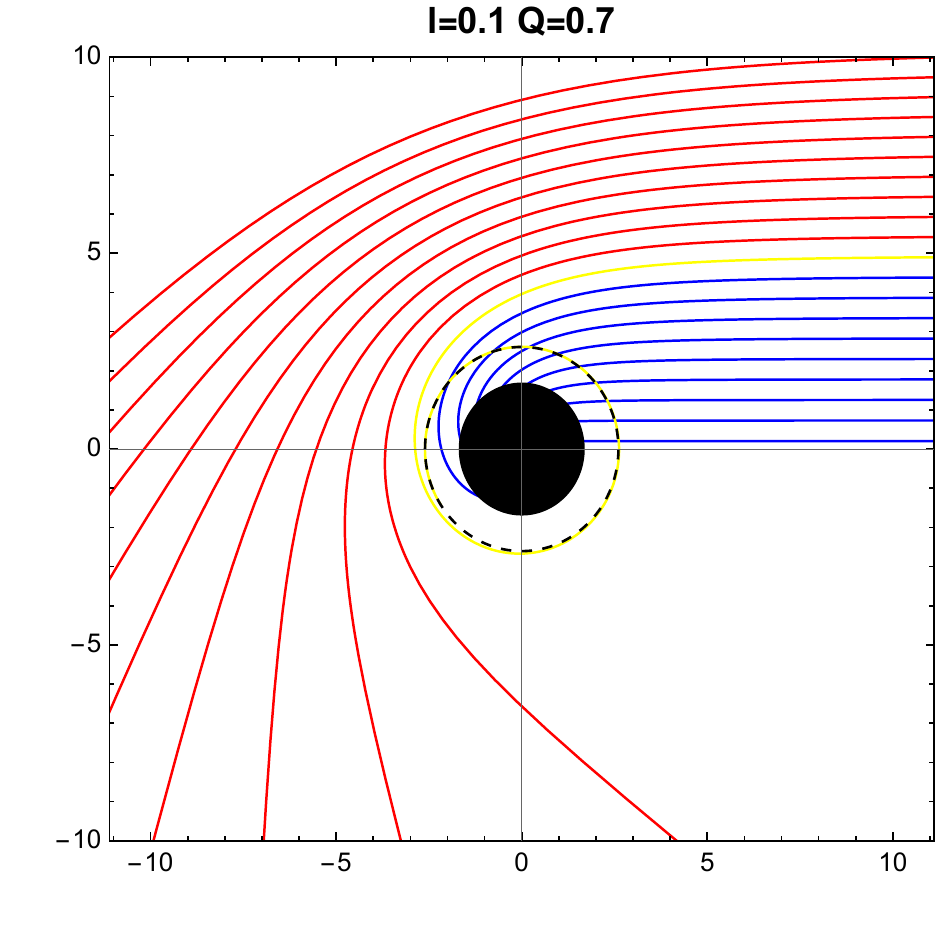}
\includegraphics[width=4.4cm]{tra3.pdf}
\includegraphics[width=4.4cm]{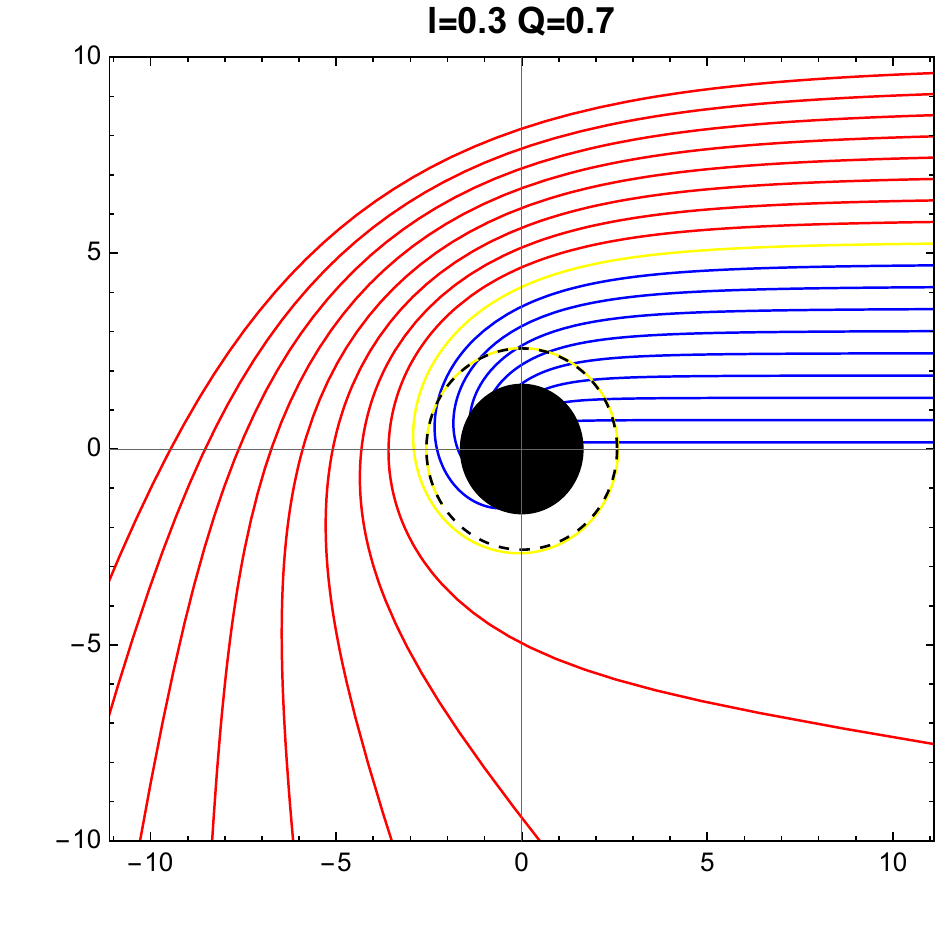}
\caption{The trajectories of photons around BCBHs of different LV parameter $l$ in the polar coordinates ($r$, $\phi$). We choose $Q=0.7$.}
\label{fig325}
\end{figure}

Next, we investigate the photon trajectories around the BCBHs. By combining Eqs.(\ref{eq322}-\ref{eq323}), the equation of motion for photons is derived as:
\begin{equation}
\left(\frac{dr}{d\phi}\right)^2=r^4\left(\frac{1}{(1+l)b^2}-\frac{1-\frac{2M}{r}+\frac{2(1+l)Q^{2}}{(2+l)r^{2}}}{(1+l)r^2}\right).
\label{eq329}
\end{equation}
Considering the parameter transformation $u = \frac{1}{r}$, Eq.(\ref{eq329}) can be rewritten as:
\begin{equation}
\left(\frac{du}{d\phi}\right)^2=\frac{1}{(1+l)b^2}-\frac{u^2\left(1-2Mu+\frac{2(1+l)Q^{2}u^2}{(2+l)}\right)}{(1+l)}.
\label{eq3210}
\end{equation}
Based on the above equations, we plot the photon trajectories around the BCBHs, as shown in Fig.\ref{fig324} and Fig.\ref{fig325}. The black disk represents the BH (i.e., the projection of the event horizon), and the black dashed line corresponds to the photon sphere ($r = r_{ph}$). From the figures, it is evident that for the BCBHs, as the LV parameter $l$ and the charge parameter $Q$ increase, the black disk gradually shrinks. Additionally, it can be observed that for photons with $b < b_{ph}$ (blue curve), the photons move inward and eventually fall into the BH; for photons with $b > b_{ph}$ (red curve), they encounter a potential barrier, approach the BH before reaching the turning point, and then escape to infinity. The yellow curves represent the photon trajectories when $b = b_{ph}$, where the photons asymptotically approach the photon sphere and move in unstable circular orbits without perturbation. Obviously, as the parameters $l$ and $Q$ increase, fewer photons originating from infinity are captured by the BCBHs, resulting in a smaller shadow seen by distant observers. Thus, this result provides a potential observational approach to distinguish between GR and the bumblebee gravity theory. 

Above, we showed the photon trajectories along geodesics around the BCBHs. However, in the actual physical processes, it has been found that there are usually accretion flows around BHs, as demonstrated by the M87* and Sgr A* observed by the EHT. This finding suggests that in studies of BH shadow, we must not only consider the background spacetime geometry but also the accretion details surrounding the BH. Next, we will focus on analyzing the effect of three types of accretion disk models on the optical appearance of the BHs when they are surrounded by optically thin accretion disks.

\section{$\text{Shadow and rings of BCBH illuminated by the thin accretion disks}$}

Due to the presence of a large amount of accreting material around BHs, the BH seen by distant observers will appear as a dark disk surrounded by a luminous region. This luminous region represents the light emitted by the accretion disk, which is seen by the observer after being affected by the BH. Apparently, photons emitted from the accretion disk may orbit the BH several times before escaping, thereby forming various rings. Investigating the optical appearance of BHs is crucial for exploring the spacetime geometry in strong gravitational fields, and further analyzing the nature of gravity. In this section, we employ relevant accretion disk models to investigate the observational features of BCBHs. Specifically, we assume that the material surrounding the BH can be viewed as an optically and geometrically thin accretion disk. The accreting material serves as the light source illuminating the BH.

\subsection{$\text{Three types of rays emitted by the accretion disk around BCBH}$}

In the following analysis, we assume that the accretion disk is located on the equatorial plane and that its emission is isotropic. A distant observer is positioned in the north pole direction facing the accretion disk. Referring to the method suggested in literature \cite{4.1}, we briefly describe the classification of the light around a BH and apply it to the BCBH. Using the orbital equation (\ref{eq329}) to calculate the total change in azimuthal angle of photon for $b<b_{ph}$, we derive:
\begin{equation}
\phi=\int_{0}^{u_{+}} \sqrt{\frac{1}{\frac{B(u)}{b^{2} A(u)}-u^{2} B(u)}} d u ,
\label{eq411}
\end{equation}
where $u_+ = \frac{1}{r_h}$; and for photon with $b>b_{ph}$, the total change in the azimuthal angle is:
\begin{equation}
\phi=2\int_{0}^{u_{t}} \sqrt{\frac{1}{\frac{B(u)}{b^{2} A(u)}-u^{2} B(u)}} d u,
\label{eq412}
\end{equation}
where $u_t = \frac{1}{r_t}$, and $r_t$ is the light ray’s radial minimal distance from its trajectory to the BCBH. To further analyze the orbital behavior of photons, the following parameter is introduced to consider classifying the photons orbiting the BH \cite{4.1}:
\begin{equation}
n(b)=\frac{\phi}{2 \pi},
\label{eq413}
\end{equation}
$n$ represents the total number of photon orbits. Based on the range of $n$ (corresponds to the number of intersections of light emitted by the disk-shaped accretion flow with the equatorial plane of the accretion disk), light rays can be classified into the following three categories: (1) when $0<n(b)<3/4$, the light rays only intersect the accretion disk once, corresponding to the direct emission of light ray; (2) when $3/4<n(b)<5/4$, the light rays intersect with the thin disk twice, corresponding to the lensed ring; (3) when $n(b)>5/4$, it is the photon ring, in which the light rays intersect the accretion disk at least three times.
\begin{figure}[H]
\centering
\includegraphics[width=8cm]{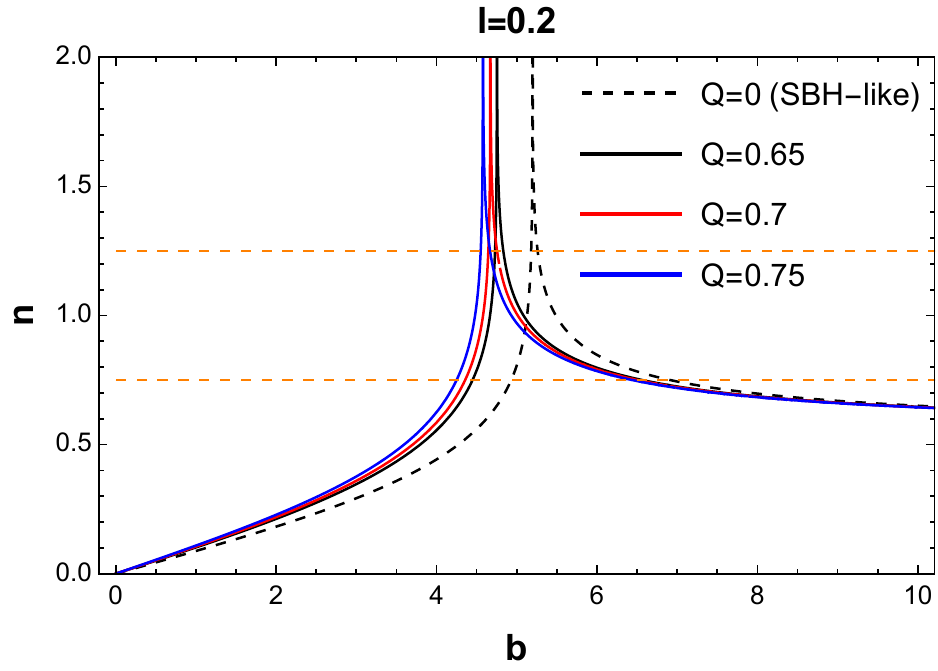}
\includegraphics[width=8cm]{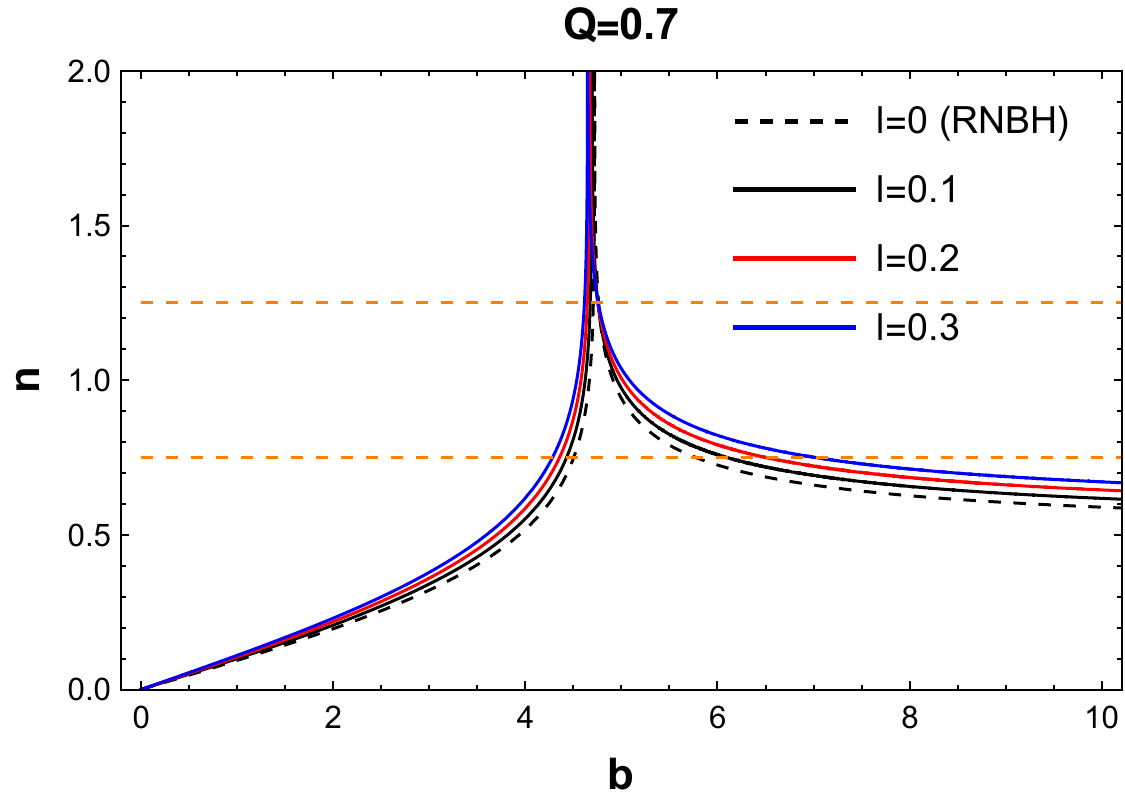}
\caption{The total number of photon orbits $n$ as a function of the impact parameter $b$ for BCBH. The left panel corresponds to $l=0.2$, and the right panel corresponds to $Q=0.7$.}
\label{fig411}
\end{figure}

Fig.\ref{fig411} shows the variation of the photon orbital number $n$ around BCBHs as a function of the impact parameter $b$. From the figure, it can be observed that when the impact parameter $b$ approaches the critical impact parameter $b_{ph}$, the orbital number $n(b)$ tends to infinity. According to Fig.\ref{fig411}, we mark the intersections of the $n(b)$ curve with $n = 0.75$ and $1.25$ as $b_2^-$, $b_3^{-}$, $b_3^{+}$, and $b_2^{+}$ from left to right, and categorize the intervals of the impact parameter as follows:

(1) Direct emission: $0 < n(b) < 3/4$, corresponding to $b \in \left(0, b_{2}^{-}\right) \cup \left(b_{2}^{+}, \infty\right)$,

(2) Lensed ring: $3/4 < n(b) < 5/4$, corresponding to $b \in \left(b_{2}^{-}, b_{3}^{-}\right) \cup \left(b_{3}^{+}, b_{2}^{+}\right)$,

(3) Photon ring: $n(b) > 5/4$, corresponding to $b \in \left(b_{3}^{-}, b_{3}^{+}\right)$.

\begin{figure}[H]
\centering
\includegraphics[width=4.4cm]{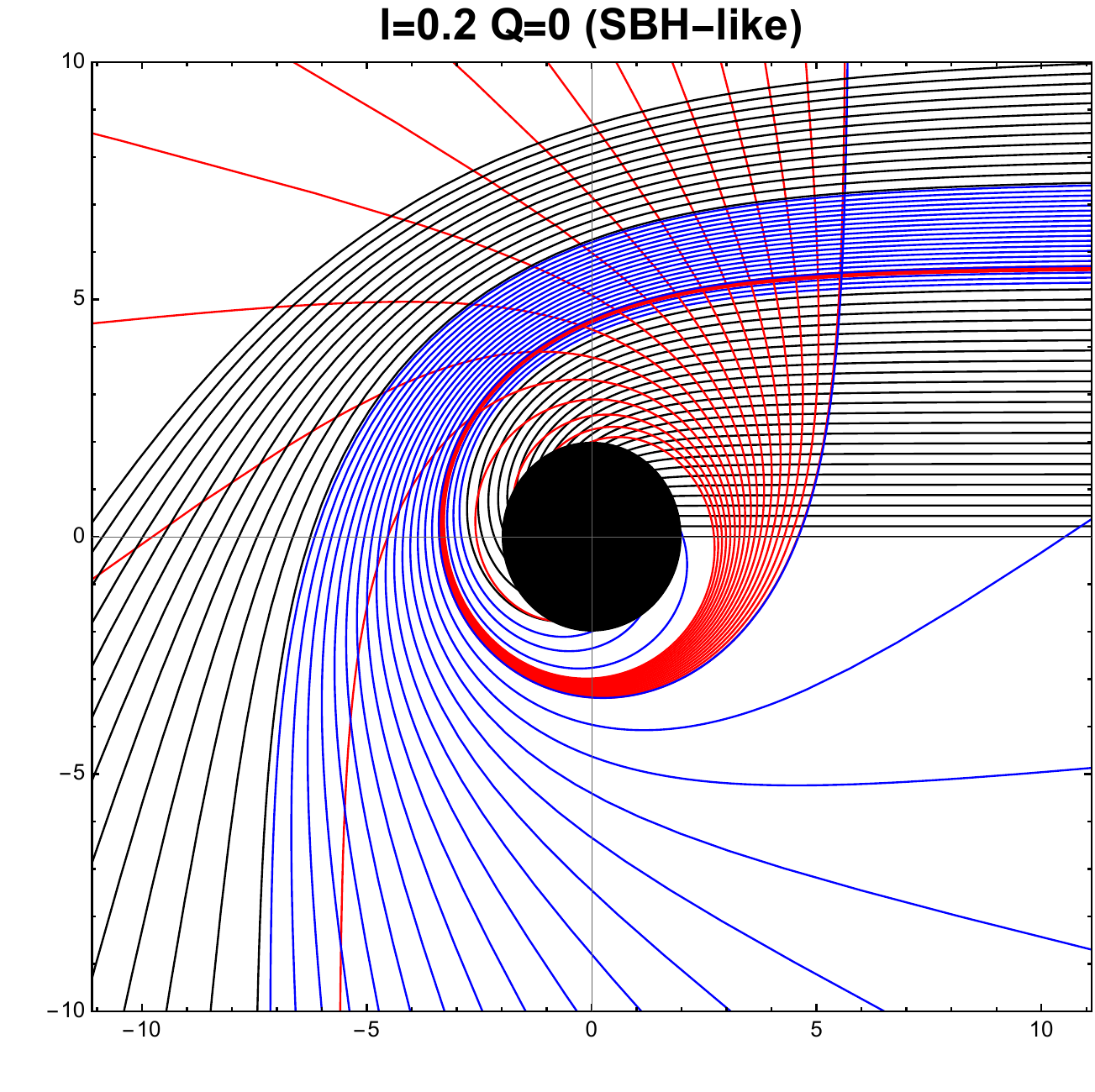}
\includegraphics[width=4.4cm]{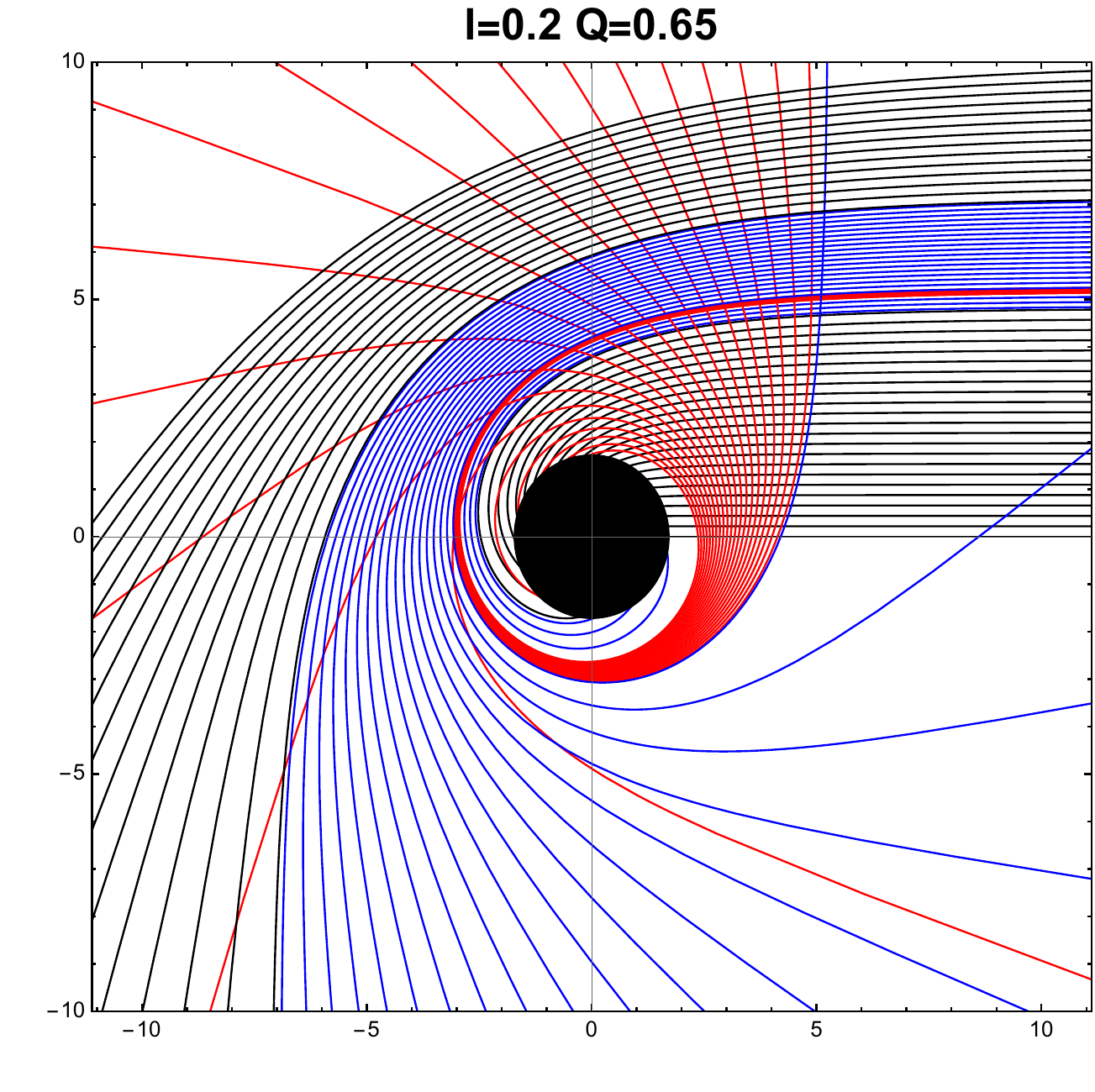}
\includegraphics[width=4.4cm]{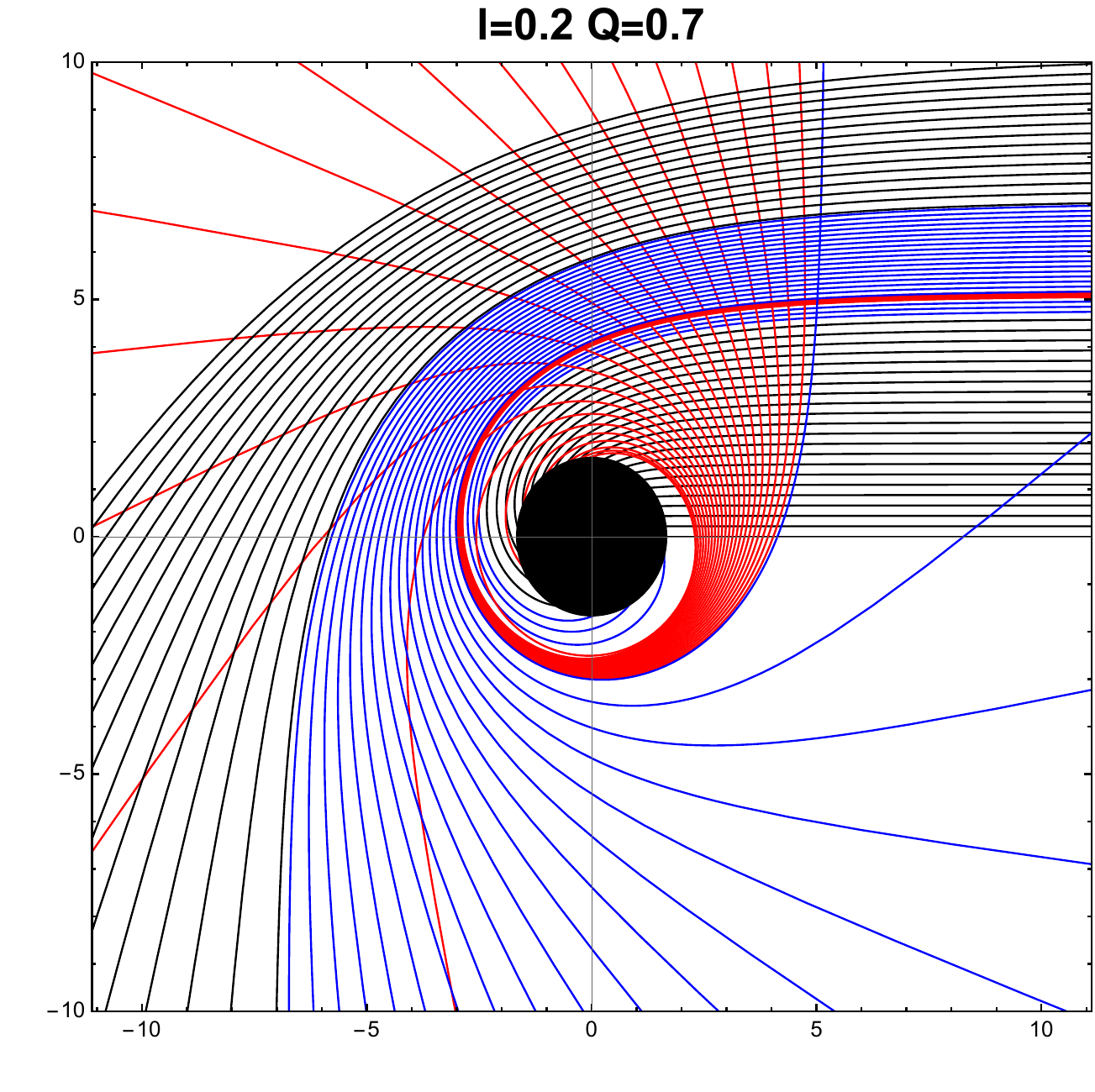}
\includegraphics[width=4.4cm]{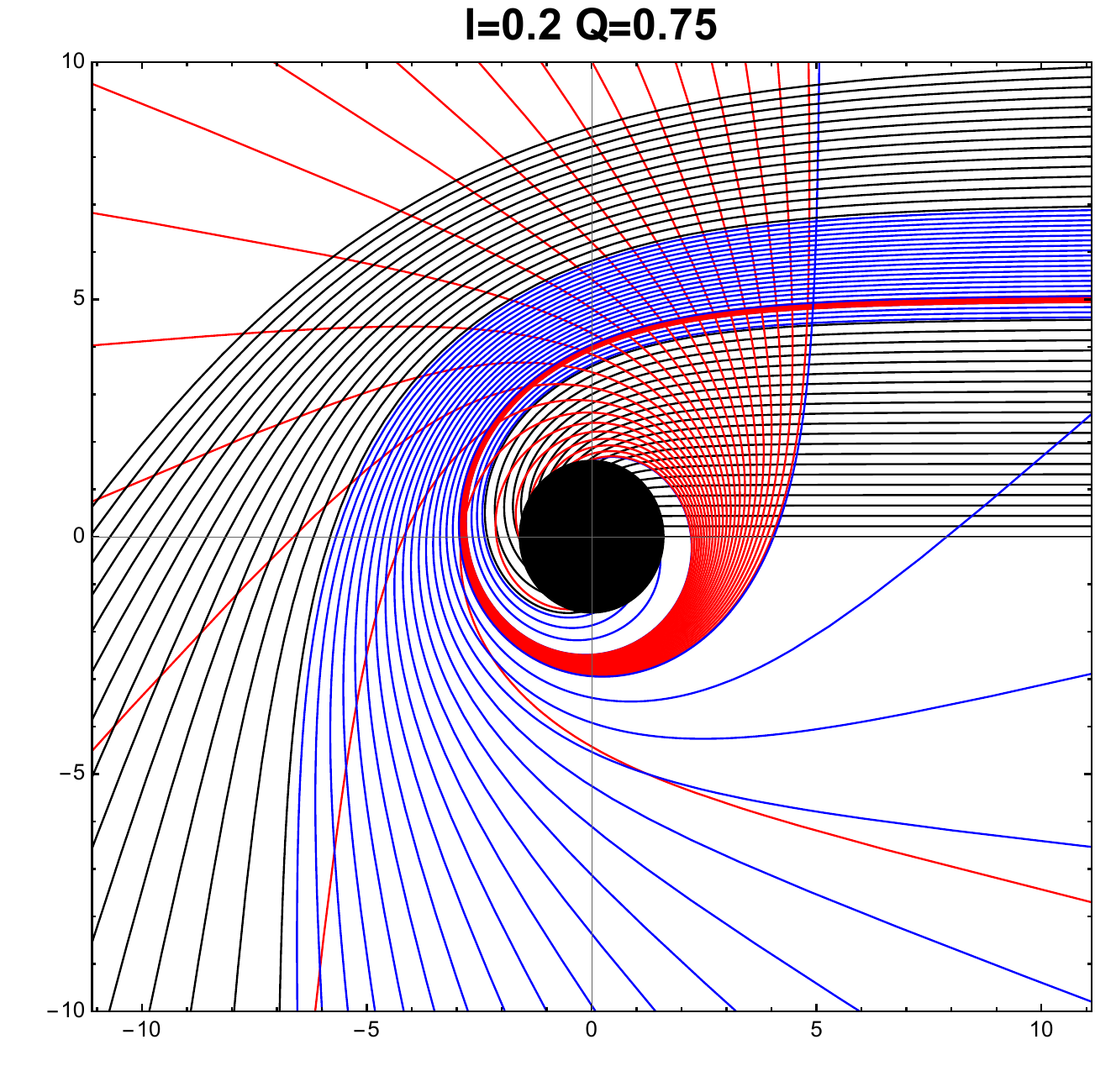}
\caption{The photon trajectories around BCBHs in the polar coordinates ($r$, $\phi$) for different charge parameter $Q$, with $l=0.2$. The black curves represent direct emission, the blue curves represent lensed ring, and the red curves represent photon ring.}
\label{fig412}
\end{figure}

\begin{figure}[H]
\centering
\includegraphics[width=4.4cm]{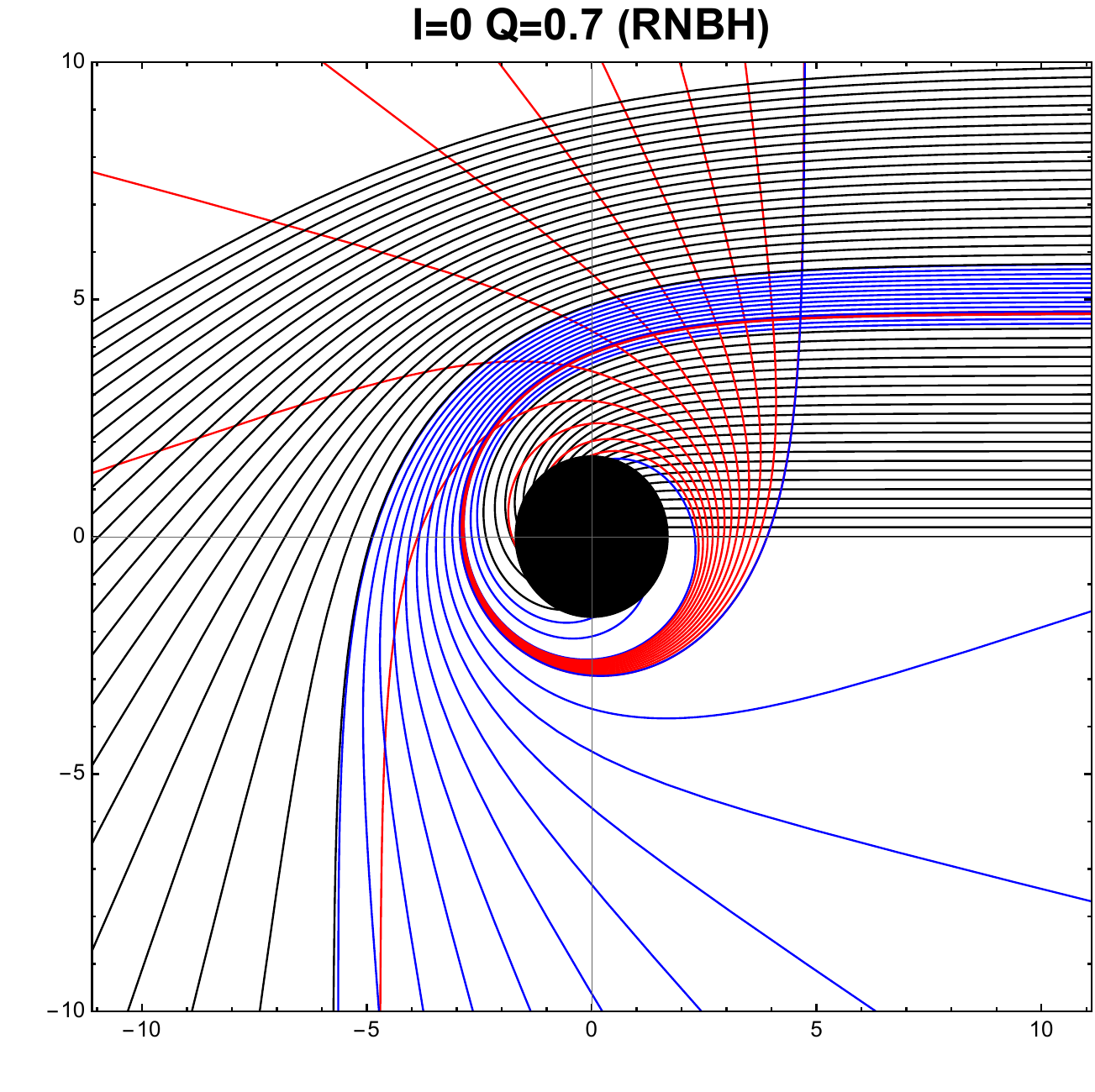}
\includegraphics[width=4.4cm]{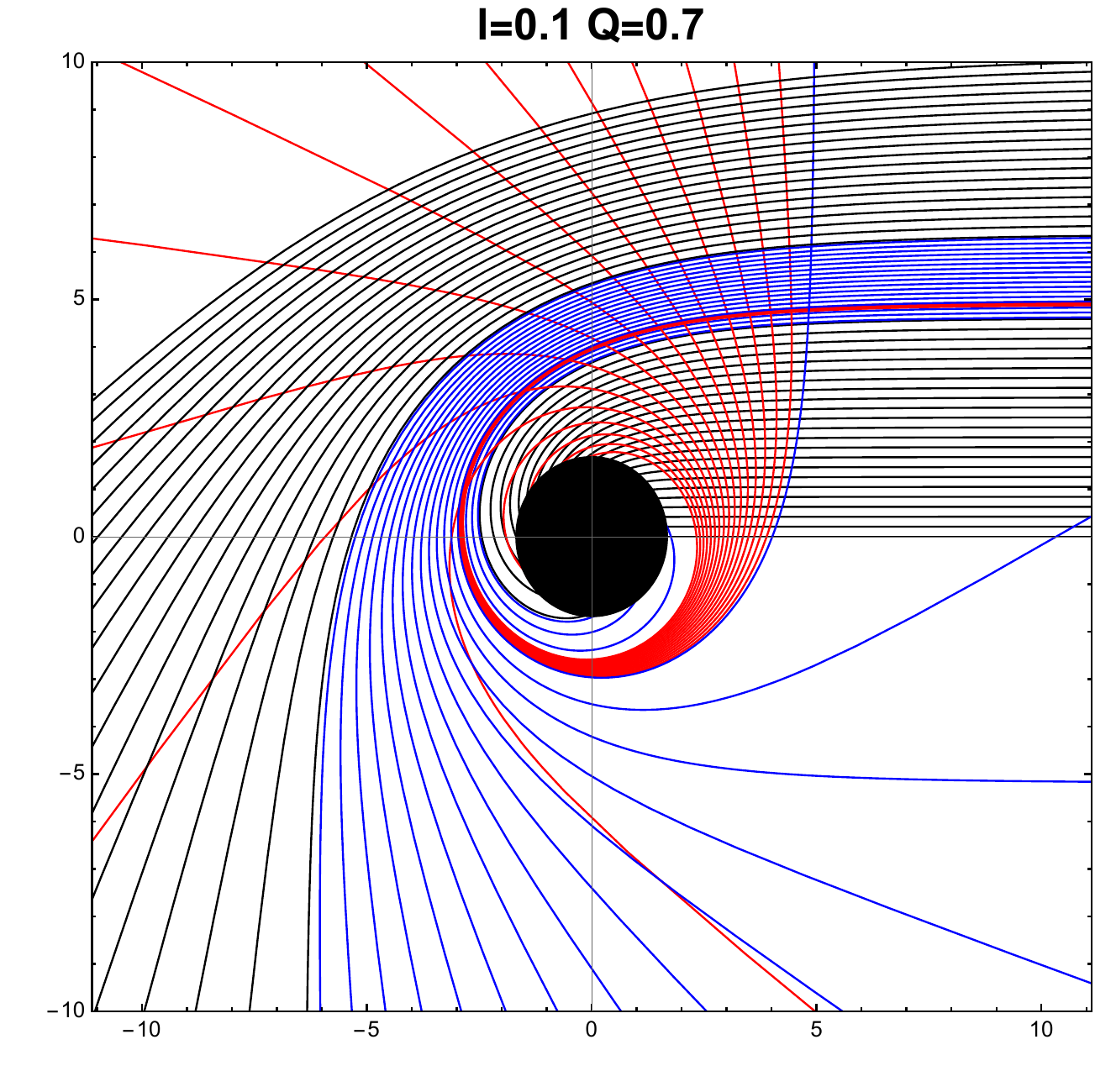}
\includegraphics[width=4.4cm]{T3.pdf}
\includegraphics[width=4.4cm]{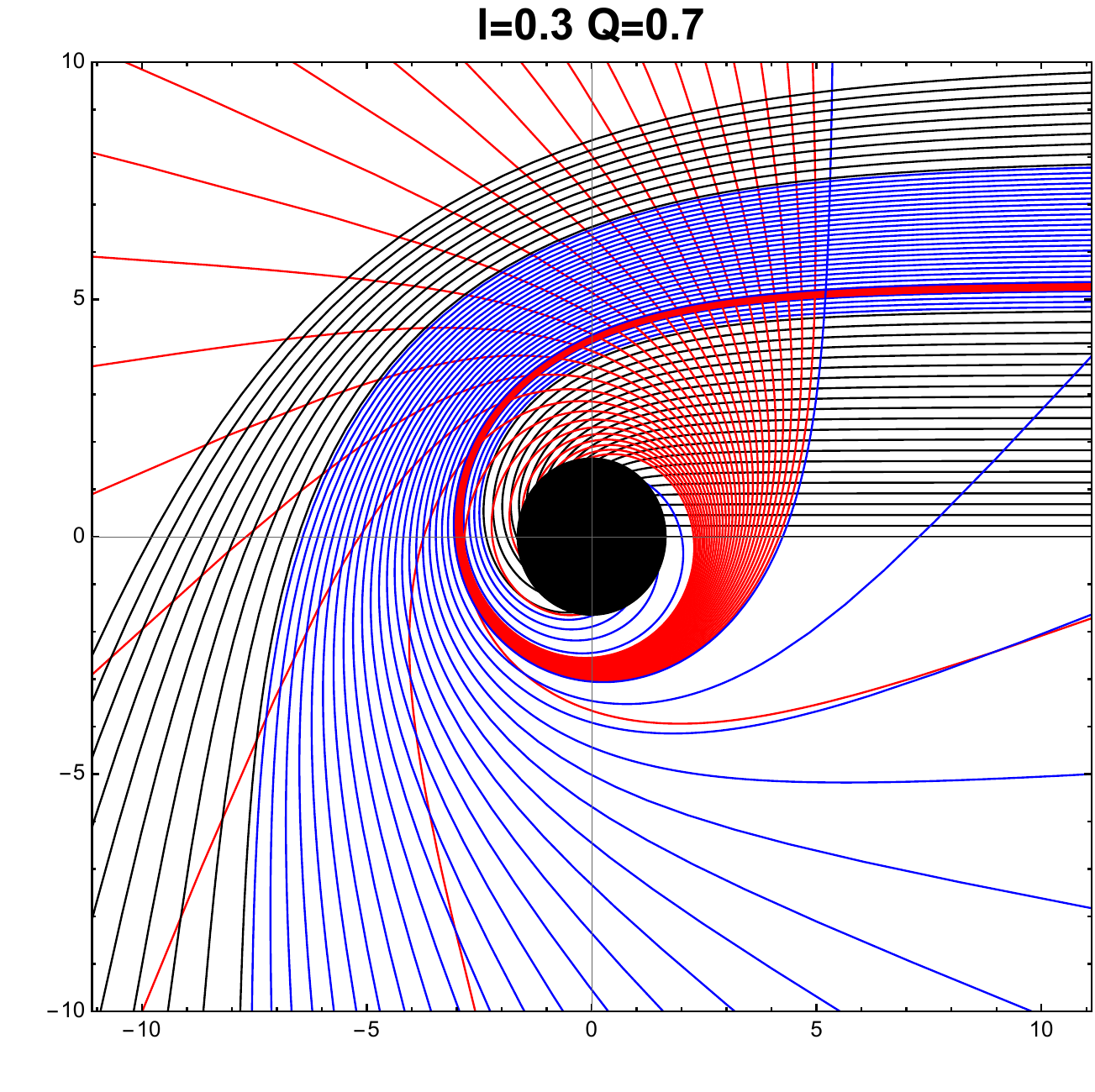}
\caption{The photon trajectories around BCBHs in the polar coordinates ($r$, $\phi$) for different LV parameter $l$, with $Q=0.7$. The black curves represent direct emission, the blue curves represent lensed ring, and the red curves represent photon ring.}
\label{fig413}
\end{figure}

\noindent Table \ref{T411} and Table \ref{T412} present the numerical calculation results of the relevant parameter values for different BH parameters. These results indicate that the ranges of impact parameter corresponding to the lensed ring and photon ring are significantly smaller than the range of impact parameter for direct emission. This suggests that the emission width of the lensed ring and the photon ring around the BCBH is much narrower than the emission width of direct emission.
\begin{table}[h]
\centering
\small
\caption{The boundary values of impact parameter, event horizon radius, photon sphere radius, and BH shadow radius of BCBHs, for $l=0.2$ and different $Q$.}
\begin{tabular}{|c|c|c|c|c|c|c|c|c|}
\hline
\multicolumn{2}{|c|}{BCBHs} & $r_h$ & $r_{ph}$ & $b_{ph}$ & $b_{2}^-$ & $b_{3}^-$ & $b_{3}^+$ & $b_{2}^+$ \\
\hline
\multirow{4}{*}{$l=0.2$} & $Q=0$ (SBH-like) & 2.000 & 3.000 & 5.196 & 4.930 & 5.179 & 5.260 &6.913 \\
\cline{2-9}
& $Q=0.65$ & 1.734 &2.652  & 4.752 & 4.448 &  4.729 & 4.832  & 6.565  \\
\cline{2-9}
& $Q=0.7$ & 1.682 & 2.586 & 4.670 & 4.365 & 4.645 & 4.754 & 6.506 \\
\cline{2-9}
& $Q=0.75$ &1.621  & 2.511  & 4.578  &  4.249 & 4.550  &4.667   &6.441   \\
\hline
\end{tabular}
\label{T411}
\end{table}
\begin{table}[h]
\centering
\caption{The boundary values of impact parameter, event horizon radius, photon sphere radius, and BH shadow radius of BCBHs, for $Q=0.7$ and different $l$.} 
    \begin{tabular}{|c|c|c|c|c|c|c|c|c|}
    \hline
    \multicolumn{2}{|c|}{BCBHs} & $r_h$ & $r_{ph}$ & $b_{ph}$ & $b_{2}^-$ & $b_{3}^-$ & $b_{3}^+$ & $b_{2}^+$ \\
    \hline
    \multirow{4}{*}{$Q=0.7$} &$ l=0$ (RNBH) & 1.714 & 2.626& 4.720 & 4.503 & 4.707 & 4.763 & 5.775 \\
    \cline{2-9}
    & $l=0.1$ & 1.697 & 2.606& 4.694&4.429 & 4.676&4.755 &6.098 \\
    \cline{2-9}
    & $l=0.2$ & 1.682 & 2.586 & 4.670 & 4.356 & 4.645 & 4.754 & 6.506 \\
    \cline{2-9}
    & $l=0.3$ &1.667  &2.568 &4.648 &4.283 &4.615 &4.759 &7.020 \\
    \hline
    \end{tabular}
        \label{T412}
\end{table}
Fig.\ref{fig412} and Fig.\ref{fig413} show the trajectory distributions of the three types of photons emitted by the accretion disk around the BCBHs. As shown in the figures, the isotropic rays emitted by the accretion disk, after intersecting the accretion disk for different numbers of times, can be observed by a distant observer located in the north polar direction. Clearly, the brightness of the three types of rays varies according to the number of intersections with the accretion disk. In the figures, we represent the trajectories of direct emission, lensed ring, and photon ring with black, blue, and red curves, respectively. From the figures, it can be seen that the LV parameter $l$ significantly affects the distribution of the three types of rays. As the value of $l$ increases, the ranges of the lensed ring and photon ring around the BCBHs are significantly expanded.

\subsection{$\text{Optical observational signatures of BCBH}$}

In this section, we simulate the images of the photon rings and shadows around the BCBHs as seen by distant observers under different model parameters. As previously mentioned, isotropic light rays emitted by the accretion disk will deflect as it moves around the BH and intersect with the accretion disk $m$ times ($m = 1, 2, \dots$), depending on the range of the impact parameter $b$. Photons with specific ranges $b$ are received by observers located in the polar direction. Since the rays will pick up additional brightness as they pass through the accretion disk, the total observed specific intensity should be the sum of these intensities \cite{4.1}.

\begin{figure}[H]
\centering
\includegraphics[width=8cm]{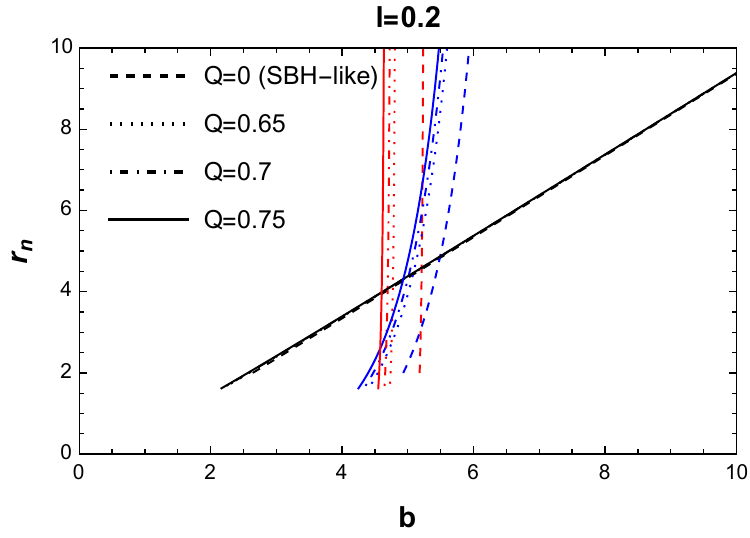}
\includegraphics[width=8cm]{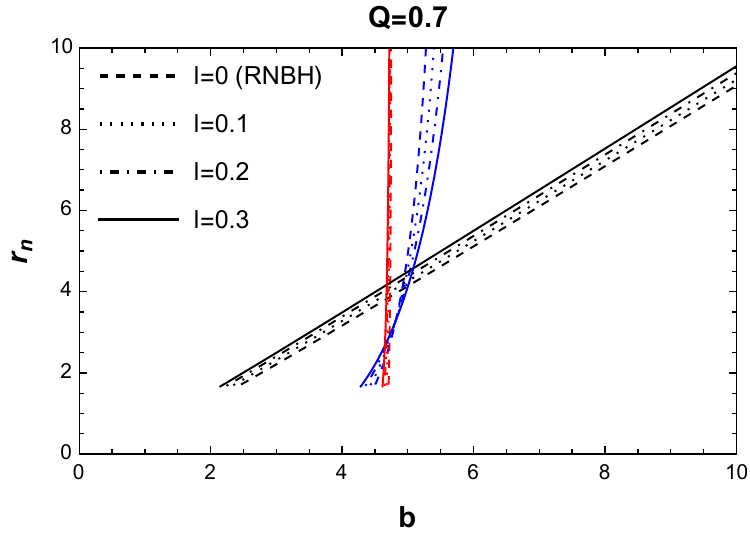}
\caption{For thin accretion disk emission, the first three transfer functions of BCBH spacetime for different BH parameters values. The left panel corresponds to $l=0.2$, and the right panel corresponds to $Q=0.7$. $r_1(b)$corresponds to direct emission, where the slope $\frac{dr}{db}$ is almost 1 (the profile image of direct emission is essentially the profile of the gravitational redshift source); $r_2(b)$ corresponds to lensed ring, where the observer will see the backside of a highly demagnified accretion disk in which demagnification is determined by the slope of the profile; $r_3(b)$ corresponds to the photon ring, when one sees an extremely demagnified image of the front of the accretion disk.}
\label{fig421}
\end{figure}

For isotropic photons emitted by the accretion disk in the rest frame, the specific intensity received by the observer is related to the specific intensity of the emitted photons $I_{em}$ as follows:
\begin{equation}
I_{obs}(r)=g_s^3I_{em}(r),
\label{eq421}
\end{equation}
where $g_s \equiv \frac{\nu_{obs}}{\nu_{em}} = \sqrt{A(r)}$ is the redshift factor, and $\nu_{em}$ and $\nu_{obs}$ represent the emission frequency and the observed frequency of the photons, respectively. Considering the BCBHs, the specific intensity of photons around it is:
\begin{equation}
I_{o b s}(r)=A(r)^{3 / 2} I_{e m}(r)=\left(1-\frac{2M}{r}+\frac{2(1+l)Q^{2}}{(2+l)r^{2}}\right)^{3/2} I_{e m}(r).
\label{eq422}
\end{equation}
By integrating over all the observed frequencies, the total observed specific intensity is:
\begin{equation}
I_{o b s}^{total}(r)=\int I_{o b s}(r) d \nu_{o b s}=\left(1-\frac{2M}{r}+\frac{2(1+l)Q^{2}}{(2+l)r^{2}}\right)^{2} I_{e m}^{total}(r),
\label{eq423}
\end{equation}
where $I_{em}^{total} = \int I_{em}(r) d\nu_{em}$ is the total emitted specific intensity from the accretion disk. Additionally, using the transfer function $r_m(b)$, the total observed intensity $I_{obs}^{total}$ can be written as a function of the impact parameter $b$ :
\begin{equation}
I_{o b s}^{t o t a l}(b)=\left.\sum\left(1-\frac{2M}{r}+\frac{2(1+l)Q^{2}}{(2+l)r^{2}}\right)^{2} I_{e m}^{t o t a l}(r)\right|_{r=r_{m}(b)}.
\label{eq424}
\end{equation}
The transfer function reveals the correlation between the impact parameter $b$ of photons and the radial coordinate $r$ when they intersect the accretion disk for the $m$-th time. Considering different values of the BCBH parameters, we plot the first three transfer functions of the BCBH in Fig.\ref{fig421}. As can be easily seen from the figure, the charge parameter $Q$ has a relatively small effect on the first transfer function (black curve). Within the considered parameter ranges, all the curves almost overlap. In contrast, the LV parameter $l$ has a more significant impact on the first transfer function. For the second transfer function (blue curve), at the same $b$ value, as the charge parameter $Q$ increases, the observed intensity is increasingly demagnified, whereas, as the LV parameter $l$ increases, the observed intensity is less demagnified. Furthermore, the red curves correspond to the transfer functions for photon ring, where its slope tends toward infinity, indicating that the observed intensities are extremely demagnified.

Considering the thin accretion disk as the primary source of illumination for the BH, it is evident that the total observed intensity depends on the radial coordinate $r$ of the emitted from the accretion disk. Therefore, studying the effect of the accretion disk emission profile on the observed appearance of the BH is important. Specifically, we consider three toy models for the emission profile of the accretion disk. With the help of the above mentioned transfer function and Eq.(\ref{eq424}), we analyze the optical appearance of the BCBHs for these accretion disk emission models.

\begin{figure}[H]
\centering
\includegraphics[width=5cm]{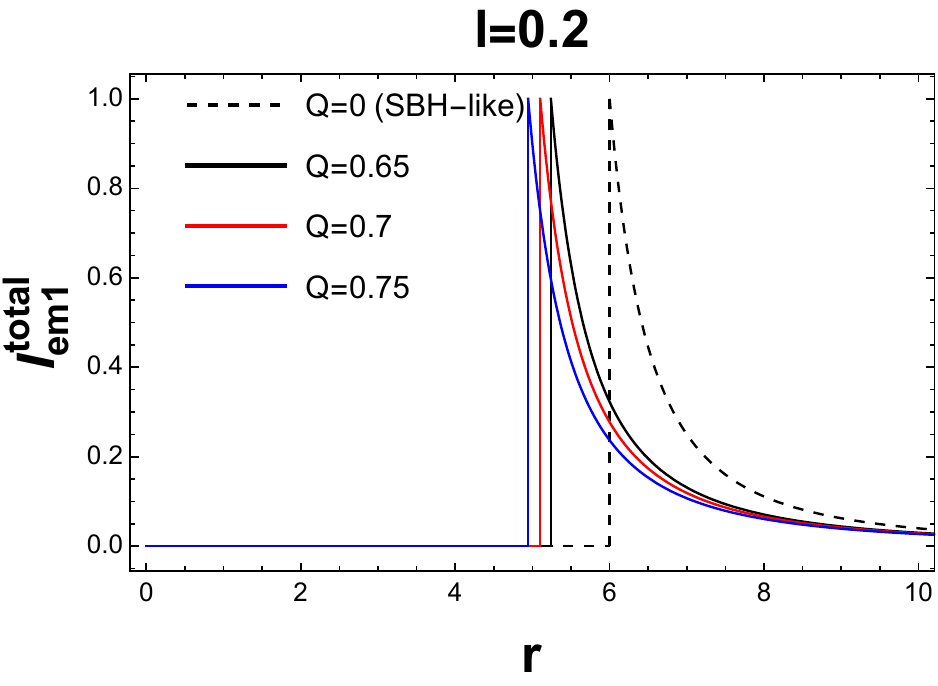}
\includegraphics[width=5cm]{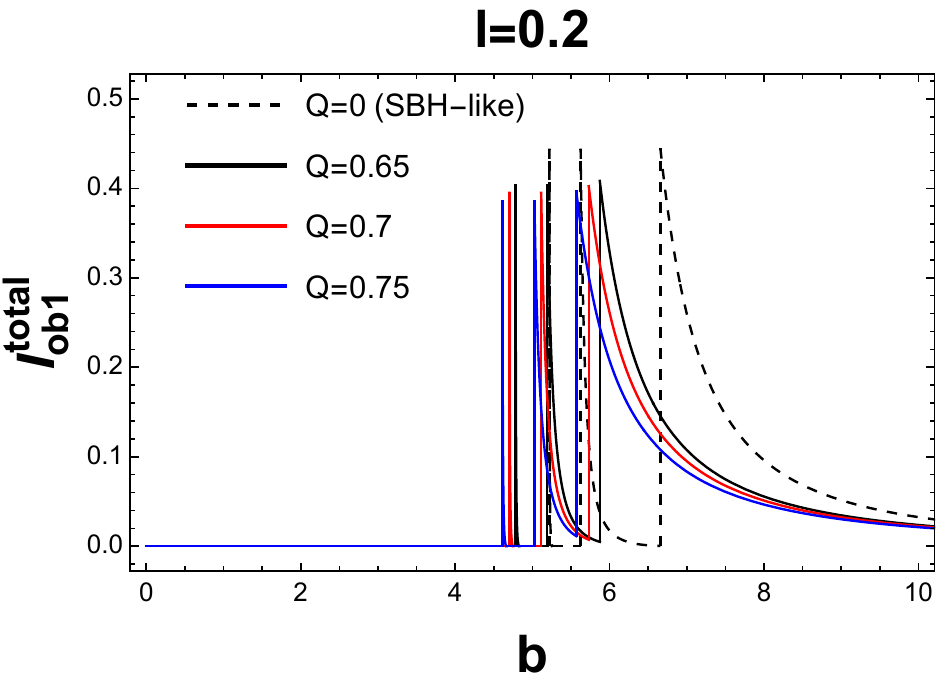}
\includegraphics[width=3.4cm]{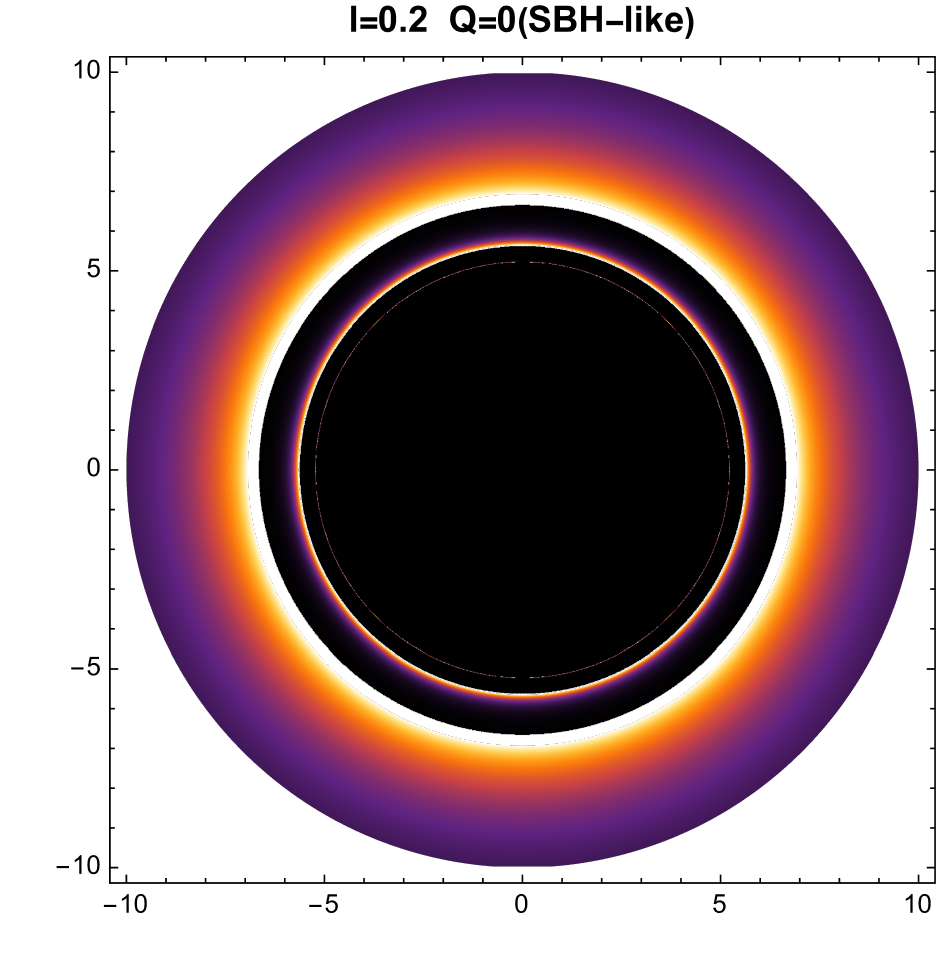}
\includegraphics[width=3.4cm]{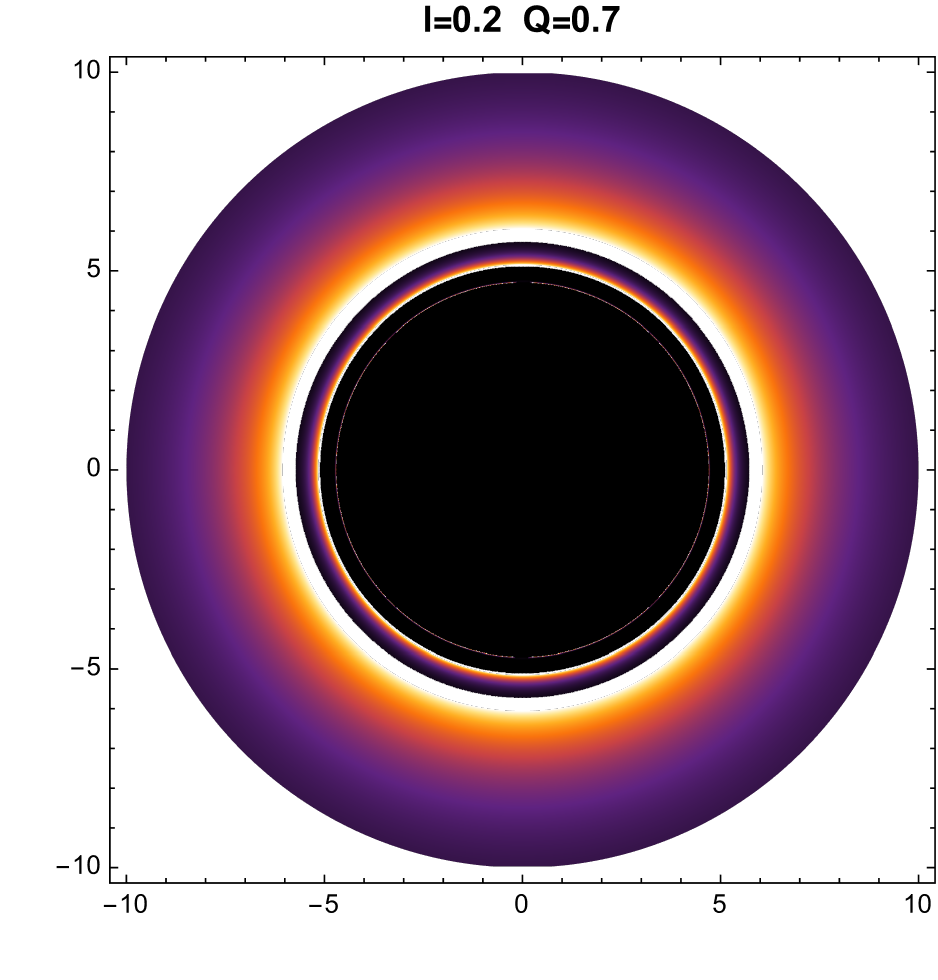}
\caption{In emission model I, the effect of different charge parameter $Q$ on the optical appearance of BCBHs. From left to right: the total emission intensity $I_{em1}^{total}$ of the thin accretion disk; the observed total intensity as a function of the impact parameter $b$; the optical appearance of the BH for $l=0.2$, $Q=0$ (Schwarzschild-like BH); and the optical appearance of the BCBH for $l=0.2$, $Q=0.7$.}
\label{fig422}
\end{figure}

\begin{figure}[H]
\centering
\includegraphics[width=5cm]{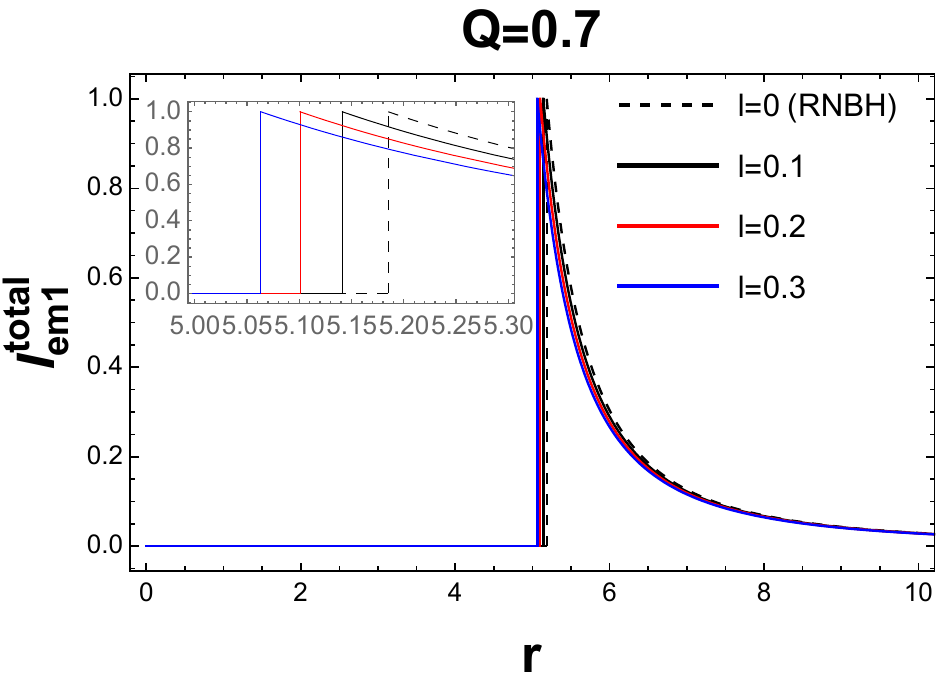}
\includegraphics[width=5cm]{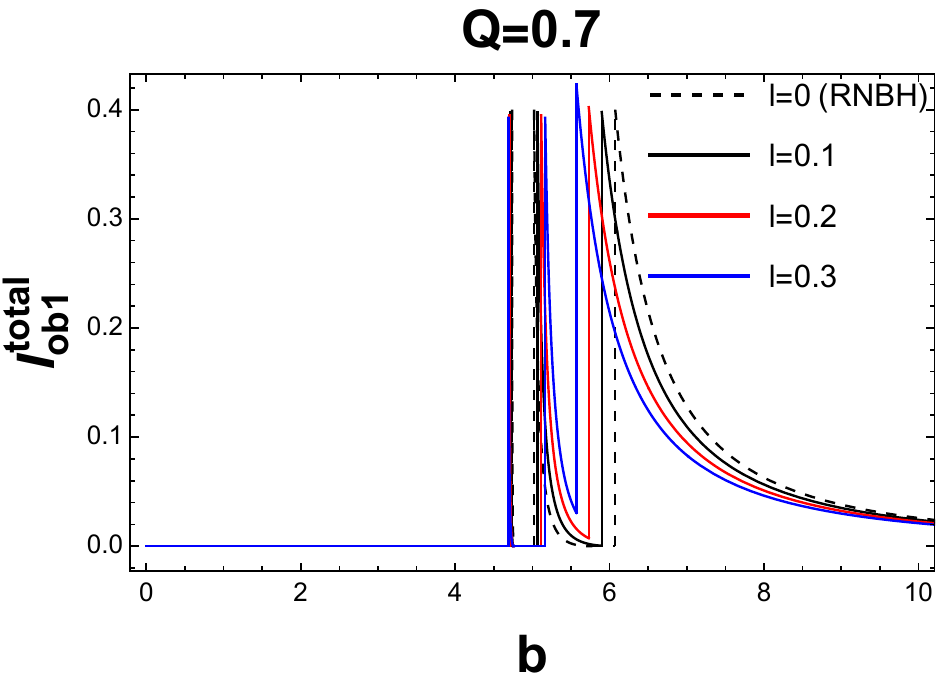}
\includegraphics[width=3.4cm]{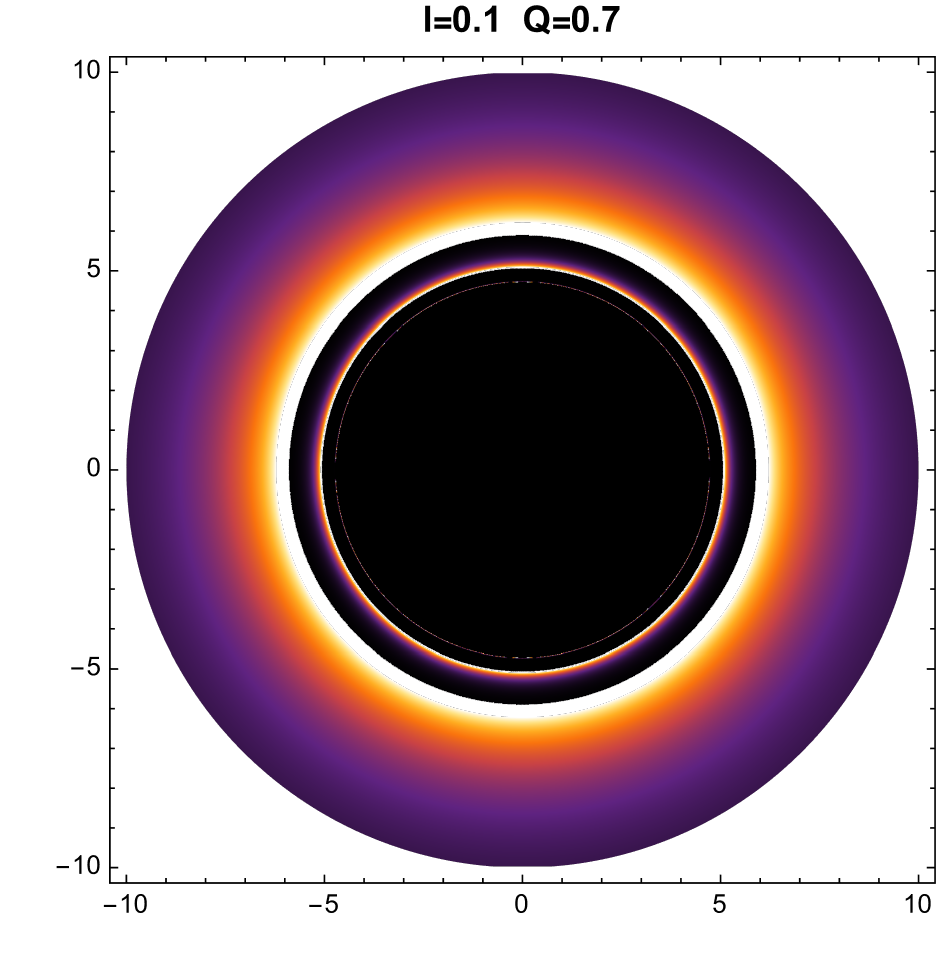}
\includegraphics[width=3.4cm]{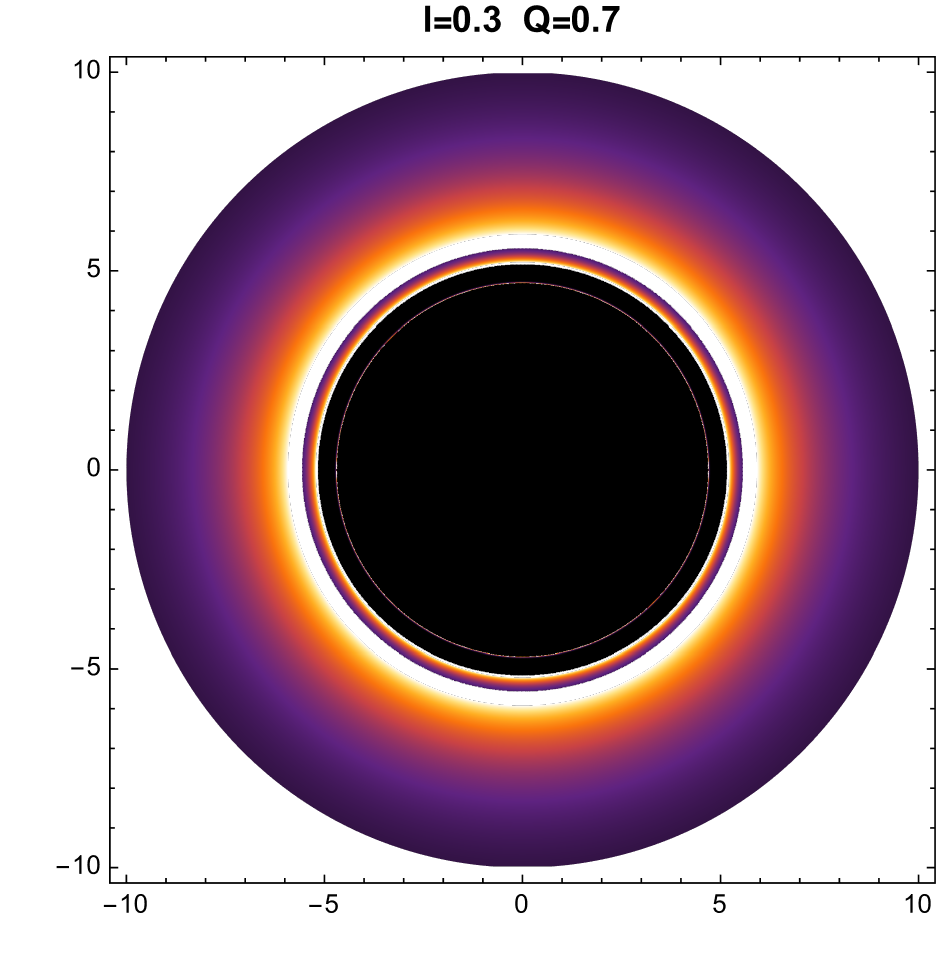}
\caption{In emission model I, the effect of different LV parameter $l$ on the optical appearance of the BCBH. From left to right: the total emission intensity $I_{em1}^{total}$ of the thin accretion disk; the observed total intensity as a function of the impact parameter $b$; the optical appearance of the BCBH for $l=0.1$, $Q=0.7$; and the optical appearance of the BCBH for $l=0.3$, $Q=0.7$.}
\label{fig423}
\end{figure}

It is well known that the ISCO is the boundary between the particles moving around the BH and the particles captured by the BH \cite{2112.11227}. First, we assume that the first emission model for the accretion disk is as follows: the emitted light rays start from the ISCO, and the emission intensity decreases as the radial coordinate increases. The expression for this emission model is given by:
\begin{equation}
I_{em1}^{total}=\left\{\begin{array}{cc}
\left[\frac{1}{r-\left(r_{ISCO}-1\right)}\right]^{2} & \quad r>r_{ISCO}, \\
0 &  \quad r \leq r_{ISCO},
\end{array}\right. 
\label{eq425}
\end{equation}
where ISCO radius of the BCBH can be determined by Eq.(\ref{eq316}). Fig.\ref{fig422} shows the effect of the charge parameter $Q$ on the optical appearance of the BCBHs when the LV parameter is $l=0.2$. From the figure, it is evident that when the parameters are $l=0.2$ and $Q=0.7$, the photon ring peak is distributed within the range 4.701-4.754, while the lensed ring peak is in the range 5.113-5.729. Due to the narrow observation region of the photon ring and lensed ring, direct emission contributes to the main of the observed intensity. As $Q$ increases, we find that the peak of the observed intensity gradually decreases, and the emission region of the lensed ring and photon ring becomes wider. Fig.\ref{fig423} shows the effect of the LV parameter $l$ on the optical appearance of the BCBH when the charge parameter is $Q=0.7$. The observed intensity also shows multiple peaks, with direct emission still dominating.

\begin{figure}[H]
\centering
\includegraphics[width=5cm]{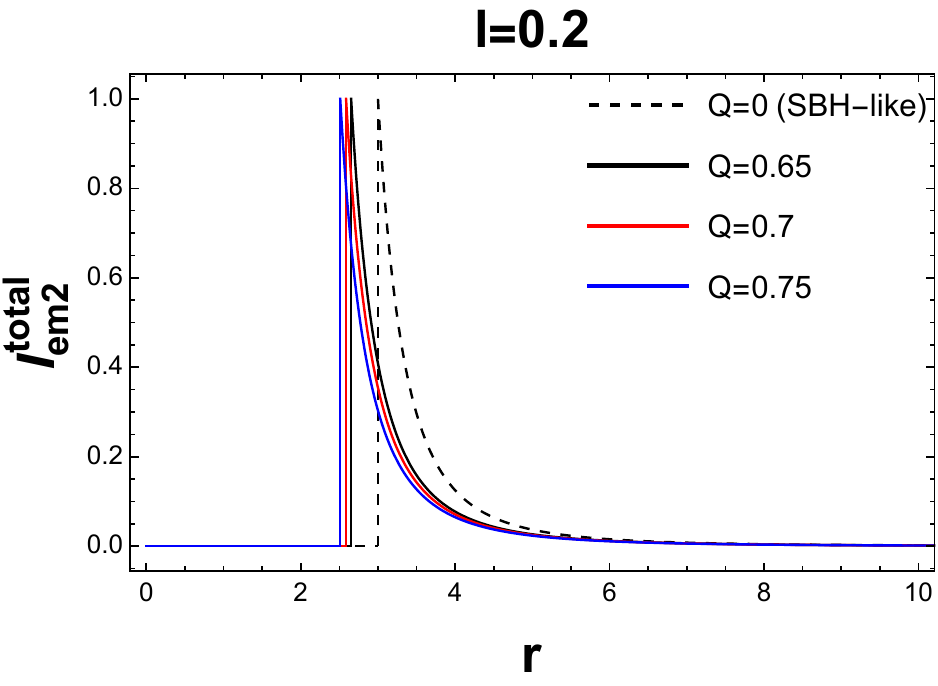}
\includegraphics[width=5cm]{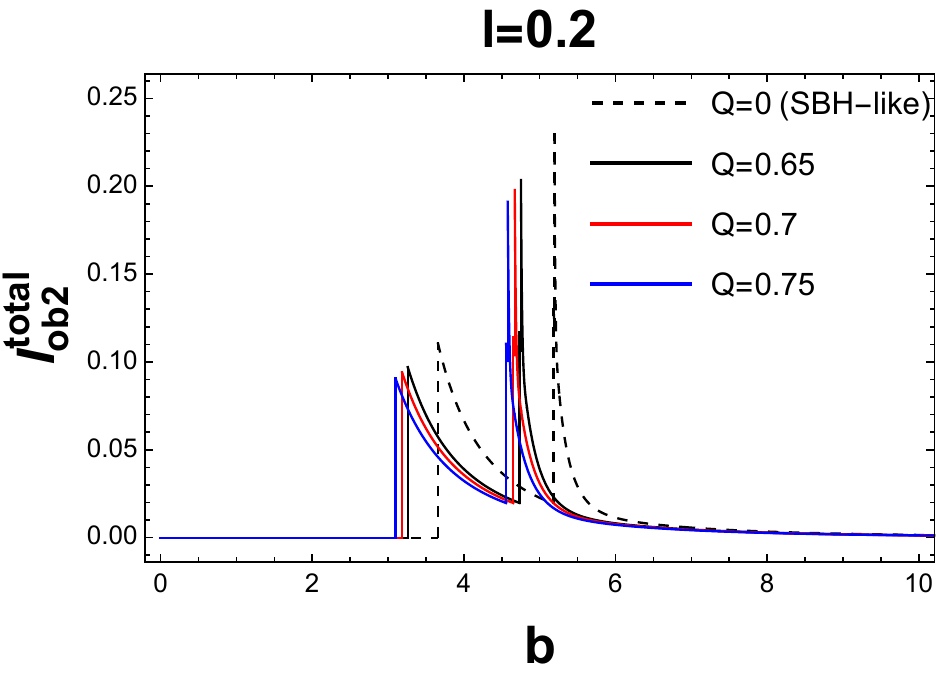}
\includegraphics[width=3.4cm]{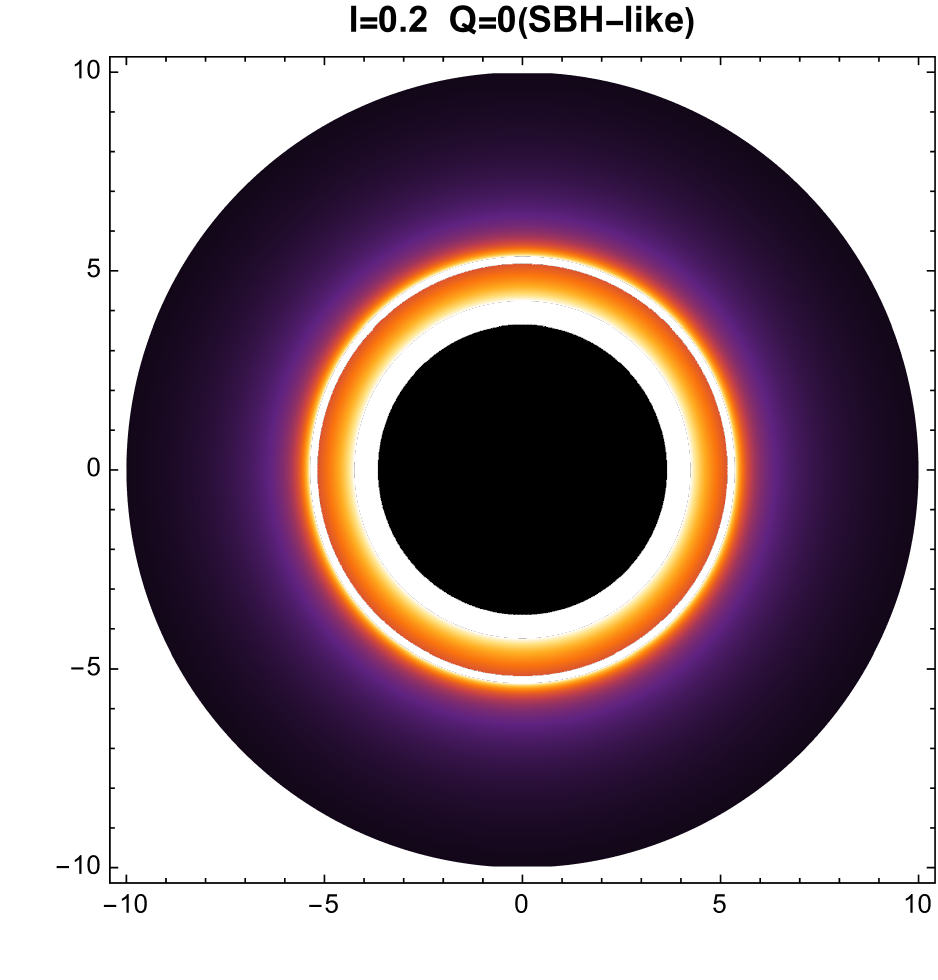}
\includegraphics[width=3.4cm]{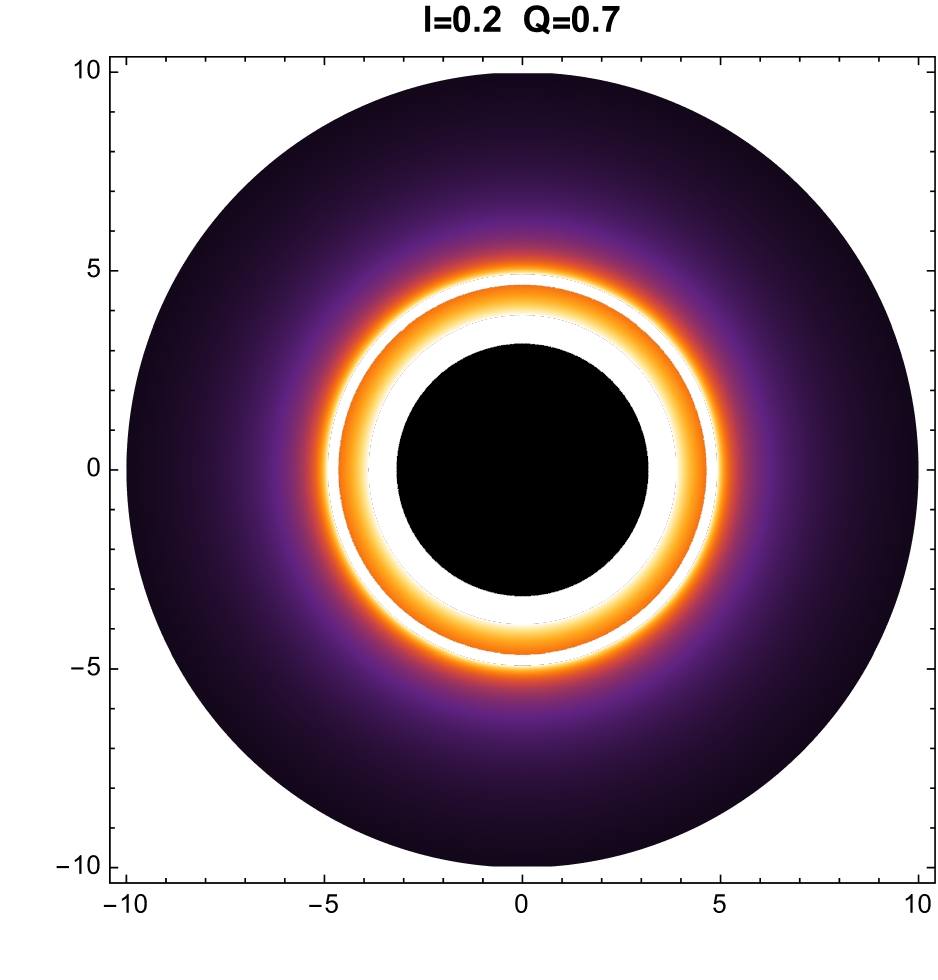}
\caption{In emission model II, the effect of different charge parameter $Q$ on the optical appearance of the BCBH. From left to right: the total emission intensity $I_{em2}^{total}$ of the thin accretion disk; the observed total intensity as a function of the impact parameter $b$; the optical appearance of BH for $l=0.2$, $Q=0$ (Schwarzschild-like BH ); and the optical appearance of BCBH for $l=0.2$, $Q=0.7$.}
\label{fig424}
\end{figure}

\begin{figure}[H]
\centering
\includegraphics[width=5cm]{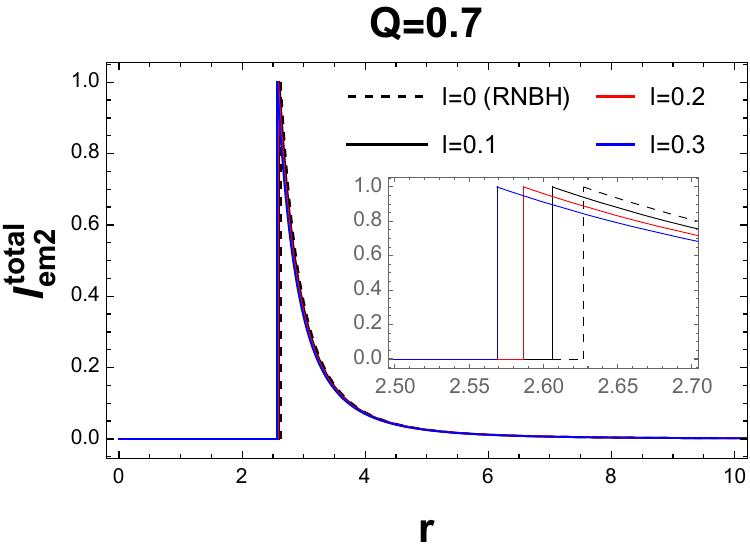}
\includegraphics[width=5cm]{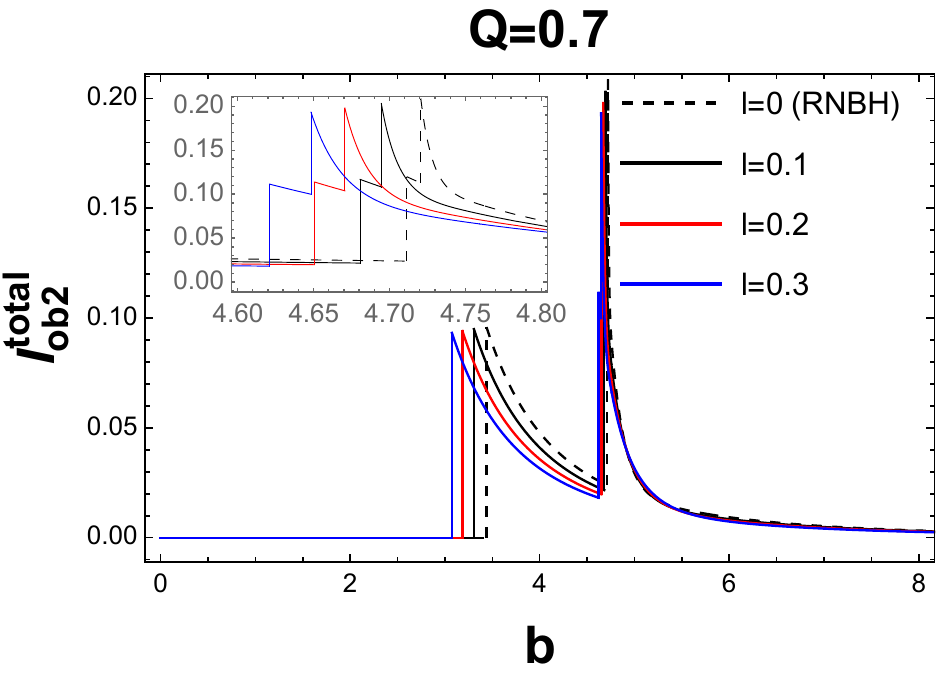}
\includegraphics[width=3.4cm]{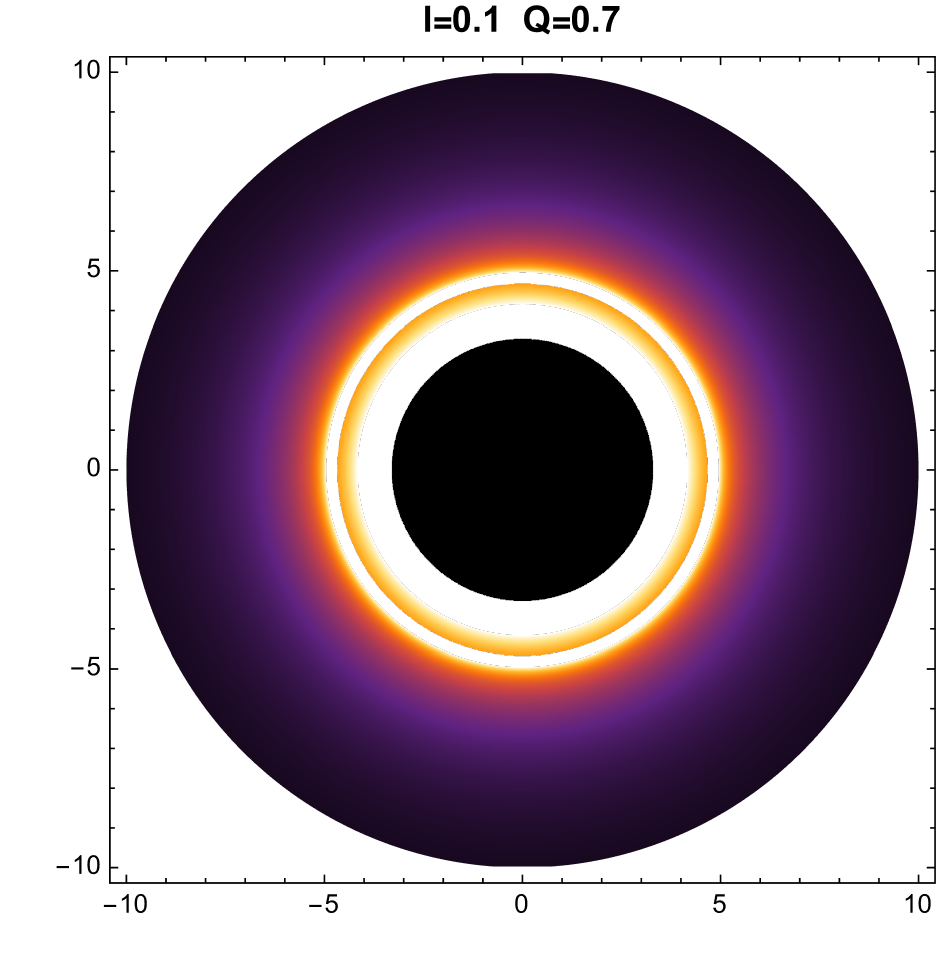}
\includegraphics[width=3.4cm]{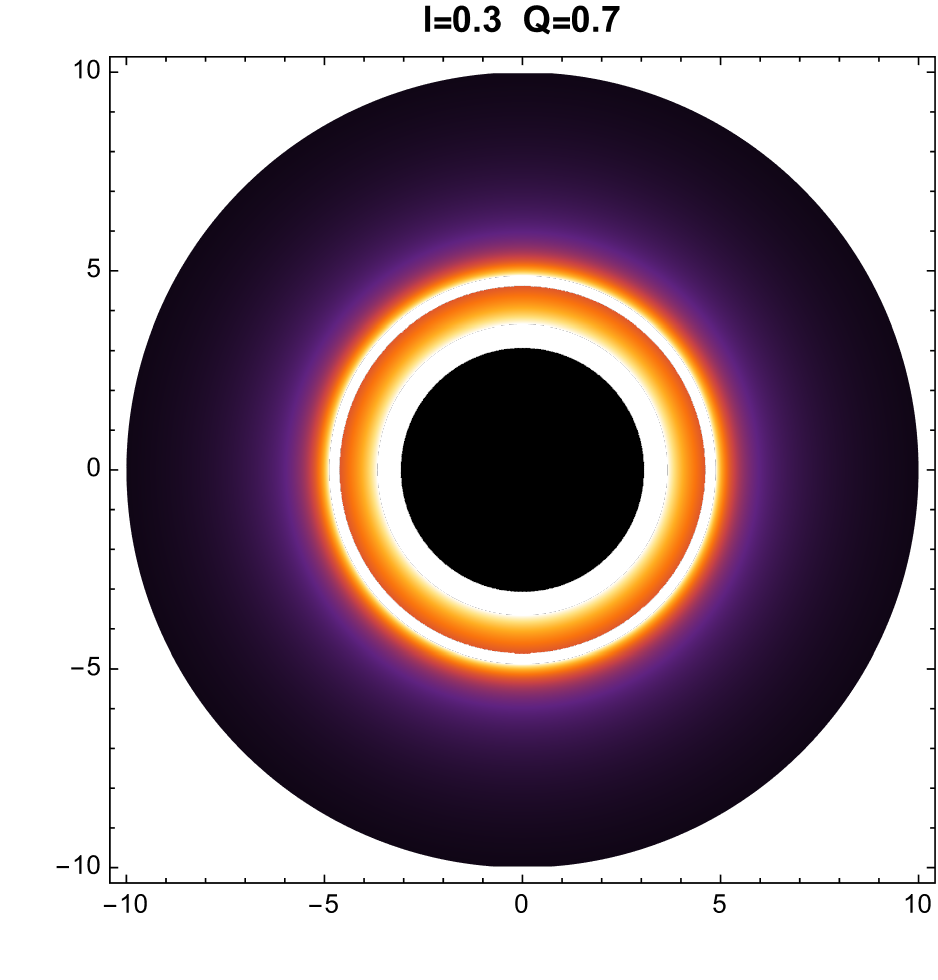}
\caption{In emission model II, the effect of different LV parameters $l$ on the optical appearance of the BCBH. From left to right: the total emission intensity $I_{em2}^{total}$ of the thin accretion disk; the observed total intensity as a function of the impact parameter $b$; the optical appearance of BH for $l=0.1$, $Q=0.7$; and the optical appearance of BCBH for $l=0.3$, $Q=0.7$.}
\label{fig425}
\end{figure}

In the second model, the emission of light is considered to start from the position of radius of the photon sphere, and decreases in the form of a power function as the radial coordinate increases:
\begin{equation}
 I_{e m 2}^{{total }}=\left\{\begin{array}{cc}
{\left[\frac{1}{r-\left(r_{p h}-1\right)}\right]^{3}} &\quad r>r_{p h}, \\
0 &\quad r \leq r_{p h}.
\end{array}\right.
\label{eq426}
\end{equation}
For this model, the effects of the LV parameter and the charge parameter on the optical appearance of the BCBH are shown in Fig.\ref{fig424} and Fig.\ref{fig425}, respectively. The first peak of the observed intensity is caused by direct emission. When $l=0.2$ and $Q=0.7$, the lensed ring exhibits an intensity peak in the range 4.754-6.506, while the photon ring shows a narrower peak in the range 4.670-4.754. These two peaks nearly overlap, leading to a significant increase in the total observed intensity in this region. However, due to the narrow emission regions of both the lensed ring and the photon ring, direct emission still contributes primarily to the optical appearance of the optical image. Further, it is shown that as the LV parameter $l$ or the charge parameter $Q$ increases, the intensity peak gradually decreases.
 
\begin{figure}[H]
\centering
\includegraphics[width=5cm]{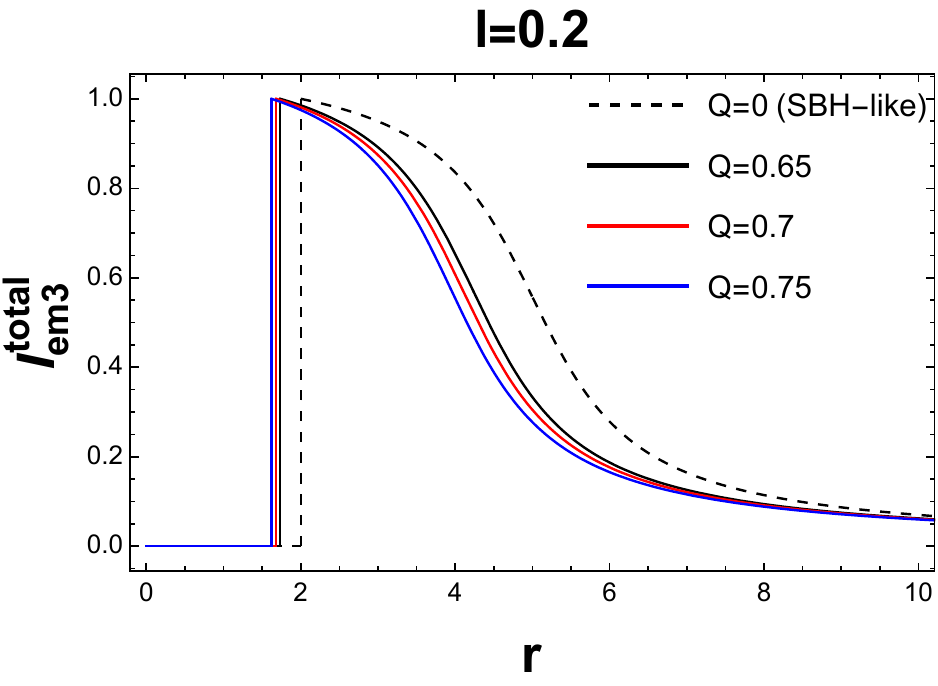}
\includegraphics[width=5cm]{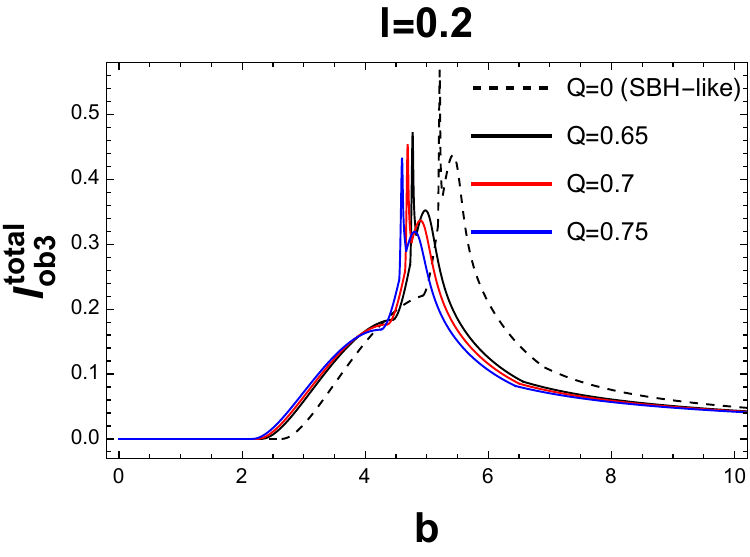}
\includegraphics[width=3.4cm]{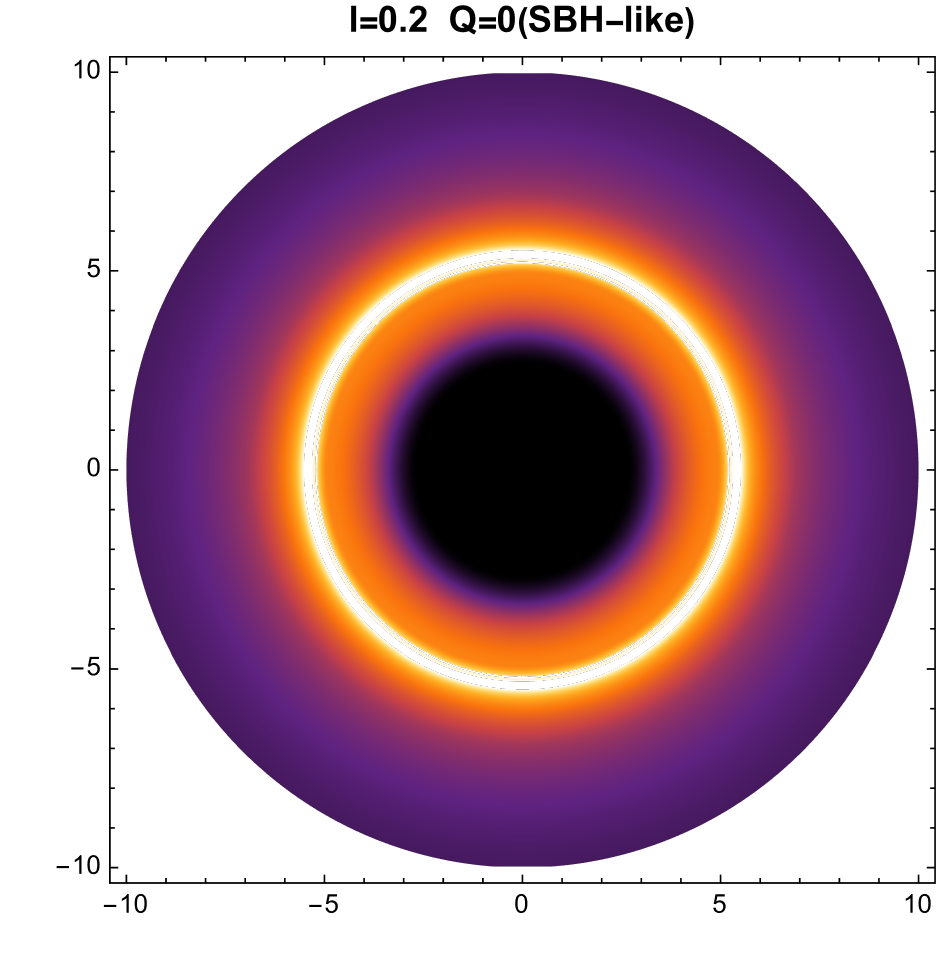}
\includegraphics[width=3.4cm]{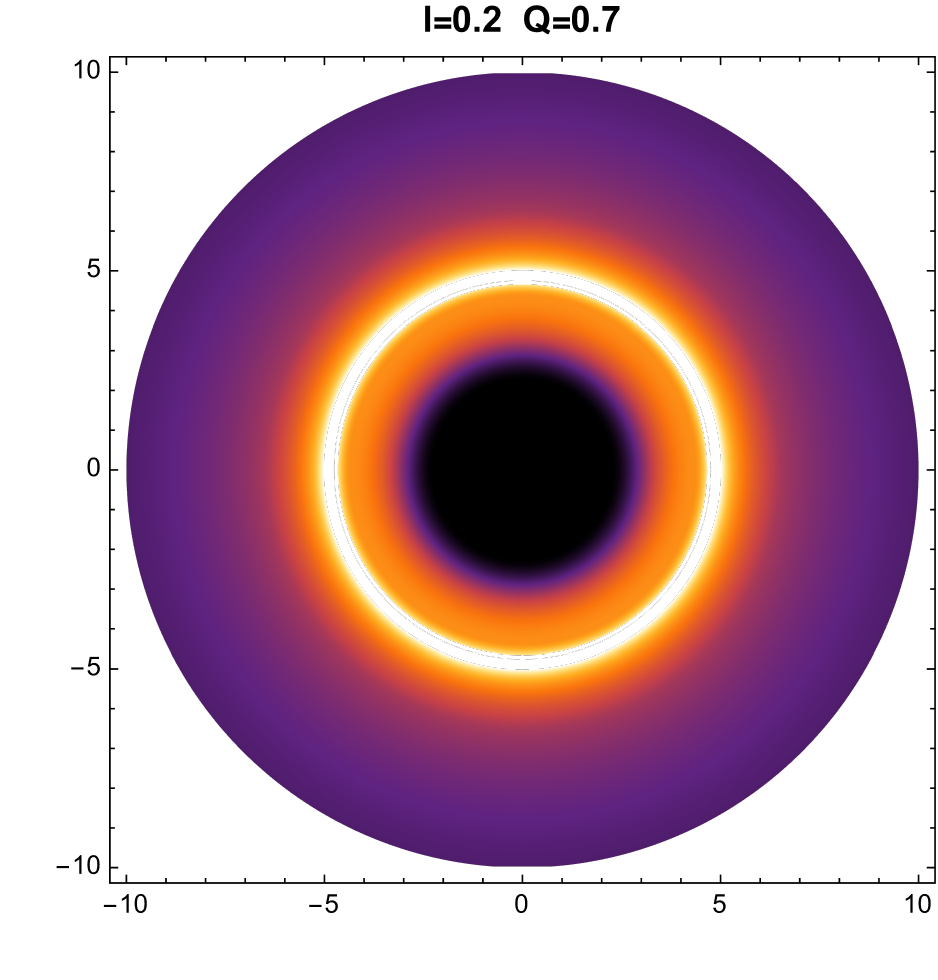}
\caption{In emission model III, the effect of different charge parameter $Q$ on the optical appearance of the BCBH. From left to right: the total emission intensity $I_{em3}^{total}$ of the thin accretion disk; the observed total intensity as a function of the impact parameter $b$; the optical appearance of BH for $l=0.2$, $Q=0$ (Schwarzschild-like BH); and the optical appearance of BCBH for $l=0.2$, $Q=0.7$.}
\label{fig426}
\end{figure}

\begin{figure}[H]
\centering
\includegraphics[width=5cm]{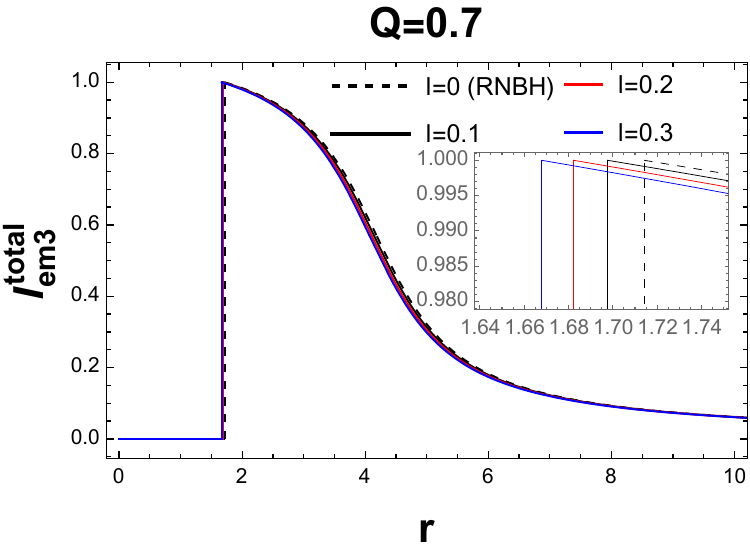}
\includegraphics[width=5cm]{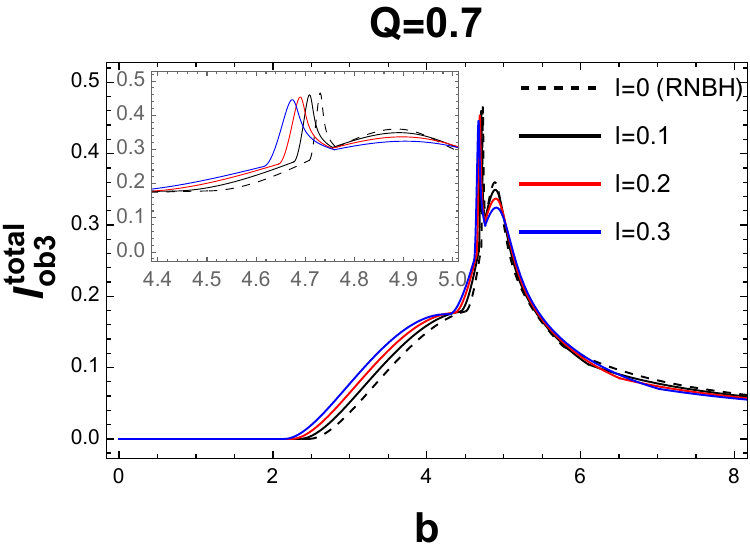}
\includegraphics[width=3.4cm]{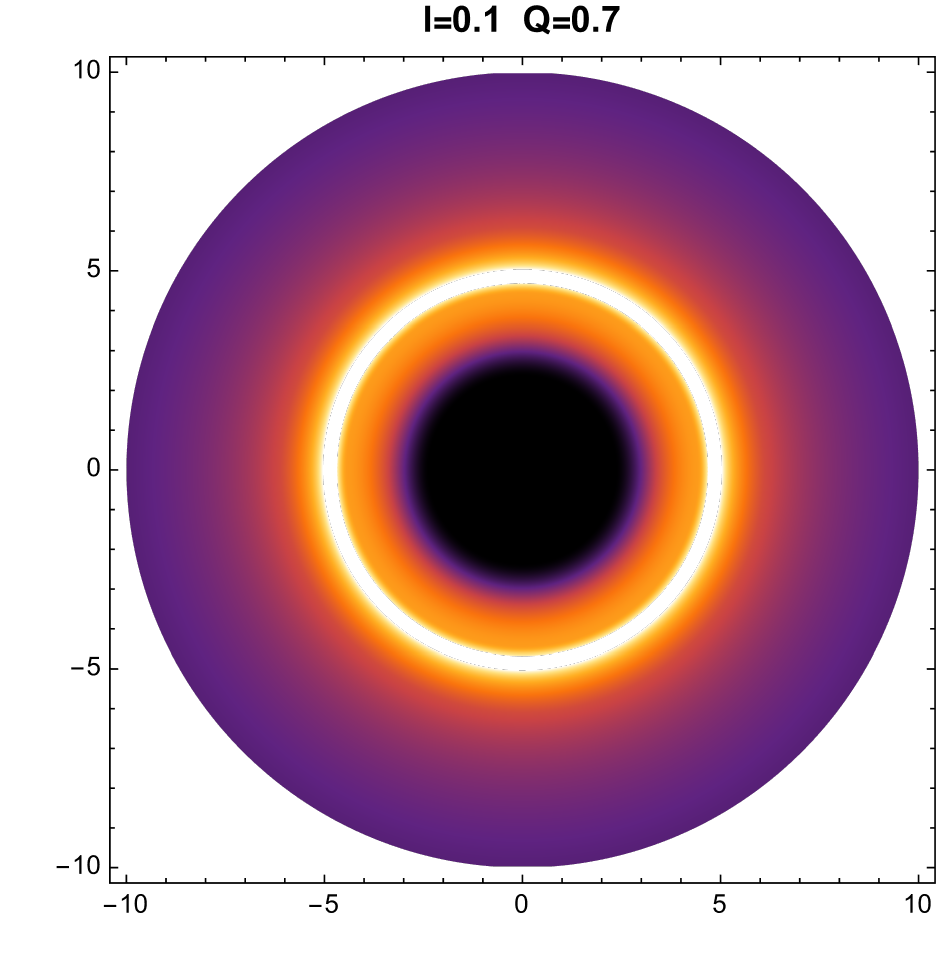}
\includegraphics[width=3.4cm]{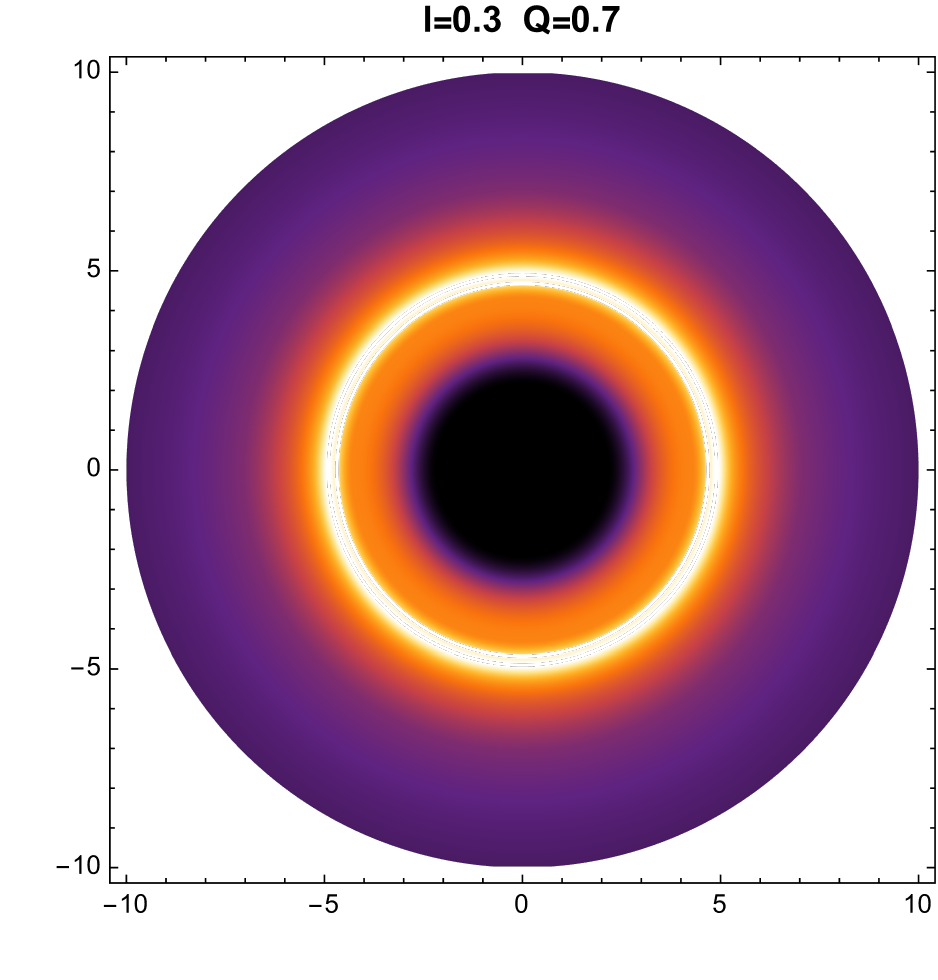}
\caption{In emission model III, the effect of different LV parameter $l$ on the optical appearance of the BCBH. From left to right: the total emission intensity $I_{em3}^{total}$ of the thin accretion disk; the observed total intensity as a function of the impact parameter $b$; the optical appearance of BH for $l=0.1$, $Q=0.7$; and the optical appearance of BCBH for $l=0.3$, $Q=0.7$.}
\label{fig427}
\end{figure}

In the third model, the emission function owns a form:
\begin{equation}
 I_{e m 3}^{{total }}=\left\{\begin{array}{cc}
\frac{\frac{\pi}{2}-\arctan \left[r-\left(r_{I S C O}-1\right)\right]}{\frac{\pi}{2}-\arctan \left[r_{h}-\left(r_{I S C O}-1\right)\right]} &\quad r>r_{h}, \\
0 &\quad r \leq r_{h}.
\end{array}\right.
\label{eq427}
\end{equation}
It is clear that in this model the emission intensity has a peak at the event horizon and decreases at a relatively slow rate with radial coordinate. As shown in Fig.\ref{fig426} and Fig.\ref{fig427}, the observed intensity increases sharply in the photon ring region and reaches a peak. When $l=0.2$ and $Q=0.7$, the lensed ring contributes a significant intensity peak at $b \approx 4.896$, while the observed intensity subsequently decreases with increasing $b$. Analyzing the optical appearance, the bright rings consist of the photon ring, lensed ring, and direct emission together. However, the primary contribution to the observed intensity is still from direct emission, while the contribution from the photon ring is negligible.

In summary, for the BCBHs under thin accretion disk radiation and the three considered disk emission models, the observed intensity is primarily dominated by direct emission, with the contribution of lensed ring to the total flux being small. Although the brightness of the photon ring is very high, its contribution is insignificant due to its high demagnification. It is important to note that the peak of observed intensity does not fully characterize the total observed intensity $I_{obs}^{total}$. One knows, Figs.\ref{fig422}-\ref{fig427} are plotted based on Eq.(\ref{eq424}). According to this equation, $I_{obs}^{total}$ varies with the impact parameter $b$, and is influenced by several factors: the number of times photons intersect the accretion disk (as photons pick up energy at each intersection with the disk), the redshift factor (associated with the metric function $A(r)$), and the total emission intensity (e.g. determined by the initial emission position and the emission profile). All of these factors can be related to the parameters $l$ and $Q$. For the three different emission models considered in this work, Figs.\ref{fig422}-\ref{fig427} show the dependence of the total observed intensity on the parameters $b$ and $Q$ (or $l$), from which the position and magnitude for peak of $I_{obs}^{total}$ can be directly read. For instance, when $l=0.2$, for the same emission model, the observed intensity peak of each photon ring decreases with increasing charge parameter $Q$. Compared to RNBH, as the LV parameter $l$ increases, both the photon ring and lensed ring thickness increase. Moreover, the optical appearance of the BCBH exhibits significant differences for different emission models. This indicates that the emission model and the BH model parameters together determine the characteristics of the BCBH optical appearance.

\section{$\text{Conclusion}$}

In the bumblebee gravity framework, we systematically investigated the geodesic structure and optical appearance of charged BHs. First, we introduced the static spherically symmetric charged BH solutions, which described the not asymptotically flat spacetime. It was shown that as the LV parameter $l$ and the charge parameter $Q$ increase, the event horizon radius of the BCBH decreases and was always smaller than the corresponding values for the RNBH and the Schwarzschild-like BH. By studying the timelike geodesics, we found that the charge parameter $Q$ had a positive effect on the peak of the effective potential, while the LV parameter $l$ weakened the peak. When the BH model parameters were set to $l = 0.2$ and $Q = 0.7$, the circular orbit motion of the particles existed only under the condition of angular momentum $L \gtrsim 3.175$. Additionally, through numerical calculations, we investigated the ISCO radius of massive particles and plotted its variation with respect to the parameters. The results showed that as $l$ and $Q$ increase, the ISCO radius decreases. We also analyzed the conserved quantities of the particle orbit (energy and angular momentum) and the variation of the Keplerian frequency with respect to the radial coordinate.

Next, we made an in-depth study of the null geodesics and calculated the photon sphere radius $r_{ph}$ and the BH shadow radius $b_{ph}$. The results indicated that as the parameters $l$ and $Q$ increase, both $r_{ph}$ and $b_{ph}$ decrease. Furthermore, we constrained the LV parameter $l$ and the charge parameter $Q$ using the shadow radius data released by the EHT. In the numerical calculations presented in this paper, the chosen parameters satisfied these constraint results. Based on the orbital equations, we further explored the photon trajectories around the BCBHs. As the parameters $l$ and $Q$ increase, fewer photons are absorbed by the BCBH, resulting in a smaller BH shadow observed by a distant observer.

Finally, we analyzed the shadow images and rings features around the BCBHs when thin accretion disk with different emission profiles. The study showed that, in the optical appearance of the BCBH, the lensed ring always appeared near the photon ring. Compared to the RNBH, the increase in the LV parameter $l$ resulted in an increase in both the thickness of the photon ring and the lensed ring. However, due to the very narrow range of both the photon ring and the lensed ring, their contribution to the total observed intensity was minimal, meaning the total observed intensity was mainly dominated by direct emission. Furthermore, when the parameter $l = 0.2$, for the same emission model, the peak observed intensity of the photon ring decreases as the charge parameter $Q$ increases, and were always lower than the corresponding values for a Schwarzschild-like BH. Through three different emission models, we found that an increase in the BH model parameters led to a reduction in the BCBH shadow. Therefore, we can distinguish the BCBH from the RNBH and Schwarzschild-like BH based on their optical appearances. In conclusion, this paper explored the optical appearance of BCBHs. These results contribute to the potential for future experiments of BCBHs.

\textbf{\ Acknowledgments }
 The research work is supported by the National Natural Science Foundation of China (12175095,12075109 and 12205133), and supported by  LiaoNing Revitalization Talents Program (XLYC2007047).

\end{document}